\pdfoutput=1
\documentclass[graphicx,amssymb,amsmath,enumerate,11pt]{article}
\usepackage{fancyhdr}
\fancyheadoffset[]{0.1cm}
\usepackage{graphicx}
\usepackage{geometry}
\usepackage{amsfonts, amssymb, amsthm, amstext}
\usepackage[backref]{hyperref}

\usepackage{epsfig}
\usepackage{url}
\usepackage{upgreek}

\graphicspath{{Figures/}}

\newcommand{\bspec}{\mbox{$\upbeta$-spectrum}}
\newcommand{\belec}{\mbox{$\upbeta$-electron}}
\newcommand{\bdec}{\mbox{$\upbeta$-decay}}
\newcommand{\ev}{\mbox{\,eV}}
\newcommand{\evtwo}{\mbox{\,eV$^2$}}
\newcommand{\kr}{\mbox{$\rm ^{83m}Kr$}}
\newcommand{\rb}{\mbox{$\rm ^{83}Rb$}}
\newcommand{\rbkr}{\mbox{$\rm ^{83}Rb/^{83m}Kr$}}
\newcommand{\etal}{{\it et al.}\ }
\newcommand{\prespectrometer}{pre-spec\-tro\-meter}

\newcommand{\nui}{\mbox{$\nu_{\mathrm i}$}}

\newcommand{\mtwonui}{\mbox{$m^2(\nu_{\mathrm i})$}}

\newcommand{\mnui}{\mbox{$m(\nu_{\mathrm i})$}}

\newcommand{\mnue}{\mbox{$m(\nu_{\mathrm e} )$}}

\newcommand{\mtwonue}{\mbox{$m^2(\nu_{\mathrm e} )$}}

\newcommand{\mee}{\mbox{$m_\mathrm{ee}$}}
\newcommand{\me}{\mbox{$m_\mathrm{e}$}}

\newcommand{\mnu}{\mbox{$m_\nu$ }}
\newcommand{\ttwo}{\mbox{$\rm T_2$}}

\newcommand{\rhenium}{\mbox{$\rm ^{187}Re$}}
\newcommand{\etot}{\mbox{$E_{\rm tot}$}}
\newcommand{\detot}{\mbox{$dE_{\rm tot}$}}

\newcommand{\gfermi}{\mbox{$G_{\rm F}$}}
\newcommand{\mlep}{\mbox{$M_{\rm lep}$}}
\newcommand{\mhad}{\mbox{$M_{\rm nucl}$}}

\newcommand{\ezero}{\mbox{$E_0$}}

\newcommand{\mtwolep}{\mbox{$|M^2_{\rm lep}|$}}
\newcommand{\mtwohad}{\mbox{$|M^2_{\rm nucl}|$}}

\newcommand{\vj}{\mbox{$V_{\rm j}$}}
\newcommand{\pj}{\mbox{$P_{\rm j}$}}
\newcommand{\erec}{\mbox{$E_{\rm rec}$}}
\newcommand{\be}{\begin{equation}}
\newcommand{\ee}{\end{equation}}
\newcommand{\bea}{\begin{eqnarray}}
\newcommand{\eea}{\end{eqnarray}}

\begin{document}      

\rhead{\bfseries Current Direct Neutrino Mass Experiments}

\title{\bf Current Direct Neutrino Mass Experiments}

\author{
G.~Drexlin$^{a1}$, V.~Hannen$^b$, S.~Mertens$^a$ and C.~Weinheimer$^b$\\[2mm]
{\small $^a$Institut f\"ur Experimentelle Kernphysik, Karlsruhe Institute of Technology, 76021 Karlsruhe, Germany}\\[2mm]
{\small $^b$Institut f\"ur Kernphysik, Westf\"alische Wilhelms-Universit\"at M\"unster, 48149 M\"unster, Germany}\\[2mm]
{\small $^1$Corresponding author, e-mail: guido.drexlin@kit.edu}
}

\maketitle

\begin{abstract}

In this contribution we review the status and perspectives of direct neutrino mass experiments. These experiments investigate the kinematics of $\beta$-decays of specific isotopes ($^3$H, $^{187}$Re, $^{163}$Ho) to derive model-independent information on the averaged electron (anti-) neutrino mass, which is formed by the incoherent sum of the neutrino mass eigenstates contributing to the electron neutrino. We first review the kinematics of $\beta$-decay and the determination of the neutrino mass, before giving a brief overview of past neutrino mass measurements (SN1987a-ToF studies, Mainz and Troitsk experiments for $^3$H, cryo-bolometers for $^{187}$Re). We then describe the Karlsruhe Tritium Neutrino (KATRIN) experiment which is currently under construction at Karlsruhe Institute of Technology. The large-scale setup will use the MAC-E-Filter principle pioneered earlier to push the sensitivity down to a value of 200~meV~(90\%~C.L.). KATRIN faces
many technological challenges that have to be resolved with regard to source intensity and stability, as well as precision energy analysis and low background rate close to the kinematic endpoint of tritium $\beta$-decay at 18.6~keV. We then review new experimental approaches such as the MARE, ECHO and Project8 experiments, which offer the promise to perform an independent measurement of the neutrino mass in the sub-eV region. This variety of methods and the novel technologies developed in all present and future experiments demonstrate the great potential of direct neutrino mass experiments in providing vital information on the absolute mass scale of neutrinos.

\end{abstract}

\section{Introduction}
\begin{figure}[t!]
\includegraphics[width=\textwidth]{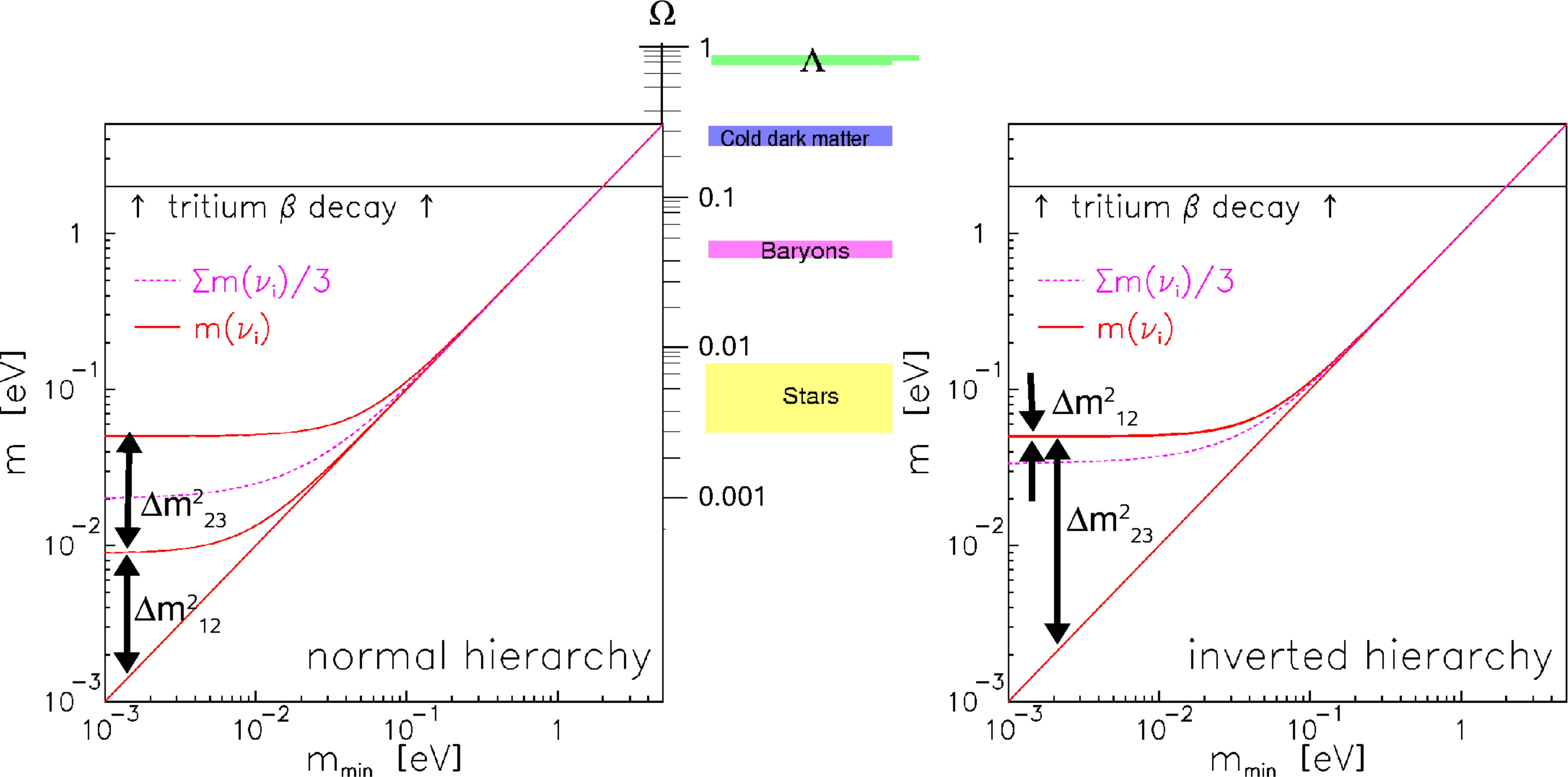}
\caption{Neutrino mass eigenvalues
 \mnui\ (solid lines) and one third of the cosmologically relevant sum of the three neutrino mass eigenvalues $\sum m(\nu_i) /3$ (dashed line)
as a function of the smallest neutrino mass eigenvalue $m_{\rm min}$ for normal hierarchy $m(\nu_3) > m(\nu_2) > m(\nu_1) = m_{\rm min}$
(left) and inverted hierarchy  $m(\nu_2) > m(\nu_1) > m(\nu_3) = m_{\rm min}$ (right).
The upper limit from the tritium \bdec\ experiments at Mainz and Troitsk
on \mnue\  (solid line), which holds in the degenerate neutrino mass region for each
\mnui , and for $\sum \mnui /3$ (dashed line) is also marked.
We plot here the third of the sum of the neutrino mass eigenstates because it coincides with the
mass \mnui\ of the individual neutrino mass states in the case of quasi-degenerate neutrino masses
(for $\mnui > 0.1$~eV).
The temperature of the cosmic microwave background photons together with the different decoupling times of the relic photons and the relic neutrinos after the big bang yields a relic neutrino density of 336/cm$^3$ \cite{03-cosmos}. Using this number
the hot dark matter contribution $\Omega_\nu$ of neutrinos to the matter/energy density of the universe
relates directly to the average neutrino mass $\sum \mnui /3$. This hot dark matter component $\Omega_\nu$ is indicated by the right scale
of the normal hierarchy plot and compared to all other known
matter/energy contributions in the universe (middle).
Thus, the laboratory neutrino mass limit from tritium \bdec\ $\mnue < 2 \ev$
 corresponds to a maximum allowed neutrino matter contribution
in the universe of
$\Omega_\nu < 0.12$.}
\label{fig-03:nu_scenarios}
\end{figure}

The various experiments with atmospheric, solar, accelerator and reactor neutrinos \cite{03-atmos, 03-solar, 03-SBL, 03-LBL, 03-reactor} provide compelling evidence that neutrino flavor states are non-trivial superpositions of neutrino mass eigenstates and that neutrinos oscillate from one flavor state into another during flight. By these neutrino oscillation experiments we can determine the neutrino mixing angles and the differences between the squares of neutrino masses. In the case of the so-called {\it solar or reactor mass splitting} $\Delta m^2_{\rm 12}$ we not only know the modulus of this difference but also its sign.
Clearly these findings prove that neutrinos have non-zero masses, but neutrino oscillation experiments being a kind of {\it interference experiment}
cannot determine absolute masses. We may parameterize our missing knowledge by a free parameter $m_{\rm min}$, the mass of the smallest
neutrino mass eigenstate (see fig. \ref{fig-03:nu_scenarios}).

We should note that throughout this paper we will not distinguish between the mass of a neutrino and the mass of an antineutrino, which should be the same if the CPT theorem holds. Therefore, we will use the term neutrino when we speak of neutrinos and of antineutrinos. But we will explain for each measurement whether the result is obtained for neutrinos or antineutrinos.

The absolute value of the neutrino masses is very important for astrophysics and cosmology because of the role of neutrinos in structure formation due to the
huge abundance of {\it relic neutrinos} left over in the universe from the big bang (336/cm$^3$) \cite{03-cosmos}.
In addition, the key role of neutrino masses in understanding, which of the possible extensions or new theories beyond the Standard Model of particle physics is the right one \cite{03-nature_neutrinos, 03-origin_mass},
makes the quest for the absolute value of the neutrino mass among of the most urgent questions of nuclear and particle physics.

Three different approaches can lead to the absolute neutrino mass scale:
\begin{itemize}
\item Cosmology\\
Today's visible structure of the universe has been formed out of fluctuations of the very early universe.  Due to the large abundance of relic neutrinos and their low masses they acted as hot dark matter: Neutrinos have smeared out fluctuations
at small scales. How small or large these scales are is described by the free streaming length of the neutrinos which depends on their mass. By determining the early fluctuations imprinted on the cosmic microwave background with the WMAP satellite \cite{03-wmap}
and mapping out today's structure of the universe by large galaxy surveys like SDSS \cite{03-sdss} conclusions on the sum of the neutrino masses $\sum \mnui$ can be drawn.
Up to now, only upper limits on the sum of the neutrino masses have been obtained around
  $\sum \mnui < 0.5~\ev$ \cite{03-hannestad12}, which are to some extent
model- and analysis dependent \cite{03-cosmos}.
\item Neutrinoless double \bdec ($0\nu\beta\beta$)\\
  A neutrinoless double {\bdec} (two \bdec s in the same nucleus at the same time with emission of two \belec s (positrons) while the antineutrino (neutrino)
  emitted at one vertex is absorbed at the other vertex as a neutrino (antineutrino))
  is forbidden in the Standard Model of particle physics. It could exist, if the neutrino is its own antiparticle (``Majorana-neutrino'' in
  contrast to ``Dirac-neutrino'') \cite{03-dbd}. Furthermore, a finite neutrino mass is the most natural explanation
  to produce in the chirality-selective interaction a neutrino with a small component of opposite handedness on
  which this neutrino exchange subsists. Then the decay rate will scale with the absolute square of the so-called effective Majorana neutrino mass,
  which takes into account the neutrino mixing matrix $U$:
  \be\ \label{eq-03:mee}
    \Gamma_{0\nu\beta\beta} \propto \left| \sum U^2_{\rm ei} \mnui \right|^2 := \mee^2
  \ee
Here \mee\ represents the sum of the neutrino masses \mnui  contribution coherently to the $0\nu\beta\beta$-decay. Hence this coherent sum carries their relative phases
  (the usual CP-violating phase of an unitary $3 \times 3$ mixing matrix plus two so-called Majorana-phases). A significant additional uncertainty which enters the relation of \mee\ and the decay rate is the nuclear matrix element of the neutrinoless double \bdec\
  \cite{03-dbd} .
  There is one claim for evidence at $\mee \approx 0.3 ~\ev$ by part of the Heidelberg-Moscow collaboration \cite{03-klapdor06}, which is being challenged
  by limits from different experiments in the same range, e.g. very recently by the EXO-200 experiment \cite{03-exo200_2012}.
\item Direct neutrino mass determination\\
  The direct neutrino mass determination is based purely on kinematics without further assumptions. Essentially, the neutrino mass is    determined by using the relativistic energy-momentum-relationship $E^2 = p^2 + m^2$.
  Therefore it is sensitive to the neutrino mass squared $m^2(\nu)$.
  In principle there are two methods: time-of-flight measurements and precision investigations of weak decays.
  The former requires very long baselines and therefore
  very strong sources, which only cataclysmic astrophysical events like a core-collapse supernova could
  provide. The supernova explosion SN1987a in the Large Magellanic Cloud gave limits of 5.7~\ev\  (95~\%~C.L.) \cite{03-loredo02} or
  of 5.8~\ev\ (95~\%~C.L.) \cite{03-pagliaroli10} on the neutrino mass depending somewhat on the underlying supernova model.
  Unfortunately nearby supernova explosions are too rare
  and seem to be not well enough understood to allow to compete with the laboratory direct neutrino mass experiments.

  Therefore, aiming for this sensitivity, the investigation of the kinematics of weak decays and more explicitly the
  investigation of the endpoint region of a \bdec\ spectrum (or an electron capture) is
  still the most sensitive model-independent and direct method
  to determine the neutrino mass.
  Here the neutrino is not observed but the charged decay products are precisely measured. Using energy and momentum conservation
  the neutrino mass can be obtained. In the case of the investigation of a \bspec\ usually the ``average electron neutrino mass''
  \mnue\ is determined (see equation (\ref{eq-03:definition_mnue}) in the next subsection):
  \be\ \label{03-eq:define_mnue}
    \mnue^2 := \sum |U^2_{\rm ei}| \mnui^2
  \ee
In contrast to \mee\ in neutrinoless double \bdec\ (see equation (\ref{eq-03:mee})), this sum averages over all neutrino mass states \mnui\ contributing to the electron neutrino and no  phases of the neutrino mixing matrix $U$ enter. The decay into the different neutrino mass eigenstates \nui\ add incoherently, which we will discuss in more detail for the neutrino mixture to sterile neutrinos in section \ref{sec-03:allowed_transitions}.
\end{itemize}

\begin{figure}[tb]
\centerline{\includegraphics[width=0.49\textwidth]{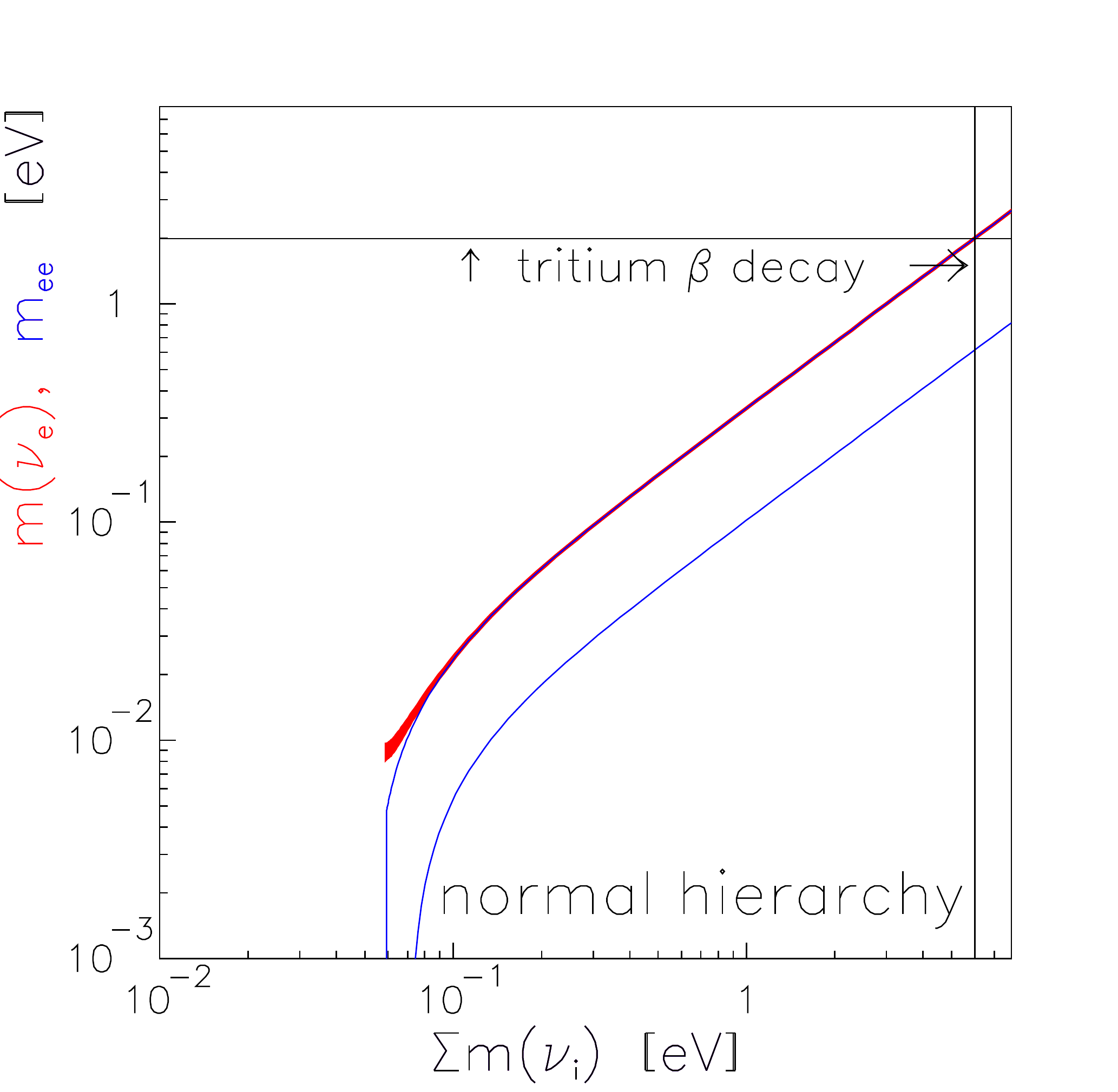}
\includegraphics[width=0.49\textwidth]{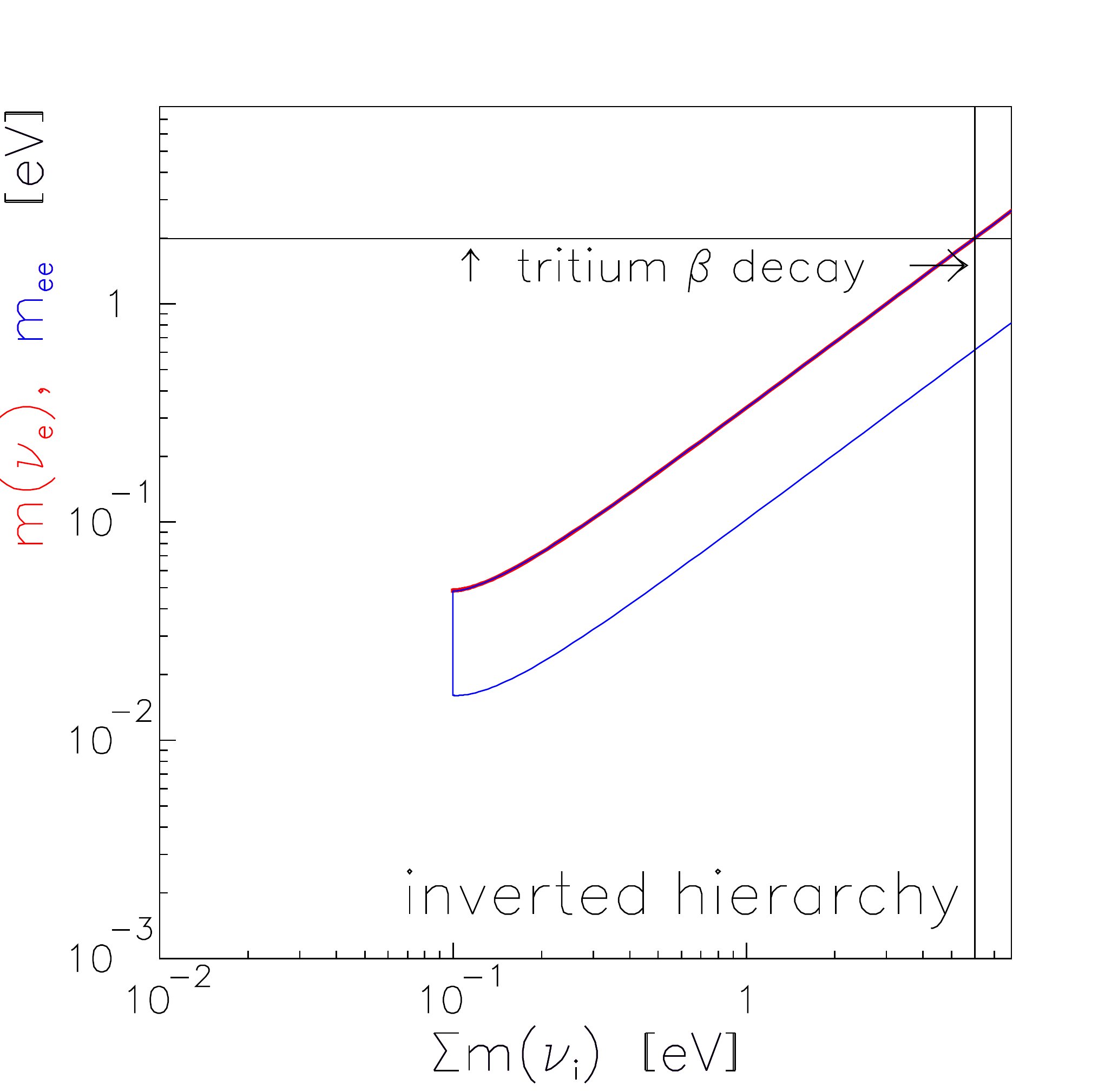}}
\caption{Observables of neutrinoless double \bdec\ \mee\ (open blue band) and of direct neutrino mass determination by single \bdec\ \mnue\
(red) versus the cosmologically relevant sum of neutrino mass eigenvalues
$\sum \mnui$ for the case of normal hierarchy (left) and of inverted hierarchy (right).
The width of the bands/areas is caused by the experimental uncertainties ($2 \sigma$)
of the neutrino mixing angles \cite{03-fogli12}
and in the case of \mee\ also by the completely unknown Majorana- and CP-phases.
Uncertainties of the nuclear matrix elements, which enter the determination of \mee\ from the  measured values of half-lives or of half-live limits of neutrinoless double \bdec , are not considered.
\label{fig-03:comparison_methods}}
\end{figure}

Figure \ref{fig-03:comparison_methods} demonstrates that the different methods are complementary to each other and
compares them.

This chapter is structured as follows:
In subchapter 2 the neutrino mass determination from the kinematics of \bdec\ is described.
Subchapter 3 presents the analysis of the spectrum of neutrinos from supernova SN1987a and the
recent \bdec\ experiments in search for the neutrino mass scale.
In subchapter 4  an overview of the present KATRIN experiment
is given. New approaches to directly determine the neutrino mass are presented in subchapter 5.
This article closes with a conclusion in subchapter 6.
For a more detailed and complete overview on this subject we would like to refer to the reviews
\cite{03-robertson_rev88,03-holzschuh_rev92,03-wilkerson_rev01,03-weinheimer_rev03, 03-Otten08, 03-giuliani_rev12}.

\section{\texorpdfstring{$\upbeta$-decay and $\nu$-mass}{Beta-decay and neutrino mass}}
\label{sec-03:beta_spec}

According Fermi's Golden Rule the decay rate for a \bdec\ is given by the square of the transition matrix element $M$
summed and integrated over
all possible discrete and continuous final states $f$  (We use the convention $\hbar = 1 = c$ for simplicity):
\begin{equation}
\label{eq-03:fermisgoldenrule}
  \Gamma = 2\pi \sum \hspace*{-1.3em} \int | M^2 | df
\end{equation}
Let us first calculate the density of the final states.
The number of different final states $dn$ of outgoing particles inside a normalization volume $V$
into the solid angle $d\Omega$ with momenta between $p$ and $p+dp$,
or, respectively, with energies in the corresponding
interval around the total energy \etot, is:
\begin{equation}
\label{03-eq:nstates}
  dn = \frac{V \cdot p^2 \cdot dp \cdot d\Omega}{h^3}
     = \frac{V \cdot p^2 \cdot dp \cdot d\Omega}{(2 \pi)^3}
     = \frac{V \cdot p \cdot \etot \cdot \detot \cdot d\Omega}{(2 \pi)^3} \quad .
\end{equation}
This gives a state density per energy interval and solid angle of
\begin{equation}
\label{03-eq:ndensity}
  \frac{dn}{\detot d\Omega} = \frac{V \cdot p \cdot \etot}{(2 \pi)^3} .
\end{equation}

Since the mass of the nucleus is much larger
than the energies of the two emitted leptons we can use for the next steps the following
simplification: The nucleus takes nearly no energy but balances all momenta\footnote{We will consider the recoil energy of the nucleus later.}.
Therefore we need to count the state density of the electron and the neutrino only
\begin{eqnarray}
\label{eq-03:ndensityall}
  \rho(E_{\rm e}, E_\nu, d\Omega_e, d\Omega_\nu)
          & =  & \frac{dn_{\rm e}}{dE_{\rm e} d\Omega_{\rm e}} \cdot  \frac{dn_{\nu}}{dE_\nu d\Omega_\nu}
          = \frac{V^2 \cdot p_{\rm e} \cdot E_{\rm e} \cdot p_\nu \cdot E_\nu}{(2 \pi)^6}\\
          & = &\frac{V^2 \cdot \sqrt{ E^2_{\rm e} - \me^2} \cdot E_{\rm e} \cdot \sqrt{E^2_\nu - m^2(\nu_{\rm e})} \cdot E_\nu}{(2 \pi)^6}. \nonumber
\end{eqnarray}

The transition matrix element $M$ can be divided into a leptonic part,
\mlep, and a nuclear one, \mhad . Usually the
coupling is written separately and expressed in terms of
Fermi's coupling constant \gfermi\ and the Cabibbo angle $\Theta_{\rm C}$:
\begin{equation}
\label{eq-03:matrixelement}
  M = \gfermi \cdot \cos{\Theta_{\rm C}} \cdot \mlep \cdot \mhad
\end{equation}

\subsection{Allowed and superallowed transitions}
\label{sec-03:allowed_transitions}
We first discuss the case of allowed or superallowed decays like that of tritium. Here, none of the leptons has to carry away angular momentum. Hence, the leptonic part \mtwolep\ essentially results in the
probability of the two leptons to be found at the nucleus, which is $1/V$
for the neutrino and $1/V \cdot F(E,Z')$ for the electron, yielding
\begin{equation}
\label{eq-03:matrixelementlep}
  \mtwolep = \frac{1}{V^2} \cdot F(E,Z') .
\end{equation}
The Fermi function $F(E,Z')$
takes into account the final
electromagnetic interaction of the emitted \belec\ with the
daughter nucleus of nuclear charge $(Z')$.
The Fermi function is approximately given by
\cite{03-holzschuh_rev92}
\begin{equation}
  F(E,Z') = \frac{2\pi \eta}{1 - exp( -2\pi \eta )}
\end{equation}
with the Sommerfeld parameter $\eta = \alpha Z' / \beta$.

For an allowed or super-allowed transition the nuclear matrix element  \mhad\
is independent of the kinetic energy of the electron. The coupling of the lepton spins to the nuclear spin is usually contracted into the nuclear matrix element.
This nuclear matrix element of an allowed or super-allowed transition
 can be divided into a vector current or
Fermi part ($\Delta I_{\rm nucl} = 0$)
and into an axial current or Gamov-Teller part
($\Delta I_{\rm nucl} = 0, \pm 1$ but no $I_{\rm nucl} =0 \rightarrow I_{\rm nucl}=0$).
In the former case, the spins of electron and neutrino couple to $S=0$, in the latter
case to $S = 1$.
 What remains is an angular correlation of the two out-going leptons. Since charge current weak interactions like \bdec\ maximally violate parity they prefer -- depending on  velocity -- negative helicities for particles and positive helicities for antiparticles. Thus the momenta or directions of the leptons are correlated with respect to their spins, and therefore to each other. This results in an
($\beta,\nu$) angular correlation factor
\begin{equation} \label{eq-03:angular_correlation}
  1 + a \cdot (\vec \beta  ~ \vec \beta_\nu)
\end{equation}
with the electron velocity $\beta = v/c$ and the neutrino velocity $\beta_\nu = v_\nu /c$. The angular correlation coefficient $a$ amounts to $a=1$ for pure Fermi transitions and to $a=-1/3$ for pure Gamov-Teller transitions within the Standard Model \cite{03-severijns06}.

The phase space density (\ref{eq-03:ndensityall}) is distributed over a surface in the two-particle phase space which is defined by a $\delta$-function
conserving the decay energy. With this prescription, we can integrate (\ref{eq-03:fermisgoldenrule}) over the continuum states and get the partial decay
rate into a single channel; for instance, the ground state of the daughter system with probability $P_0$:
\bea \label{eq-03:total_rate}
  \Gamma_0 = P_0 & \cdot & \int_{E_{\rm e}, E_{\rm \nu}, \Omega_\mathrm{e}, \Omega_\nu}
 \frac{G_{\rm F}^2 \cdot \cos^2\Theta_{\rm C} \cdot \mtwohad}{(2\pi)^5} \cdot F(E ,Z')\\ \nonumber
    & \cdot & \sqrt{ E^2_{\rm e} - \me^2} \cdot E_{\rm e} \cdot  \sqrt{E^2_\nu - m^2(\nu_{\rm e})} \cdot E_\nu \cdot
      \left( 1 + a \cdot (\vec \beta ~ \vec \beta_\nu) \right)\\ \nonumber
   & \cdot & \delta(Q + \me - E_{\rm e} - E_{\rm \nu} - \erec )~
     dE_{\rm e}~ dE_{\rm \nu}~ d\Omega~ d\Omega_\nu
\eea

In direct neutrino mass measurements usually the formulas are given in terms of the kinetic energy of the electron $E$:
\be
  E := E_\mathrm{e} - \me
\ee
The maximal kinetic energy of the electron for the case of zero neutrino mass zero is called endpoint energy \ezero\ which is defined by a vanishing neutrino energy $E_\nu$:
\be \label{eq-03:ezero}
  \ezero := \mathrm{max}(E) = \mathrm{max}(E_\mathrm{e} - \me)
\ee

A correct integration over the unobserved neutrino variables in (\ref{eq-03:total_rate}) has to respect the
($\beta,\nu$) angular correlation factor (\ref{eq-03:angular_correlation}), which also has to be considered in calculating the exact recoil energy of the nucleus $E_\mathrm{rec}$.
If we consider that the \belec s of interest have a certain minimal kinetic energy $E_\mathrm{min}$ then we can calculate the range of recoil energies of
  the daughter nucleus of mass $m_{\rm daughter}$: The recoil energy \erec\ is bound upwards by the case, in which the out-going electron takes the maximum kinetic energy \ezero\ and downwards by the case, in which the electron of kinetic energy $E_\mathrm{min}$ is emitted opposite to the direction of the neutrino, which
  has in this case a momentum $p_\nu = E_\nu = \ezero-E_\mathrm{min}$ (neglecting for a moment the non-zero value of the neutrino mass):
\be \label{eq-03:recoil_energy}
\frac{(p_e - p_\nu)^2}{2 m_\mathrm{daughter}}  =
  \frac{\left(\sqrt{E^2_\mathrm{min} + 2 E_\mathrm{min} m_e} - (\ezero - E_\mathrm{min}) \right) ^2}{2 m_\mathrm{daughter}}
      	\leq \erec \leq E_\mathrm{rec,~max} = \frac{ p^2_{\rm max}}{2 m_{\rm daughter}}
                                = \frac{E^2_0 + 2E_0 m_e}{2 m_{\rm daughter}}.
\ee
Due to the largeness of $m_\mathrm{daughter}$ even for sizeable electron energy ranges below the endpoint \ezero ,  according to equation (\ref{eq-03:recoil_energy}) the recoil energy \erec\ does not change much  (numbers are given for the example of tritium in the next section). Therefore, for the region of interest below the endpoint \ezero\ we can apply a constant recoil energy correction $\erec = const.$ and equation (\ref{eq-03:ezero}) becomes\footnote{We should note here that
we cannot use equation(\ref{eq-03:rec_correction}) to derive the endpoint energy \ezero\ from a measured nuclear $Q$-value with the precision required for
the direct measurement of the neutrino mass due to the uncertainty of $Q$, which is  ${\cal O}(1)$~eV at best. Therefore \ezero\ has to be fitted from the \bspec\ together with the neutrino mass squared. For the case of tritium there is currently a large experimental effort to improve significantly the precision on the $Q$-value of
tritium \bdec\  by ultra-high precision ion cyclotron resonance mass spectroscopy in a multi-Penning trap setup measuring the $^3$He-$^3$H mass difference \cite{03-blaum2010} with the final goal to use the measured $Q$-value in the neutrino mass fit.}:
\be \label{eq-03:rec_correction} \ezero = Q - \erec \quad .
\ee

Further integration over the angles yields through (\ref{eq-03:angular_correlation}) an averaged nuclear matrix element, as mentioned above.
Besides integrating over the ($\beta, \nu$)-continuum, we have to sum over all other final states. For a \bdec ing atom or molecule
it is a double sum: One summation runs  over all neutrino mass
eigenstates \mnui\ with probabilities $|U^2_{\rm ei}|$ which are kinematically accessible ($\mnui \leq \ezero$).
The second summation  has to be done over all of the electronic final states of the daughter system with probabilities
$\pj$ and excitation energies \vj . These comprise excitations of the electron shell but also -- in the case of \bdec ing molecules --
rotational and vibrational excitations. These excitations are caused by the sudden change of the nuclear charge from $Z$ to $Z' = Z+1$ which requests a re-arrangement of the electronic orbitals of the daughter atom or molecule   and the interatomic distances in case of a molecule. They give rise to shifted endpoint energies. Introducing the definition
\be
  \varepsilon := (\ezero - E) ~,
\ee
the total neutrino energy now amounts for this excitation to $E_\mathrm{\nu, j} = \varepsilon - \vj$.
The \belec s are leaving the nucleus on a time scale much shorter than the typical Bohr velocities of the shell electrons of the mother isotope. Therefore, the excitation probabilities of electronic states -- and of vibrational-rotational excitations in the case of molecules --
can be calculated in the so-called sudden approximation from the overlap of
the primary electron wave function $\Psi_0$ with the wave
functions of the daughter  ion $\Psi_{\rm f,j}$
\begin{equation} \label{eq_sudden_approximation}
  \pj = |\left< \Psi_0| \Psi_{\rm f,j} \right>| ^2.
\end{equation}

Rather than in the total decay rate, we are interested in its energy spectrum $\dot N(E) := d\Gamma / dE$,
which we can read directly from (\ref{eq-03:total_rate}) without performing the second integration over the $\upbeta$-energy. Using
$\varepsilon = \ezero - E$ and summed up over the final states it reads
\bea  \label{eq-03:betaspec}
  \dot N(E)  =  & & \frac{G_{\rm F}^2 \cdot \cos^2\Theta_{\rm C}}{2 \pi^3}
              \cdot \mtwohad \cdot F(E ,Z')
              \cdot  (E + \me ) \cdot \sqrt{(E + \me)^2 - \me^2} \\ \nonumber
             & \cdot & \sum_{\rm i,j} |U^2_{\rm ei}| \cdot P_{\rm j} \cdot (\varepsilon-\vj)
                                                       \cdot \sqrt{(\varepsilon - \vj)^2 - \mtwonui}
                     \cdot \Theta(\varepsilon - \vj - \mnui).
\eea

The $\Theta$-function confines the spectral components to the physical sector $\varepsilon - \vj - \mnui >0$.
This causes a technical difficulty in fitting mass values smaller than the sensitivity limit of the data,
as statistical fluctuations of the measured spectrum might occur which can no longer be fitted within the allowed physical parameter space.
Therefore, one has to define a reasonable mathematical continuation of the spectrum into the region which leads to $\chi^2$-parabolas
around  $\mtwonui \approx 0$ (see e.g. \cite{03-weinheimer93}).

Assuming unitarity of the kinematic accessible neutrino mass states ($\sum_{\rm i} |U^2_{\rm ei}| = 1$) we can expand the second line of equation (\ref{eq-03:betaspec}) for $(\varepsilon - \vj)^2 \gg \mtwonui$:
\bea  \label{eq-03:betaspec_expansion}
  \dot N(E) & \propto & \sum_{\rm i,j} |U^2_{\rm ei}| \cdot \pj \cdot (\varepsilon-\vj)
                                                       \cdot \sqrt{(\varepsilon - \vj)^2 - \mtwonui} \\
                 & \approx & \sum_{\rm j} \pj \cdot \left( (\varepsilon-\vj)^2 - \frac{1}{2} \sum_{\rm i} |U^2_{\rm ei}| \mtwonui \right)
                 =:   \sum_{\rm j} \pj \cdot \left( (\varepsilon-\vj)^2 - \frac{1}{2} \mtwonue \right). \label{eq-03:definition_mnue}
 \eea
This average over the squared masses of the neutrino mass states \mtwonui\ in equation (\ref{eq-03:definition_mnue}) defines what we called the {\it electron neutrino mass} \mnue\ in equation (\ref{03-eq:define_mnue}). This simplification always applies, if we cannot resolve the different neutrino mass states.

Figure \ref{fig-03:beta_spec} shows the \bspec\ at the endpoint according to equation ({\ref{eq-03:betaspec}).
The influence of the neutrino mass on the \bspec\ shows only at the upper end
below \ezero , where the neutrino is not fully relativistic and can exhibit its massive character.
The relative influence decreases in proportion to
$\mtwonue / \varepsilon^2$ (see figure \ref{fig-03:beta_spec}), which leads far below the endpoint -- according to equation (\ref{eq-03:definition_mnue}) --
 to a small constant offset
proportional to $-\mtwonue$.

Concerning the various neutrino mass states we can also assume that there is one heavy neutrino mass state $m(\nu_{\rm h})$ \footnote{This heavy state might comprise more than one heavy state, which are experimentally not distinguishable.}
and one light one $m(\nu_{\rm l})$\footnote{Again this could be the sum of more than one light neutrino.}. Such a situation could arise, if 3 light active neutrinos and one heavy sterile neutrino are mixed.
With $\sum_{\rm i, l} |U^2_{\rm ei}| =: \cos^2 \theta$
and  $\sum_{\rm i, h} |U^2_{\rm ei}| = 1 - \sum_{\rm i, l} |U^2_{\rm ei}| =  \sin^2 \theta$
we can rewrite the last line of equation (\ref{eq-03:betaspec}) for this case into:
{\small \bea  \label{eq-03:betaspec_nusterile}
\dot N(E) & \propto & \sum_{\rm i,j} |U^2_{\rm ei}| \cdot \pj \cdot (\varepsilon-\vj)
                                                       \cdot \sqrt{(\varepsilon - \vj)^2 - \mtwonui} \\
   & \approx &  \sum_{\rm j} \pj \cdot \left(
                      	\sin^2\theta \cdot (\varepsilon-\vj)  \sqrt{(\varepsilon - \vj)^2 - m^2(\nu_{\rm h})}
                     + \cos^2\theta \cdot (\varepsilon-\vj)  \sqrt{(\varepsilon - \vj)^2 - m^2(\nu_{\rm l})}
                      \right)
\eea
}

\begin{figure}[t!]
\centerline{\includegraphics[width=0.6\textwidth]{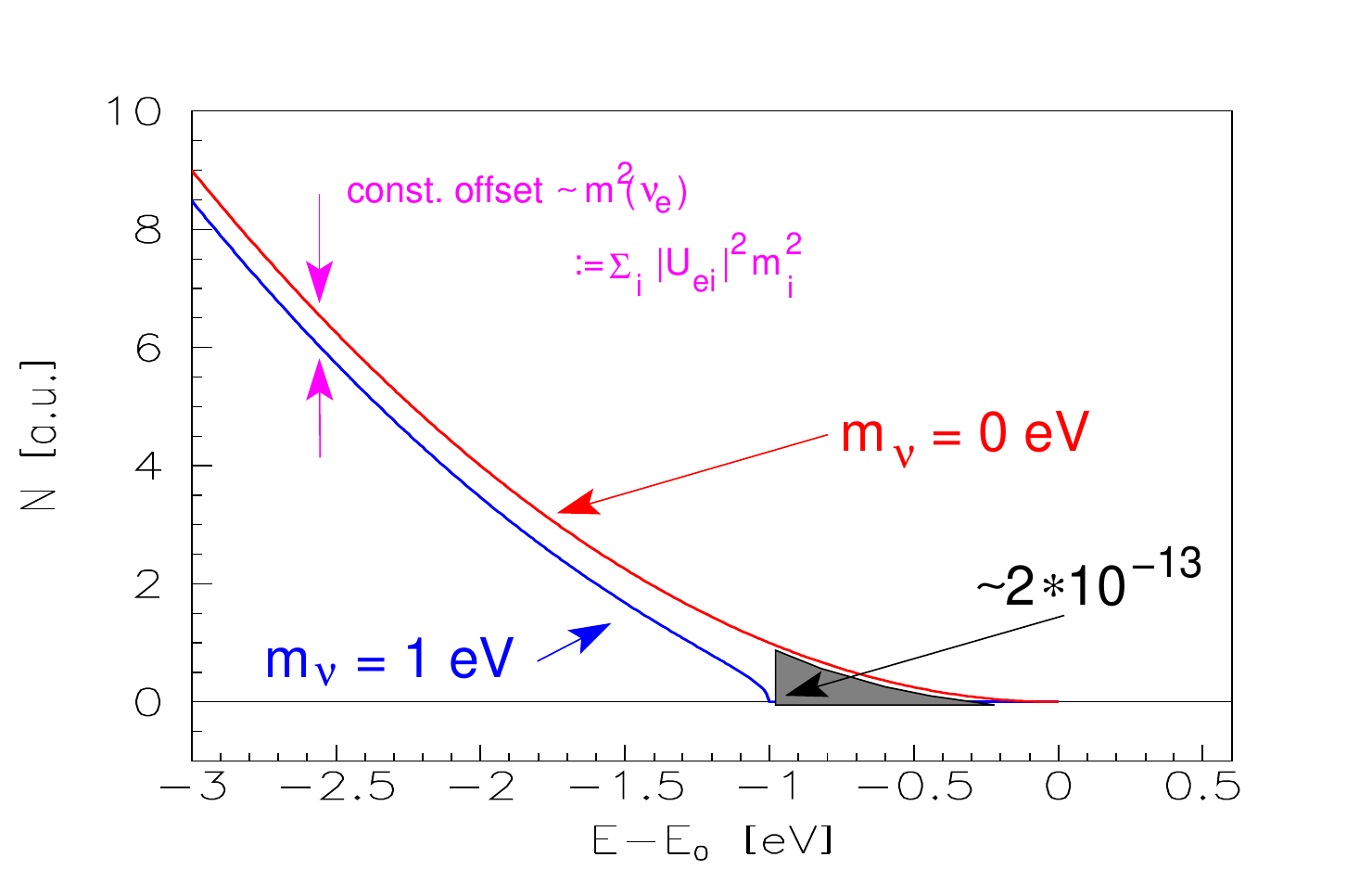}}
\caption{Expanded \bspec\ of an allowed or super-allowed \bdec\ around its endpoint \ezero\
   for $\mnue = 0$ (red line) and for an arbitrarily chosen neutrino mass
  of 1~eV (blue line).
  In the case of tritium (see section \ref{sec-03:tritium}), the gray-shaded area corresponds to a fraction of $2 \cdot 10^{-13}$ of all
  tritium \bdec s.
\label{fig-03:beta_spec}}
\end{figure}

Figure \ref{fig-03:beta_spec} defines the requirements of a direct
neutrino mass experiment which investigates a \bspec : The task
is to resolve the tiny change of the spectral shape due to the
neutrino mass in the region just below the endpoint \ezero ,
where the count rate is going to vanish. Therefore a high sensitivity experiment requires high energy
resolution, large \bdec\ source strength and
acceptance, and low background rate.

\begin{figure}[b!]
\includegraphics[width=0.33\textwidth]{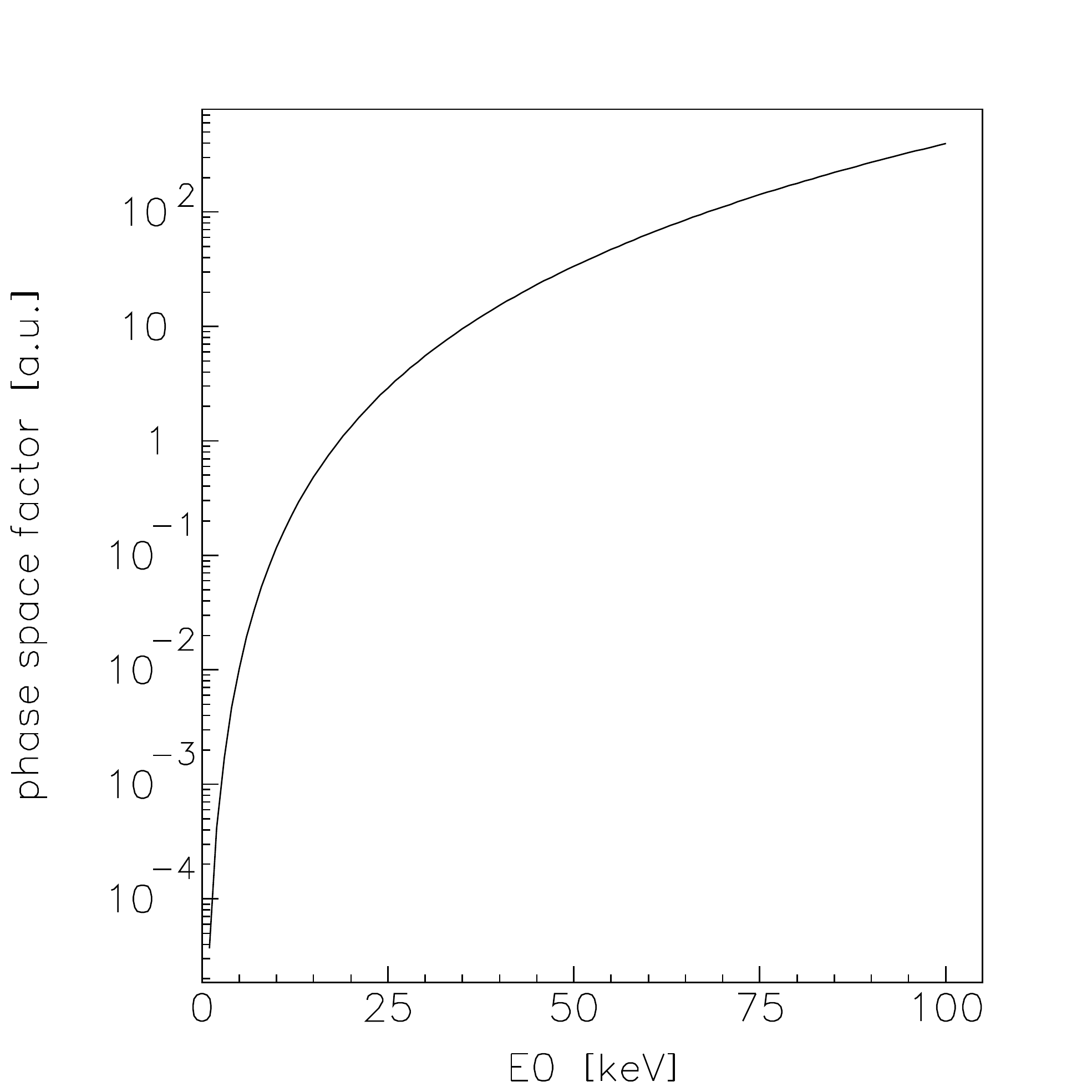}\includegraphics[width=0.33\textwidth]{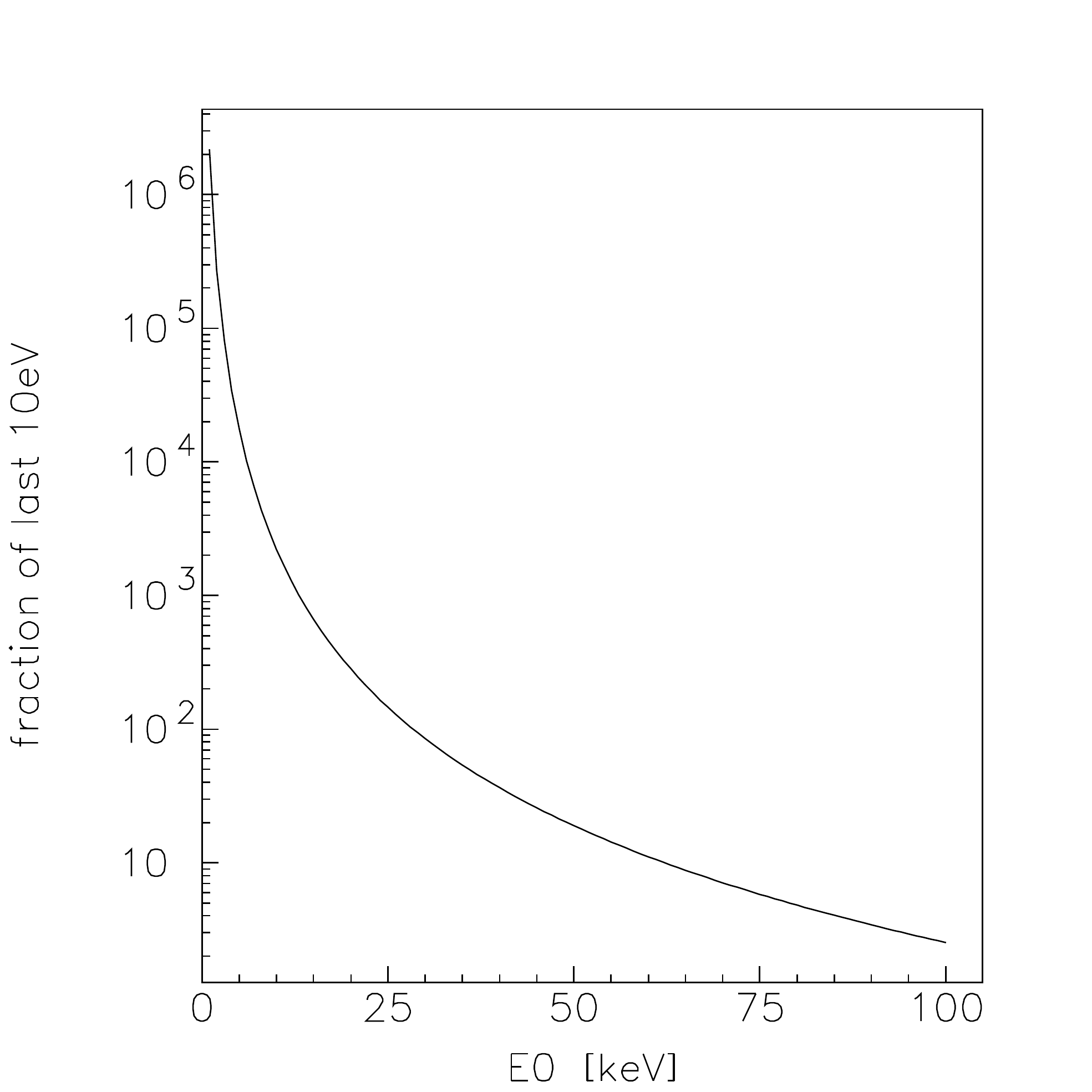}\includegraphics[width=0.33\textwidth]{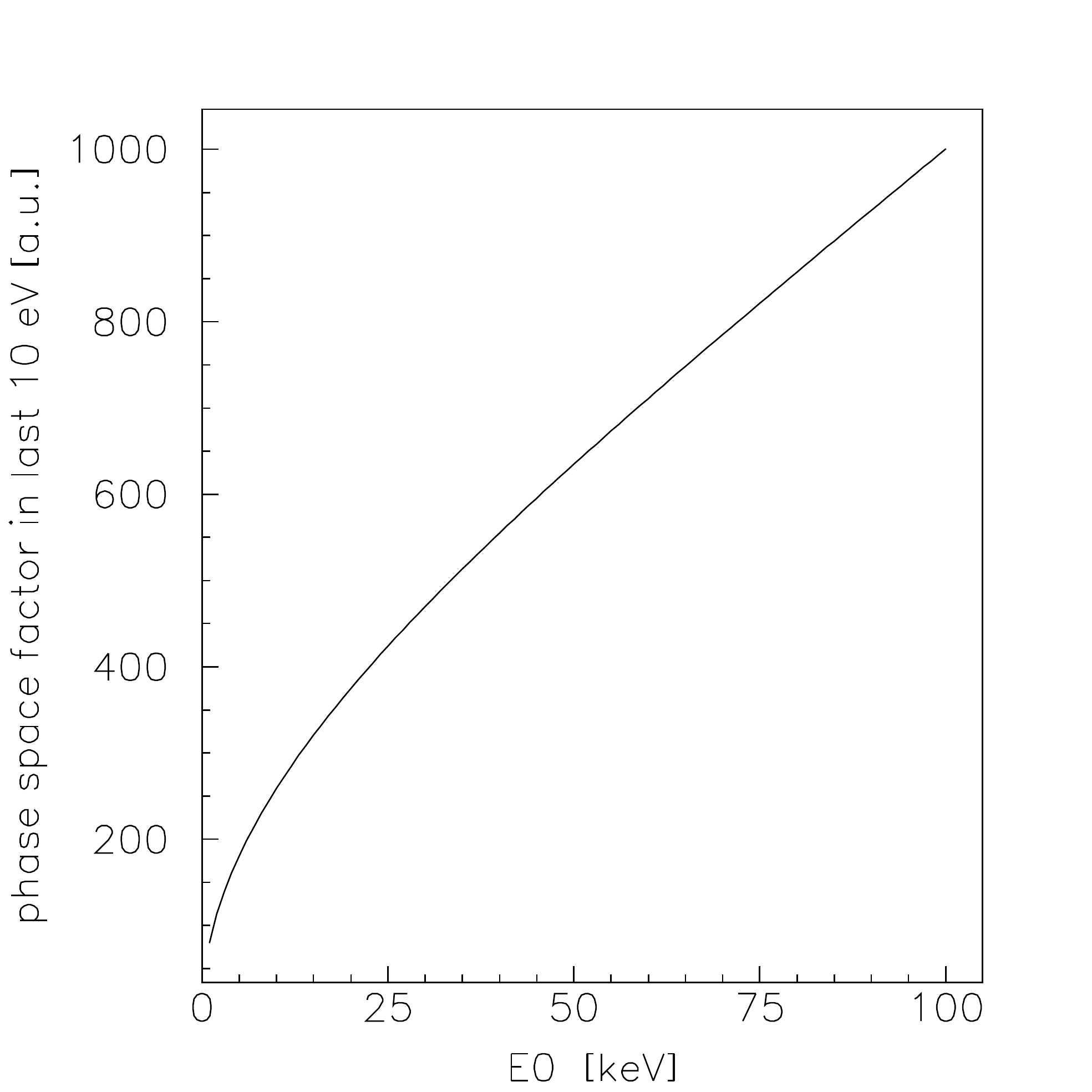}
\caption{Dependence on endpoint energy \ezero\ of total count rate (left), relative fraction in the last 10~eV below the endpoint (middle) and total count rate in the last 10~eV of a $\upbeta$-emitter (right).
  These numbers have been calculated for a super-allowed \bdec\ using (\ref{eq-03:betaspec}) for $\mnue = 0$ and neglecting possible final states as well as the Fermi function $F$.}
\label{fig-03:qvalue_dependence}
\end{figure}

Now we should firstly discuss, what is the best $\upbeta$-emitter for such a task. Figure \ref{fig-03:qvalue_dependence} shows the total count rate of
a super-allowed $\upbeta$-emitter as function of the endpoint energy. Of course, the total count rate rises strongly with \ezero , while the relative fraction
in the last 10~eV below \ezero\ decreases. Interestingly, the total count rate in the last 10~eV below \ezero , which we can take as a measure of our energy region of interest
for determining the neutrino mass, is increases with regard to \ezero . This increase is caused by the larger phase space for the \belec . From Fig. \ref{fig-03:qvalue_dependence} one might argue that the endpoint energy does not play a
significant role in selecting the right $\upbeta$-isotope, but we have to consider the fact that we need a certain energy resolution $\Delta E$ to determine the neutrino mass.
Experimentally it makes a huge difference, whether we have to achieve a certain $\Delta E$ at a low energy \ezero\ or at a higher one. Secondly, the \belec s of no interest
with regard to the neutrino mass could cause experimental problems ({\it e.g.} as background or pile-up) and again this argument favors a low \ezero .

\subsection{\texorpdfstring{Tritium $\beta$-decay}{Tritium beta-decay}}
\label{sec-03:tritium}

The heaviest of the hydrogen isotopes tritium undergoes \bdec
\begin{equation}
  ^3\mathrm{H} \rightarrow \mathrm{^3He}^+ + e^- + \bar \nu_\mathrm{e}
\end{equation}
with a half-life of 12.3~y. Tritium and Helium-3 are mirror nuclei of the same isospin doublet, therefore the decay is super-allowed.
Thus the nuclear matrix element for tritium is close to that of the \bdec\ of the free neutron and amounts to \cite{03-robertson_rev88}
\begin{equation}
\label{eq-03:matrixelementhad_tritium}
  |M^2_{\rm nucl}({\rm tritium})| = 5.55
\end{equation}

With an endpoint energy of 18.6~keV it has one of the lowest endpoints of all $\beta$ emitters together with a reasonable long half-life.
Its super-allowed shape of the \bspec\ and its simple electronic structure allow the tritium \bspec\ to be measured with small systematic uncertainties.

The recoil correction for tritium is not an issue. Up to now all tritium \bdec\ experiments used molecular tritium, which give a maximal recoil energy to the daughter molecular ion of $E_\mathrm{rec,~max}  = 1.72$~eV. Even for the most sensitive tritium \bdec\ experiment,
the upcoming KATRIN experiment (see section \ref{sec-03:intro}), the maximum variation of \erec\
over the energy interval of investigation (the last 30~eV below the endpoint) amounts to $\Delta E_\mathrm{rec} = 3.5$~meV only.
It was checked \cite{03-masood07} that this variation  can be neglected and the recoil energy can be replaced by a constant value
of $\erec = 1.72$~eV, yielding a fixed endpoint according to equation (\ref{eq-03:rec_correction}).

Furthermore, one may apply radiative corrections to the spectrum \cite{03-repco83,03-gardner04}. However, they are quite small and would influence the
result on \mtwonue\ even for the KATRIN experiment only by few a percent of its present systematic uncertainty.
One may also raise the point of whether possible contributions from right-handed currents might lead to measurable spectral anomalies
\cite{03-stephenson98, 03-ignatiev06}. It has been checked that the present limits on the corresponding right-handed boson mass
\cite{03-severijns06} rule out a sizeable contribution within present experimental uncertainties.
Even the forthcoming KATRIN experiment will hardly be sensitive to this problem \cite{03-simkovic10,03-bonn11}.

\begin{figure}[t!]
\centerline{\includegraphics[width=0.45\textwidth]{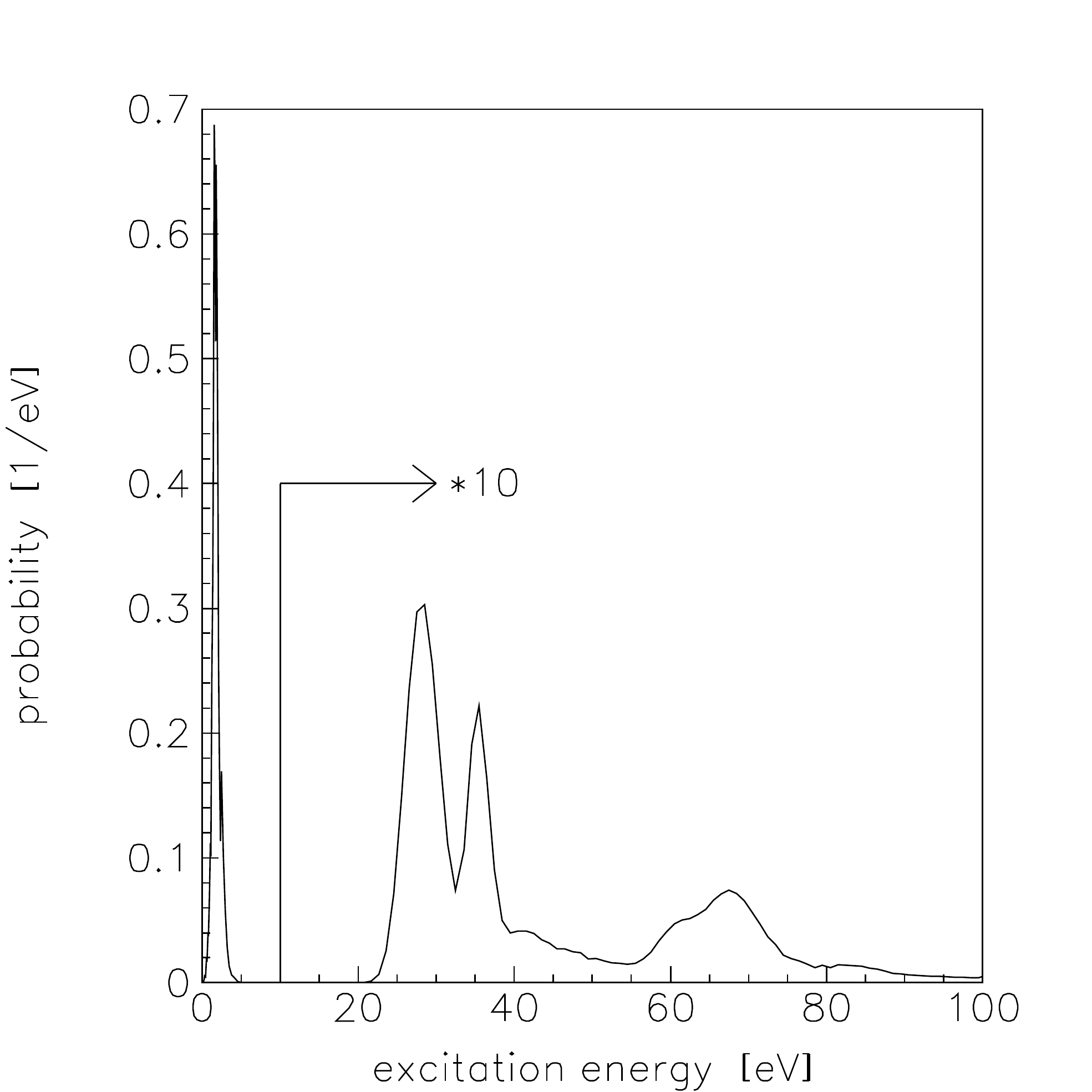}
                  \includegraphics[width=0.45\textwidth]{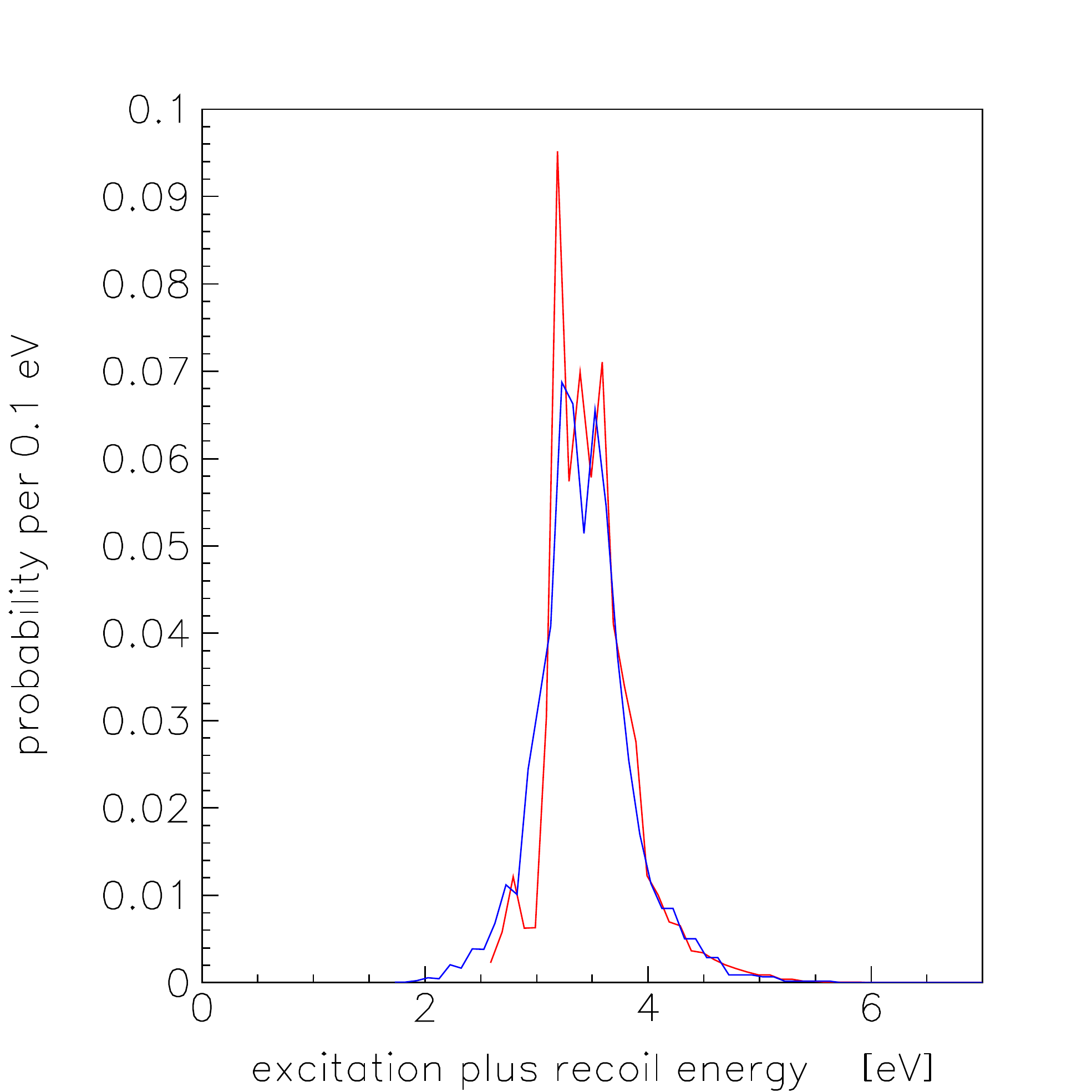}}

\caption{Excitation spectrum of the daughter
($^3$HeT)$^+$ in \bdec\ of molecular tritium (left) and rotational-vibrational excitations
of the ($^3$HeT)$^+$ molecular ion only (right, red solid curve). In comparison the rotational-vibrational excitations
of the ($^3$HeH)$^+$ molecular ion from HT \bdec\ is shown (right, blue line) \cite{03-saenz00}. Please note, that the
abscissa of the right plot reads excitation energy plus maximum recoil energy according to equation (\ref{eq-03:recoil_energy}).
} \label{fig-03:final_states_saenz}
\end{figure}

Concerning the calculation of the electronic final states according to equation (\ref{eq_sudden_approximation})
we have to consider molecular tritium since all tritium \bdec\ 	experiments so far have been using molecular tritium sources, containing the molecule \ttwo .
The wave functions of the tritium molecule are much more complicated, since in addition to two identical electrons they comprise also the description of
rotational and vibrational states, which may be excited during the \bdec\ as well.
Figure \ref{fig-03:final_states_saenz} shows a recent numerical calculation of the final states of the \ttwo\ molecule.
The transition to the electronic ground state of the $^3$HeT$^+$ daughter ion as well as the transition to higher excited electronic
states are not sharp in energy, but broadened due to rotational-vibrational
excitations. More recent calculations agree with these results \cite{03-doss06,03-doss08}.

The first group of excited electronic states starts at around $\vj = 25$~eV \cite{03-saenz00}. Therefore excited states play almost no role for the energy interval
considered for the up-coming KATRIN experiment: only the decay to the ground state of
the ($^3$HeT)$^+$ daughter molecule, which is populated with about 57~\% probability, has to be
taken into account. Due to the nuclear recoil, however, a large number
of rotational-vibrational states with a mean excitation energy of 1.7~eV and a standard deviation of
0.4~eV are populated. These values hold for a pure T$_2$ source without contamination by other hydrogen isotopes.
But a contamination of the T$_2$ molecules by DT or HT molecules does not matter in first order: The shift of the mean
rotational-vibrational excitation of HT with respect to T$_2$ is compensated by a corresponding change
of the nuclear recoil energy of HT with respect to the 1.5 times heavier T$_2$ molecule \cite{03-saenz00} (see figure \ref{fig-03:final_states_saenz} right).

Summarizing the properties of tritium for direct neutrino mass measurements: It is the standard isotope for this kind of study due to its low endpoint of 18.6 keV, its rather short half-life of 12.3 y, its super-allowed shape of the \bspec , and its simple electronic structure. Tritium \bdec\ experiments using a tritium source and a separated electron spectrometer have been performed in search for the neutrino mass for more than 60 years.
But when the electron spectrometer is identical to the $\upbeta$-source, the situation is different and a $\upbeta$-isotope with an even lower endpoint energy
\ezero\ is preferred, even if it does not have an allowed decay.

\subsection{\texorpdfstring{Forbidden transitions like \rhenium}{Forbidden transitions like 187Re}}

The isotope \rhenium\ exhibits the lowest endpoint energy with $\ezero = 2.47$~keV of all known $\upbeta$-emitters decaying to the ground-state of the daughter
nucleus\footnote{There is even a decay of the $^{115}$In into an excited nuclear $(3/2^+)$-state of the daughter nucleus $^{115}$Sn with a much lower $Q$-value
of $(155 \pm 24)$~eV \cite{03-mount09, 03-wieslander09}. But there are two reasons, why such a decay cannot be used for a direct neutrino mass measurement: this partial decay has an ultra-long half-life of $(4.1 \pm 0.6) \cdot 10^{20}$~y \cite{03-mount09} and the signature of the neutrino mass is hidden in the \bspec\ of the decay into the ground state of $^{115}$Sn.}.
The ground state of the mother isotope \rhenium\ has spin and parity $J^\pi = 5/2^+$.
The \bdec\ goes to the ground state of the daughter $^{187}$Os with spin and parity $J^\pi = 1/2^-$.
Therefore, the decay is a first unique forbidden transition. The lepton pair, electron and antineutrino, has to
carry away two units of angular momentum and has to change parity.
The two leptons will couple to spin $S=1$ in this case, one unit of orbital momentum has to be carried away by either the electron or the antineutrino. Therefore the half life  of $t_{\rm 1/2}(\rhenium) = 4.3 \cdot 10^{10}$~y
is huge and about as long as the age of the universe.
The advantage of the 7 times lower endpoint energy \ezero = 2.47~keV of \rhenium\ with respect to tritium does not compensate the fact, that one needs a large number of \rhenium\ atoms to obtain enough count rate near the endpoint to measure the neutrino mass. Therefore, a classical
${source \neq spectrometer}$ arrangement like for the tritium experiments is not feasible for \rhenium , because
the \belec s will undergo too many inelastic scattering processes within the \rhenium\ source. Secondly, the isotope \rhenium\ has a complicated
electron shell and the electronic final states might not be calculable precisely enough.

Therefore, the \bdec\ of \rhenium\
can only be exploited if the $\upbeta$-spectrometer measures  the entire
released energy, except that of the neutrino but including the energy loss by inelastic scattering processes and electronic excitations.
This situation can be realized by using a cryogenic bolometer as the $\upbeta$-spectrometer, which at the same time contains the $\upbeta$-emitter
\rhenium\ \cite{03-giuliani_rev12} (see subsection \ref{sec-03:bolometer}).

We discuss now the consequences for the $\upbeta$-spectrum. Since either the electron or the antineutrino has to be emitted with orbital angular momentum $l=1$  we cannot expand the plain wave of this lepton to zeroth order anymore as we did in equation (\ref{eq-03:matrixelementlep}), but we have to go to first order:
\bea \label{eq-03:plainwave_expansion}
   \exp{(-pR)} \approx 1 - pR
\eea
In contrast to an allowed decay, the matrix element will become dependent on energy. According to
equation (\ref{eq-03:plainwave_expansion}) an additional factor proportional to
$p_\mathrm{e} = \sqrt{(E+\me)^2-\me^2} = \sqrt{(\ezero + \me -\varepsilon)^2 - \me^2}$ or to $p_\nu = \sqrt{\varepsilon^2 - \mtwonui}$ will occur, depending on whether the electron or the antineutrino carries away the unit of orbital angular momentum. In the decay rate the square of these momenta will appear. For both cases a Fermi function $F_1$ or $F_0$ needs to be considered which describes the Coulomb interaction of the out-going electron in a $l=1$ or $l=0$ state with the remaining osmium ion \cite{03-simkovic11}:
\bea  \label{eq-03:betaspec_first_forbidden}
  \dot N(E)  =  & & \frac{G_{\rm F}^2 \cdot \cos^2\Theta_{\rm C}}{2 \pi^3}
              \cdot \mtwohad \cdot  (\ezero + \me - \varepsilon) \cdot \sqrt{(\ezero + \me -\varepsilon)^2 - \me^2} \\ \nonumber
             & \cdot & \sum_{\rm i} |U^2_{\rm ei}| \cdot  \frac{R_{\rm nucl}^2}{3} \biggl( F_1(E,Z') \cdot \Bigl( (\ezero + \me -\varepsilon)^2 - \me^2  \Bigr)
             					 + F_0(E,Z') \cdot \Bigl( \varepsilon^2 - \mtwonui \Bigr) \biggr) \\ \nonumber
             & \cdot & \varepsilon \cdot \sqrt{\varepsilon^2 - \mtwonui} \cdot \Theta(\varepsilon - \mnui),
\eea
where $R_{\rm nucl}$ is the nuclear radius.
The first term of  equation (\ref{eq-03:betaspec_first_forbidden}) proportional to $p^2$ is by 4 orders of magnitude larger than the second term proportional to $p_\nu^2$ \cite{03-simkovic11}. The nuclear matrix element $M_{\rm nucl}$ is more complex than that of an allowed \bdec .

In contrast to the case of tritium
we do not account in equation (\ref{eq-03:betaspec_first_forbidden})
for excited electronic final states, since we assume that all losses by
electromagnetic excitations will be added to the energy of the \belec\ as well as to the recoil energy in the
$ source = detector$ arrangement by the signal integration of the cryo-bolometer. Thus the \bspec\ looks simpler\footnote{We will discuss later that some excited states may live longer than the signal integration time of the cryo-bolometer and that the corresponding excitation energy may not be measured, which would cause systematic uncertainties.}.
But this is only true up to first order. In second order the electronic final states with excitation energy $\vj$ and probability $\pj$ have to be considered since the modification of the phase space of the out-going electron and the squared matrix element
($\propto (F_1 p_\mathrm{e}^3 (E+\me) p_\nu E_\nu + F_0 p_\mathrm{e} (E+\me) p^3_\nu E_\nu$)) have to be taken into account.

We will discuss the influence of electronic final states for the case of cryo-bolometers in some detail: When an electronic final state takes the excitation energy $\vj$, in the calculation for an allowed decay  (see equation (\ref{eq-03:betaspec}))
we had just shifted the effective endpoint energy  by this amount ($\ezero \rightarrow E_{\rm 0}' = \ezero - \vj$) and
multiplied this fraction of the \bspec\ with its probability $\pj$. Thus the whole \bspec\ including the phase space of the out-going leptons was calculated correctly up to possible electron energies $E_{\rm 0}'$.
When we measure  with a cryo-bolometer the sum energy $E$ in the case of an electronic excitation $\vj$, the true
kinetic energy of the electron amounts only to $E' = E-\vj$. The residual energy release $\vj$ detected in the cryo-bolometer stems from the de-excitation of the electronic excitation.
Therefore the \bdec\ probability, or the corresponding phase space factor and the squared matrix element have to be calculated for the true electron kinetic energy $E'$ and the reduced endpoint energy $E_{\rm 0}'$:
\bea \label{eq-03:modified_energies_fst}
   E' = E - \vj \qquad E_{\rm 0}' = \ezero - \vj
\eea
We can expand the relevant parameters for the phase space and squared matrix element calculation to first order
assuming $\vj \ll p$ and $\vj \ll E_e$:
\bea
  E_\mathrm{e}' & = & E' + \me = E - \vj + \me = (E + \me) \cdot \left( 1 - \frac{\vj}{E+\me} \right)
  	= E_\mathrm{e} \cdot \left( 1 - \frac{\vj}{E_e}\right)\\
  p_\mathrm{e}' & = & \sqrt{E_\mathrm{e}'^2 - \me^2} = \sqrt{(E' + \me)^2 - \me^2} = \sqrt{E'^2 + 2 \me E'} = \sqrt{(E-\vj)^2 + 2\me (E-\vj)} \nonumber \\
      & \approx & \sqrt{E^2 + 2 \me E - 2(E+\me )\vj} = \sqrt{p_\mathrm{e}^2 \left( 1 - \frac{2E_e}{p_\mathrm{e}^2} \vj \right) } \approx p_\mathrm{e} \left( 1 - \frac{E_e}{p_\mathrm{e}^2}\vj \right)\\
   E_\nu' & = &E_{\rm 0}'- E'= ( \ezero - \vj ) - (E - \vj ) = \ezero - E = E_\nu \\
   p_\nu'& = & \sqrt{E_\nu'^2 - \mtwonue}  = \sqrt{E_\nu^2 - \mtwonue} = p_\nu
\eea
Equation (\ref{eq-03:betaspec_first_forbidden}) then becomes
\bea  \label{eq-03:betaspec_first_forbidden_second order}
  \dot N(E)  =  & & \frac{G_{\rm F}^2 \cdot \cos^2\Theta_{\rm C}}{2 \pi^3}
              \cdot \mtwohad \cdot   \sum_{\rm i,j} (\ezero + \me - \varepsilon) \cdot |U^2_{\rm ei}| \cdot \pj \cdot  \frac{R_{\rm nucl}^2}{3} \\
              & \cdot & \biggl( F_1(E,Z') \cdot \Bigl( (\ezero + \me -\varepsilon)^2 - \me^2  \Bigr)^{3/2} \cdot \Bigl( 1 - 3 \frac{E_e}{p_\mathrm{e}^2}\vj \Bigr) \nonumber \\
              & &                 + F_0(E,Z') \cdot \Bigl( (\ezero + \me -\varepsilon)^2 - \me^2  \Bigr)^{1/2} \cdot \Bigl( 1 - \frac{E_e}{p_\mathrm{e}^2}\vj \Bigr)  \cdot \Bigl( \varepsilon^2 - \mtwonui \Bigr) \biggr)  \nonumber \\
             & \cdot & \varepsilon \cdot \sqrt{\varepsilon^2 - \mtwonui} \cdot \Theta(\varepsilon - \mnui). \nonumber
\eea
For a typical electronic excitation $\vj \approx 20$~eV and a typical kinetic energy of the electron $E \approx 2$~keV the correction factor
$( 1 - 3 E_e \vj / p_\mathrm{e}^2) =  0.985$ in equation ( \ref{eq-03:betaspec_first_forbidden_second order}) might have enough influence on the shape of the \bspec\ that it needs to be considered in future high precision cryo-bolometer experiments.

Secondly for a \rhenium\ atom within a crystalline environment  the  so-called {\it beta environmental fine structure} (BEFS) has to be taken into account (see subsection \ref{sec-03:rhenium}), which leads to a modulation of the \bspec .
Similarly to the extended X-ray absorption fine structure (EXAFS) this fine structure is caused by an interference of the outgoing wave of the \belec\ with waves scattered on the neighboring atoms.

\section{Past direct neutrino mass experiments}
\label{sec-03:history}

\subsection{Neutrino mass limit from supernova SN1987a}
\label{sec-03:sn1987a}

On February 23, 1987 neutrinos from the supernova SN1987a in the Large Magellanic Cloud (LMC) have been observed by the three neutrino detectors Kamiokande II \cite{03-hirata88}, IMB \cite{03-bratton88} and at Baksan \cite{03-alexeyev87} (see figure \ref{fig-03:sn1987a}). This core-collapse supernova emitted a total energy of about $3 \cdot 10^{53}$~erg = $3 \cdot 10^{46}$~J, 99~\% of this energy  was released by neutrinos \cite{03-raffelt99}. These neutrinos traveled over a distance of about $L = 50$~kpc (165,000 lyr). Although core collapse supernovae emit neutrinos and antineutrinos of all flavors at different phases of the collapse only electron antineutrinos from SN1987a have been detected
via the famous inverse \bdec\ reaction on the proton $\bar \nu_e + p \rightarrow n + e^+$. Other neutrino reactions were suppressed by too high thresholds or too low cross sections.

\begin{figure}[b!]
\centerline{\includegraphics[width=0.47\textwidth]{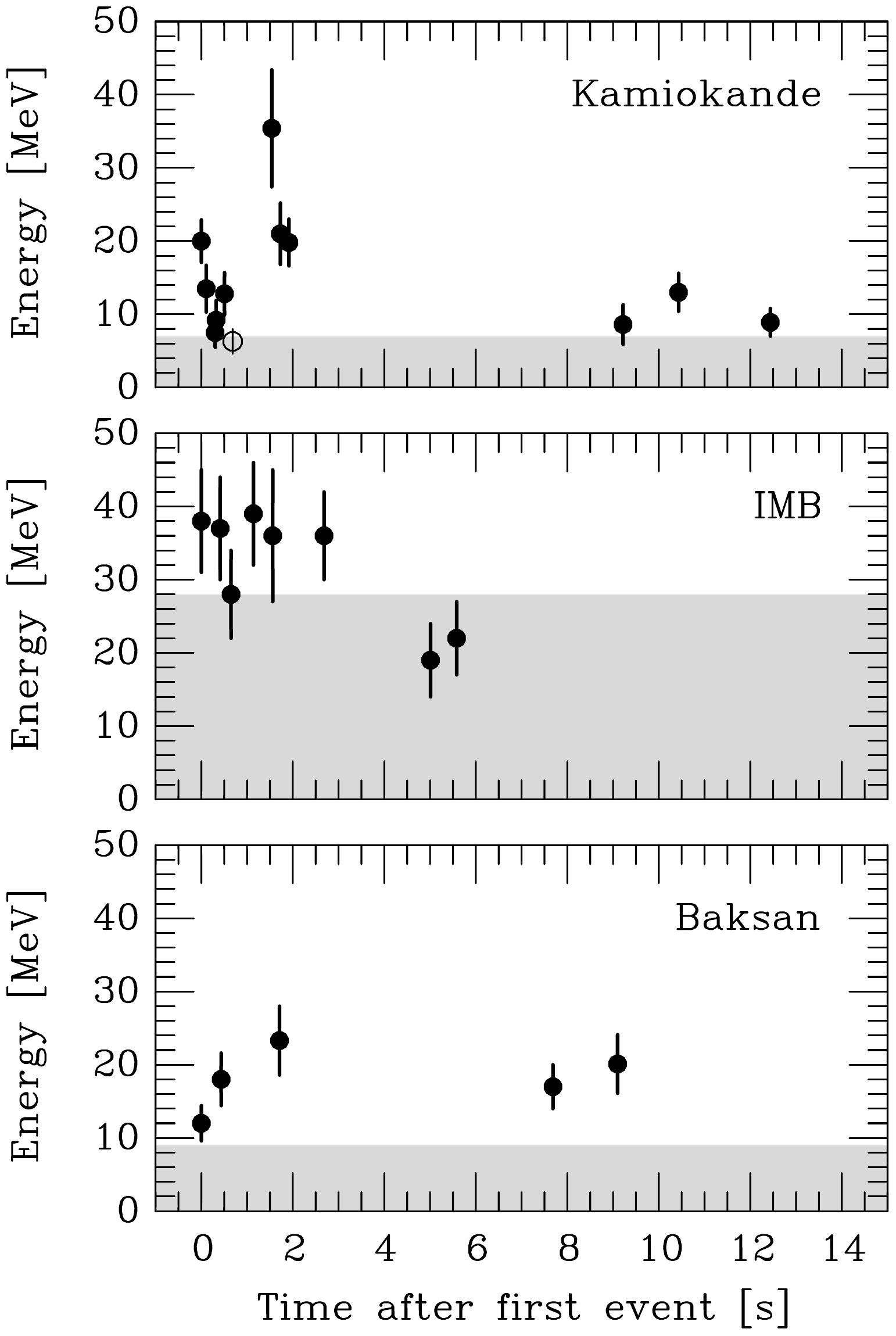}}
\caption{SN 1987A neutrino observations at the neutrino detectors Kamiokande II \cite{03-hirata88}, IMB \cite{03-bratton88} and at Baksan \cite{03-alexeyev87}.
The energies refer to the secondary positrons from the reaction $\bar \nu_e + p \rightarrow n + e^+$.
In the shaded area the trigger efficiency is less than 30~\%. The clocks of the various experiments had
unknown relative offsets. Therefore, for each detector spectrum the time was calibrated such, that the first event was recorded at $t = 0$.
In Kamiokande II, the event marked as an open circle is attributed to background
(Reprinted with permission from~\cite{03-raffelt99}, Copyright 1999, Annual Reviews).
\label{fig-03:sn1987a}}
\end{figure}

To calculate the time-of-flight we first express the velocity $\beta$ of a relativistic particle by its mass $m$ and total energy $E$:
\begin{eqnarray}
  m^2 & = & E^2 - p^2 = E^2 (1-\beta^2) = E^2 (1+\beta) (1-\beta) \approx 2 E^2 (1-\beta)\\
  \Rightarrow \beta & = & 1 - \frac{m^2}{2E^2}
\end{eqnarray}

The delay of the arrival of a supernova neutrino at earth with regard to a particle at speed of light can be expressed as function of the neutrino mass \mnu and its total energy $E_\nu$ as:
\begin{equation}
  \Delta t =  \frac{L}{\beta_\nu} -\frac{L}{c} =
   \frac{L}{1 - \frac{m_\nu^2}{2 E_\nu^2}} - L
  \approx L \cdot \left(1+\frac{m_\nu^2}{2 E_\nu^2}\right) - L = L \cdot \frac{m_\nu^2}{2E_\nu^2}  \label{eq-03:sn_tof}
\end{equation}

Therefore we would expect the neutrino energy to follow hyperbolas for each neutrino mass state $\nu_i$ as function of the square root of the neutrino arrival time.
Of course this only holds, if the neutrino emission is sharp in time with respect to the spread of arrival times on earth. The energy versus time spectra of the three experiments which detected neutrinos from supernova SN1987a do not exhibit a dependence following equation (\ref{eq-03:sn_tof}) (see fig. \ref{fig-03:sn1987a}). Therefore only upper limits
of 5.7~\ev\  (95~\%~C.L.) \cite{03-loredo02} or of 5.8~\ev\ (95~\%~C.L.) \cite{03-pagliaroli10} on the neutrino mass
can be deduced which depend somewhat on the underlying supernova model.

Nowadays more and larger neutrino detectors are online and capable to measure neutrinos from a galactic or nearby core-collapse supernova.
They are interconnected by the SuperNova Early Warning System SNEWS \cite{03-snews}.
No core-collapse supernova from our galaxy or from a satellite galaxy like the LMC  has been observed since then.
The expected rate for galactic core-collapse supernovae is 2-3 per century.
Even if a new supernova will yield many more neutrino events than supernova SN1987a it would be difficult getting a much more precise limit on the neutrino mass. The limiting factors are the time and energy spectra of the neutrino emission of a core-collapse supernova, which are not known well enough. One possibility to bypass this problem exists for core-collapse supernovae,
if the core-collapse supernova forms a black-hole.  Then
the neutrino emission will stop abruptly \cite{03-beacom00} and this stamp on the latest neutrino start time could be used in the analysis. A possible limit
of the present neutrino detectors would be in the eV range. Whether new detectors with larger masses and possibly lower energy thresholds would allow a higher sensitivity on the neutrino mass still needs to be studied.

\subsection{Laboratory direct neutrino mass limits}
\label{sec-03:labexp}

The majority of the published direct laboratory results on \mnue\
originate from
the investigation of \bdec s, which are sensitive to the average of the antineutrino mass states contributing to the electron antineutrino.
The mass of the neutrino could in principle be accessed by the investigation of $\beta^+$ decays, but measuring electron capture decays is much
more sensitive \cite{03-Ruj81}. By the investigation of the electron capture
of $^{163}$Ho two groups obtained upper limits on the average mass of the electron neutrino of
$\mnue < 225$~eV at 95\% CL~\cite{03-Spr87} and of  $\mnue < 490$~eV at 68\% CL~\cite{03-Yas94}. These experiments will be discussed in some detail in section \ref{sec-03:ec}.

An exotic way of a direct neutrino mass measurement was using bound-state $\beta^-$-decay, where the out-going \belec\ is captured into a bound electronic state:  Totally ionized $^{163}$Dy$^{66+}$ ions were circulating in a storage ring and undergoing bound-state $\beta^-$-decay although neutral $^{163}$Dy atoms are stable. The measurement of this bound state $\beta^-$-decay resulted in a limit on the neutrino mass of 410~eV at 68\%~CL \cite{03-boundstate_bdec92}.

Except \rhenium\ \bdec , on which the investigations have been started only within the last decade, all direct neutrino mass experiments using $\beta^-$-decays
were done with the
isotope tritium.
In the long history of these tritium \bdec\ experiments, about a dozen results
have been reported starting with the experiment
of Curran in the late forties yielding $\mtwonue < 1$~keV \cite{03-cur48}.

  In the beginning of the eighties a group from the Institute of Theoretical
  and Experimental Physics (ITEP) at Moscow \cite{03-itep80, 03-itep87} claimed the
  discovery of a non-zero neutrino mass of around 30~eV .
  The ITEP group used as $\upbeta$-source a
  thin film of tritiated valine combined with
  a new type of magnetic spectrometer, the {\it Tretyakov spectrometer}\footnote{A Tretyakov spectrometer uses a toroidal magnet field with a $1/r$ dependent strength like it is used for magnetic horns of neutrino beam facilities at accelerators to focus the secondary pions behind the proton target.}.
  The first tests of the ITEP claim came from the experiments at the University of Z\"urich
   \cite{03-Zuerich86} and  the Los Alamos National Laboratory (LANL)
  \cite{03-LANL87}. Both experiments applied similar Tretyakov-type
  spectrometers,  but more  advanced tritium sources with respect to the
  ITEP group.
  The Z\"urich experiment used a solid source
  of tritium implanted into carbon and later a self-assembling film
  of tritiated hydrocarbon chains.
  The LANL group was the first to develop a gaseous molecular tritium source avoiding
  solid state corrections.
  Both experiments disproved the ITEP result. The reason for the {\it mass signal} at
  ITEP was twofold: the energy loss correction was
  probably overestimated, and a  $^3$He--T mass difference
  measurement \cite{03-litmaa85} confirming the endpoint energy of the
  ITEP result, turned out only later to be significantly wrong \cite{03-vanDyck93, 03-nagy06}.

\begin{figure}[tb]
\centerline{\includegraphics[width=0.75\textwidth]{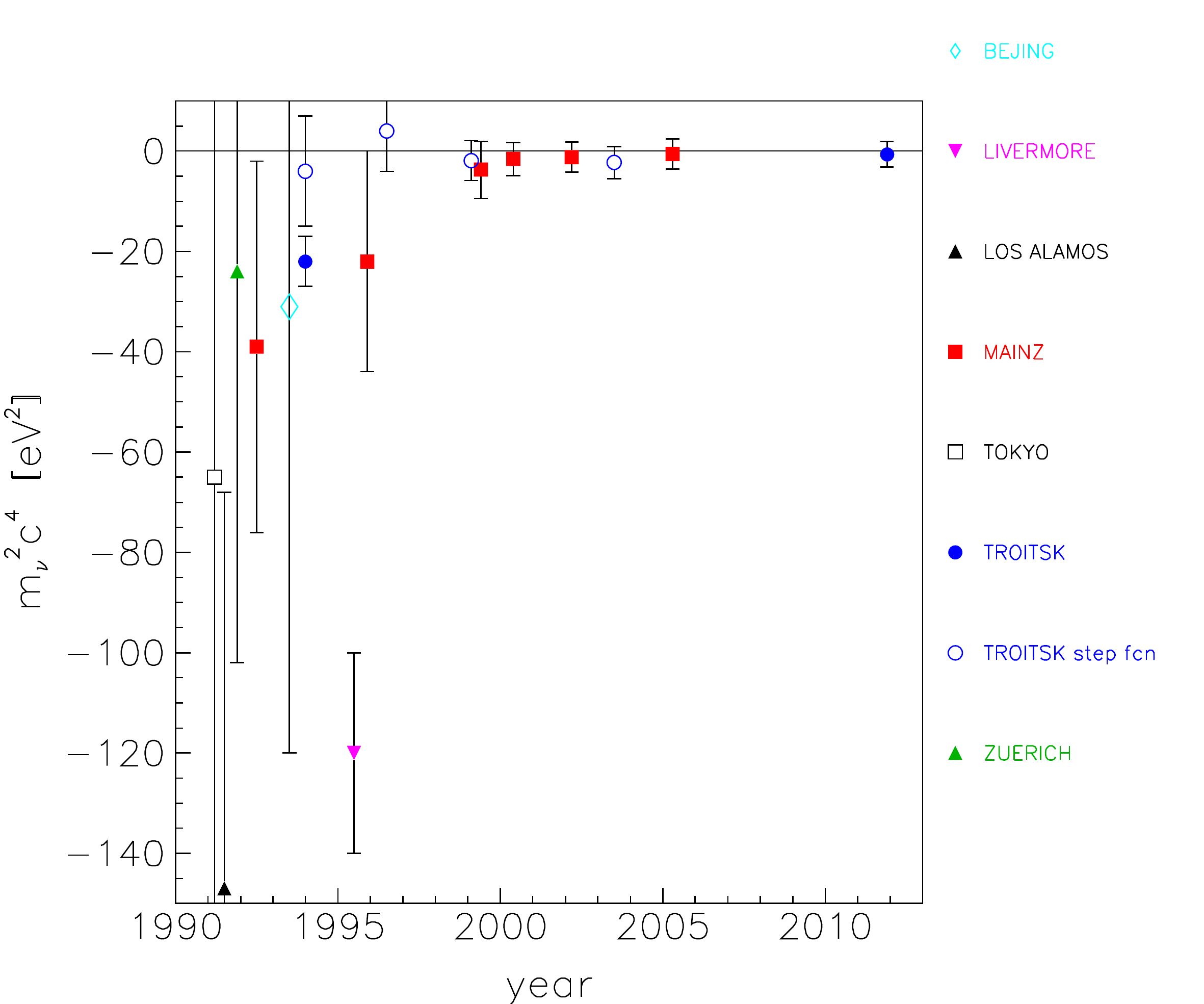}}
  \caption{Results of previous tritium \bdec\ experiments on the observable
  \mtwonue .The
   experiments at Los Alamos, Z\"urich,
   Tokyo, Beijing and Livermore \protect\cite{03-LANL91, 03-Zuerich92, 03-Tokyo91, 03-Bejing95, 03-LLNL95}
   used magnetic spectrometers, the tritium
   experiments at Mainz and Troitsk
    \protect\cite{03-weinheimer99, 03-kraus05, 03-belesev95, 03-lobashev99, 03-aseev11} applied
   electrostatic spectrometers of the MAC-E-Filter type.}
  \label{fig-03:tritium_exp}
\end{figure}

  Also in the nineties
  tritium \bdec\ experiments yielded controversially discussed results:
  Figure \ref{fig-03:tritium_exp} shows the final results of the
  experiments at LANL and Z\"urich together with the results from
  other more recent measurements with magnetic spectrometers
  at University of Tokyo, Lawrence Livermore National
  Laboratory and Beijing. The sensitivity
  on the neutrino mass had improved a lot but the values for the observable  \mtwonue\
  populated the nonphysical negative \mtwonue\ region.
  In 1991 and 1994 two new
  experiments started data taking at Mainz and at Troitsk,  which used
  a new type of electrostatic spectrometer, so called MAC-E-Filters,
  which were superior in energy resolution and luminosity with respect
  to the previous  magnetic spectrometers. However, even their early data were confirming the
  large negative \mtwonue\ values of the LANL and Livermore experiments
  when being analyzed over the last 500~eV of the \bspec\
  below the endpoint \ezero . But the large negative values of \mtwonue\ disappeared when analyzing only
  small intervals below the endpoint \ezero .
  This effect, which could only be investigated by the high luminosity MAC-E-Filters,
   pointed towards
  an underestimated or missing energy loss process, seemingly to be present
  in all experiments. The only common feature of the various experiments
  seemed to be the calculations of the electronic excitation energies
  and excitation probabilities of the daughter ions.
  Different theory groups checked these calculations in detail.
  The expansion was calculated to one order further and new interesting
  insight into this problem was obtained, but no significant changes
  were found \cite{03-saenz00, 03-doss08}.

  Then the Mainz group found the origin of the missing
  energy loss process for its experiment. The Mainz experiment
  used as tritium source a  film of molecular tritium
  quench-condensed onto aluminum or graphite substrates. Although the
  film was prepared as a homogenous thin film with flat surface, detailed studies
  showed \cite{03-fleischmann00} that the film was not stable: It underwent
  a temperature-activated roughening transition
  into an inhomogeneous film by forming micro-crystals. Thus,
  unexpected large inelastic scattering probabilities were obtained, which were not taken into account in previous analyses.
  This extra energy losses were only significant when analyzing larger energy intervals below the endpoint.

  The Troitsk experiment on the other hand
  used a windowless gaseous molecular tritium source, similar to the LANL
  apparatus. Here, the influence of large-angle scattering of electrons
  magnetically trapped in the tritium source was not considered in the
  first analysis. After correcting for this effect
  the negative values for \mtwonue\ disappeared.

  The fact that more experimental results of the early nineties populate the region of negative \mtwonue\ values
  (see  fig. \ref{fig-03:tritium_exp})
  can be  understood by the following consideration \cite{03-robertson_rev88}:
  For $\varepsilon \gg \mnue $,  eq. (\ref{eq-03:betaspec_expansion}) can be expanded into
  \begin{equation}
    \frac{dN}{dE} \propto \varepsilon^2 - \mtwonue/2.
    \label{eq-03:betaspec_expanded}
  \end{equation}
  On the other hand the convolution of a \bspec\ (\ref{eq-03:betaspec}) at $\mtwonue = 0$ with
  a Gaussian of width $\sigma$ leads to
  \begin{equation}
    \frac{dN}{dE} \propto \varepsilon^2 + \sigma^2.
    \label{eq-03:betaspec_sigma}
  \end{equation}
  Therefore, in the presence of
  a missed experimental broadening with Gaussian width $\sigma$ one expects a shift
  of the result on \mtwonue\ of
  \begin{equation}
    \Delta \mtwonue \approx - 2 \cdot \sigma^2,
    \label{eq-03:sigma_mtwonue}
  \end{equation}
  which gives rise to a negative value of \mtwonue\ \cite{03-robertson_rev88}.

  \subsubsection{MAC-E-Filter}
  \label{sec-03:mace}
  The significant  improvement in the neutrino mass sensitivity by the Troitsk and the Mainz experiments
  are due to MAC-E-Filters (\underline{M}agnetic \underline{A}diabatic \underline{C}ollimation with an
  \underline{E}lectrostatic \underline{Filter}). This new type of spectrometer
  -- based on early work by Kruit \cite{03-kruit83} -- was developed for the application to the tritium \bdec\
  at Mainz and Troitsk  independently \cite{03-pic92a, 03-Lob85}.
  The MAC-E-Filter combines high luminosity at low
  background and a high energy resolution, which are essential features
  to measure the
  neutrino mass from the endpoint region of a \bdec\ spectrum.

  \begin{figure}[tb]
  \centerline{\includegraphics[angle=0,width=0.60\textwidth]{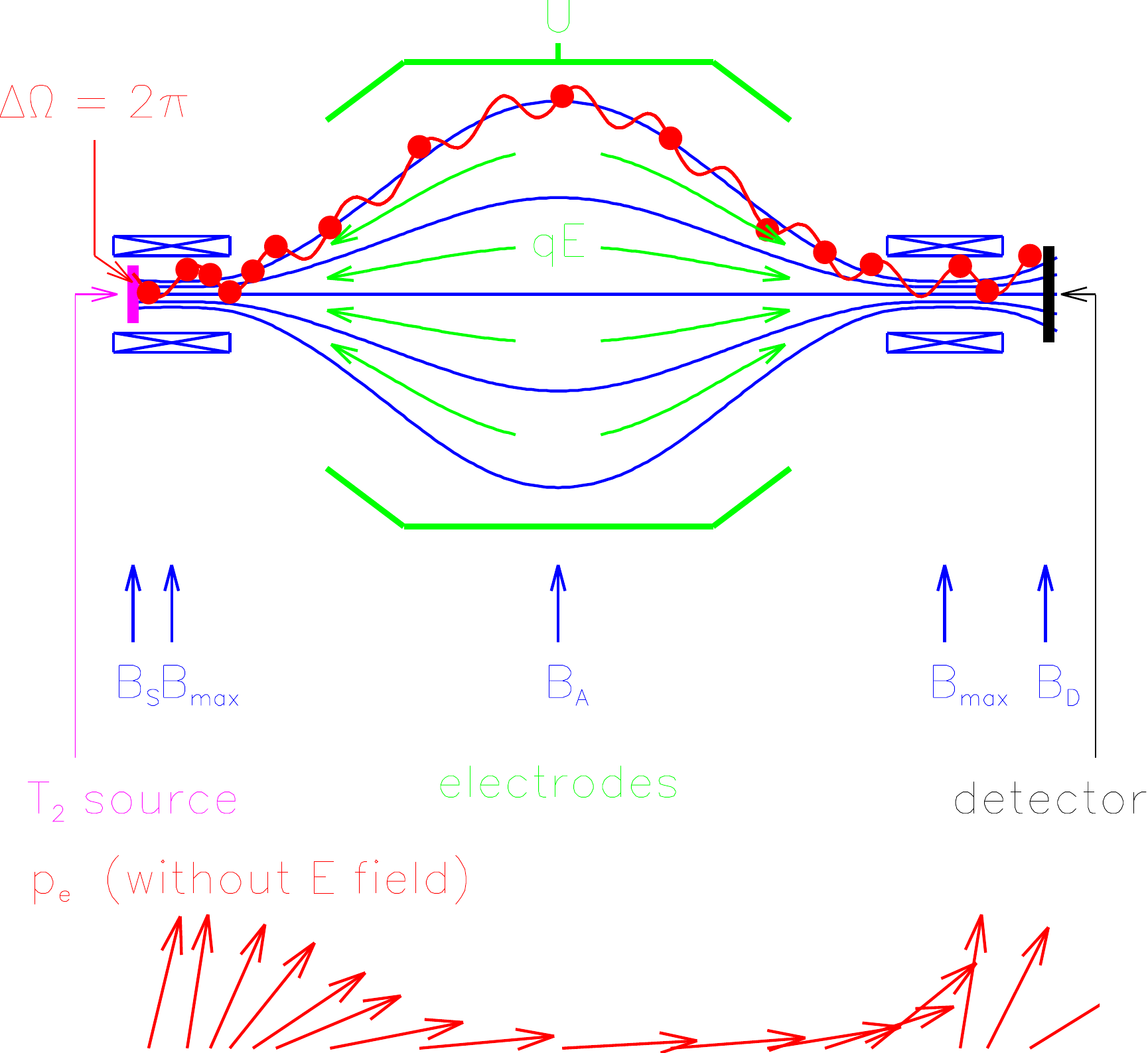}}
  \caption{Principle of the MAC-E-Filter. Top: experimental setup,
   bottom: momentum transformation due
   to adiabatic invariance of the orbital magnetic momentum $\mu$ in the
    inhomogeneous magnetic field.}
  \label{fig-03:mace}
  \end{figure}

  The main features of the MAC-E-Filter are illustrated in figure
  \ref{fig-03:mace}:
  two superconducting solenoids are producing a magnetic guiding field.
  The \belec s, starting from the tritium source
  in the left solenoid into the forward hemisphere, are
  guided magnetically on a cyclotron motion along the magnetic field lines
  into the spectrometer yielding an accepted solid
  angle of nearly $2 \pi$.
  On the way of an electron into the center of the spectrometer the magnetic
  field $B$ decreases smoothly by several orders of magnitude keeping
  the magnetic orbital moment of the electron $\mu$
  invariant  (equation given in non-relativistic approximation):
  \begin{equation} \label{eq-03:adiabatic_invariant}
    \mu = \frac{E_\perp}{B} = const.
  \end{equation}
  Therefore nearly all cyclotron energy $E_\perp$ is transformed into
  longitudinal motion (see fig. \ref{fig-03:mace} bottom)
  giving rise to a broad beam of electrons flying almost parallel to the
  magnetic field lines. This is just the opposite of the so called {\it magnetic mirror} or {\it magnetic bottle} effect.

  This parallel beam of electrons is  energetically analyzed by
  applying an
  electrostatic barrier created by a system of one or more cylindrical electrodes.
  The relative sharpness of this energy high-pass filter is only given by the ratio of the minimum
  magnetic field $B_{\rm min}$ reached at the electrostatic barrier in the so called
  analyzing plane
  and the maximum magnetic field between
  \belec\ source and spectrometer $B_{\rm max}$:
  \begin{equation} \label{eq-03:energy_resolution_mace}
    \frac{\Delta E}{E} = \frac{B_{\rm min}}{B_{\rm max}}.
  \end{equation}

  It is beneficial to place the electron source in a magnetic field $B_{\rm S}$ somewhat lower than the maximum magnetic field    $B_{\rm max}$.
  Thus the {\it magnetic mirror} effect  based on the adiabatic invariant (\ref{eq-03:adiabatic_invariant})
  hinders electrons with large starting angles at the source and long paths inside the source to enter
  the MAC-E-Filter.
  Only electrons are able to pass the pinch field $B_{\rm max}$ which exhibit starting angles $\theta_{\rm S}$ at
  $B_{\rm S}$ of:
  \begin{equation} \label{eq-03:pinch}
    \sin^2(\theta_{\rm S}) \leq \frac{B_{\rm S}}{B_{\rm max}}
  \end{equation}
  In principle, the pinch magnet could also be installed between the MAC-E-Filter and the detector, which counts the electrons transmitted by the MAC-E-Filter, as long as the electron transport is always adiabatically. Such an arrangement has been realized at the KATRIN experiment.

The exact shape of the transmission function can be calculated analytically. For an isotropically emitting monoenergetic source
of particles with kinetic energy $E$ and charge  $q$ it reads as function of the retarding potential $U$:
  \begin{equation} \label{eq-03:trans}
     T(E,U) = \left\{ \begin{array}{ll}
          0 & {\rm for~} E \leq qU\\
          1 - \sqrt{1 - \frac{E-qU}{E} \cdot \frac{B_{\rm S}}{B_{\rm min}}}
              & {\rm for~} qU < E < qU + \Delta E\\
           1 - \sqrt{1 - \frac{B_{\rm S}}{B_{\rm max}}}  & {\rm for~} E \geq qU + \Delta E
     \end{array} \right.
  \end{equation}

Fig. \ref{fig-03:transmission_katrin} shows the transmission function of a MAC-E-Filter at the example of the KATRIN experiment at its default settings
(see subsection \ref{sec-03:sds}).

\begin{figure}[t!]
    \centerline{\includegraphics[angle=0,width=0.5\textwidth]{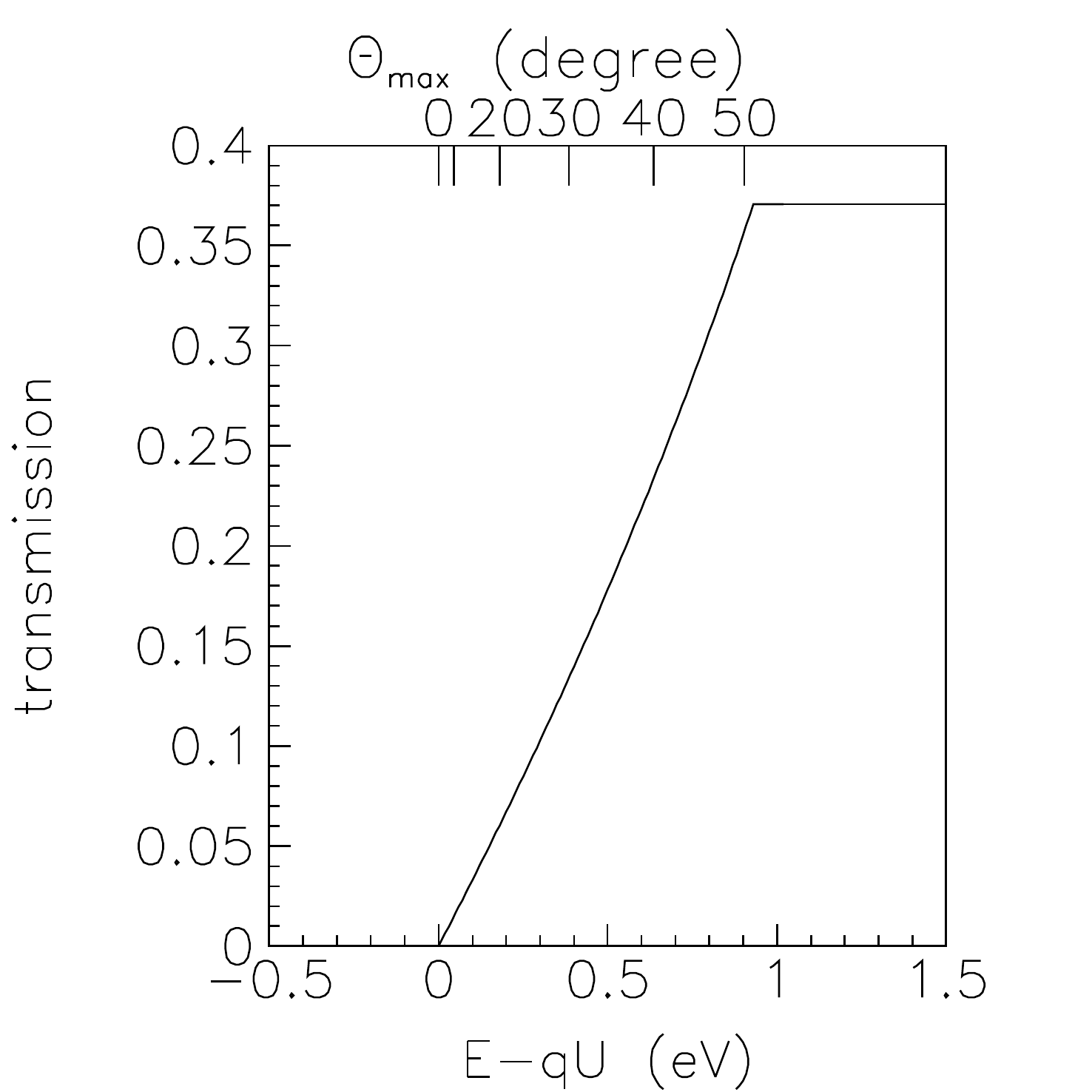}}
  \caption{Transmission function of the KATRIN experiment as a function of the surplus energy $E-qU$ according to equation (\ref{eq-03:trans}). The KATRIN default design settings \cite{03-KAT04}
were used: $B_{\rm min} = 3 \cdot 10^{-4}$~T, $B_{\rm max} = 6$~T, $B_{\rm S} = 3.6$~T. The upper horizontal axis illustrates the dependence
of the maximum starting angle $\theta_{\rm S,max}$, which is transmitted at a given surplus energy. Clearly, electrons with
larger starting angles $\theta_{\rm S} \leq \theta_{\rm S,max}$ reach the transmission condition later, since
they still have a significant amount of cyclotron energy in the analyzing plane at $B_{\rm min}$.
  \label{fig-03:transmission_katrin}}
\end{figure}

The \belec s are spiralling around the guiding magnetic field lines in zeroth approximation. Additionally, in non-homogeneous
electrical and magnetic fields they
feel a small drift $u$, which reads in first order \cite{03-pic92a}:
\begin{eqnarray}
  \vec u = \left( \frac{\vec E \times \vec B}{B^2} -\frac{(E_\perp + 2 E_{||})}{e\cdot B^3}
(\vec B \times \nabla_\perp \vec B) \right)
\label{eq-03:drift}
\end{eqnarray}

The clear advantages of the MAC-E-Filter of large angular acceptance and high energy resolution come together with the danger to store charged particles in Penning, magnetic mirror and combined traps \cite{03-mueller03}. This problem and countermeasures will be discussed later at the example of the KATRIN experiment (see subsections
\ref{sec-03:sds}, \ref{sec-03:sensitivity}).

A very interesting application is using the MAC-E-Filter in time-of-flight mode. This mode has the advantage to be non-integrating but requests to measure to or restrict the start time of the electron under investigation. The analysis can be done on cutting on the time-of-flight \cite{03-bonn99} or fully making use of the individual measured time-of-flights \cite{03-steinbrink12}.

The two  tritium \bdec\ experiments at Mainz and at Troitsk
used similar MAC-E-Filters with an energy resolution of 4.8~eV (3.5~eV) at Mainz (Troitsk).
The spectrometers differed slightly in size:
the diameter and length of the Mainz (Troitsk)
spectrometer are 1~m (1.5~m) and 4~m (7~m).
The major differences between the
two setups are the tritium sources:
The Mainz Neutrino Mass Experiment used a novel condensed solid tritium source,
whereas the experiment at Troitsk applied a windowless gaseous
molecular tritium source similar to the ones of the experiments at Los Alamos and at Livermore before.

\subsubsection{The Mainz Neutrino Mass Experiment}
\label{sec-03:mainz}

The tritium source was a thin film of molecular tritium, which was quench-condensed
on a cold graphite substrate. By laser ellipsometry the film thickness was determined. Typically film thicknesses
of 20 to 40 monolayers were applied before and of 150 monolayers after  the upgrade of the experiment in 1995-1997.
The retarding potential of the MAC-E-Filter was created by a complex system of cylindrical electrodes.
The upgrade of the Mainz setup in 1995-1997 ($\rightarrow$ {\it Mainz phase II}, see fig. \ref{fig-03:mainz_newsetup}),
includes the installation of
a new tilted pair of superconducting
solenoids between  the tritium source and the spectrometer and the use of a
new cryostat to keep the temperature of the tritium film below 2 K.
The first measure  eliminated source-correlated background and allowed
the source strength to be increased significantly.
From this upgrade on, the Mainz experiment ran with a similar signal and similar background rate as the Troitsk experiment.
The second measure
avoided the roughening transition of the homogeneously condensed tritium films with time
\cite{03-fleischmann00}, which before had given  rise to negative values of \mtwonue\ when the data
analysis used large intervals of the \bspec\ below the endpoint \ezero\ \cite{03-weinheimer93}.
The upgrade was completed by the application
of HF pulses on one of the electrodes in between measurements every
20~s, and a full  automation of the apparatus and remote control. The former improvement
lowered and stabilized the background, the latter one allowed long--term measurements.

\begin{figure}[t!]
    \centerline{\includegraphics[angle=0,width=0.8\textwidth]{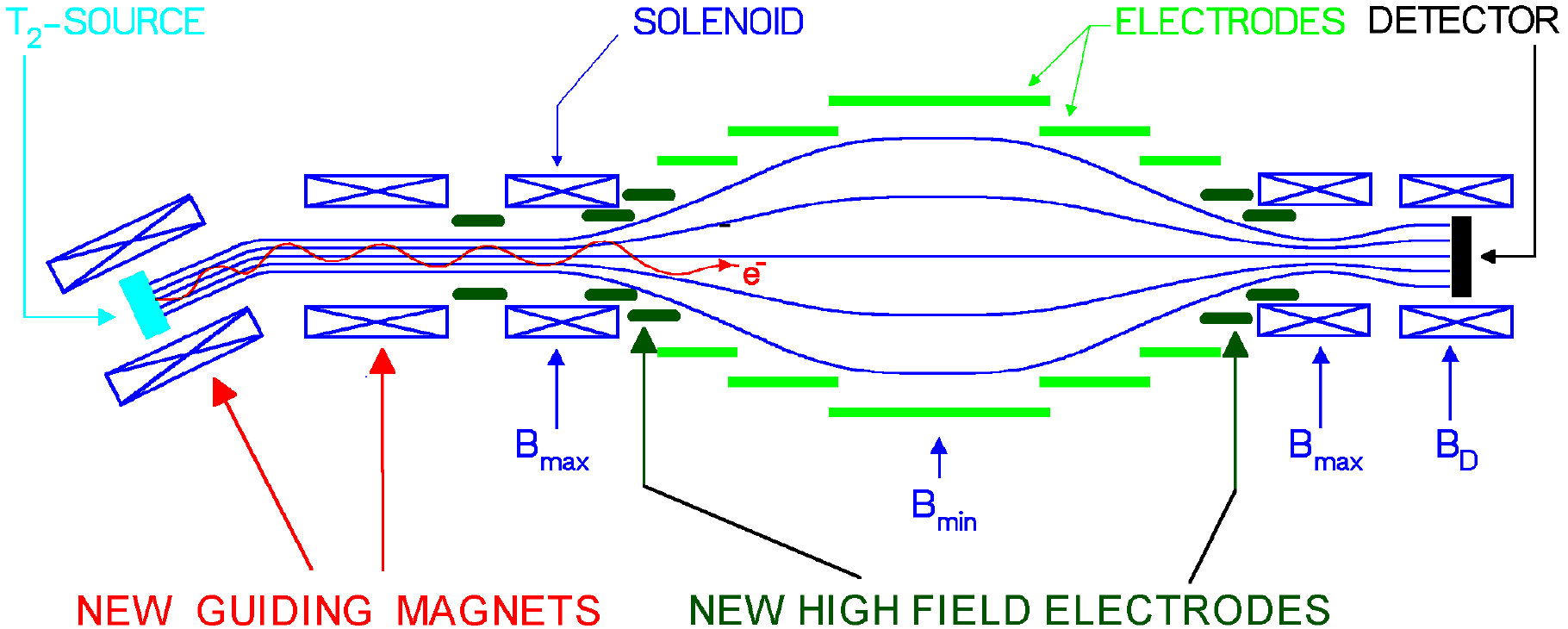}}
  \caption{The upgraded Mainz setup shown schematically.
    The outer diameter amounts to 1~m, the distance
    from source to detector is 6~m.}
  \label{fig-03:mainz_newsetup}
\end{figure}

\begin{figure}[t!]
\centerline{\includegraphics[width=0.6\textwidth]{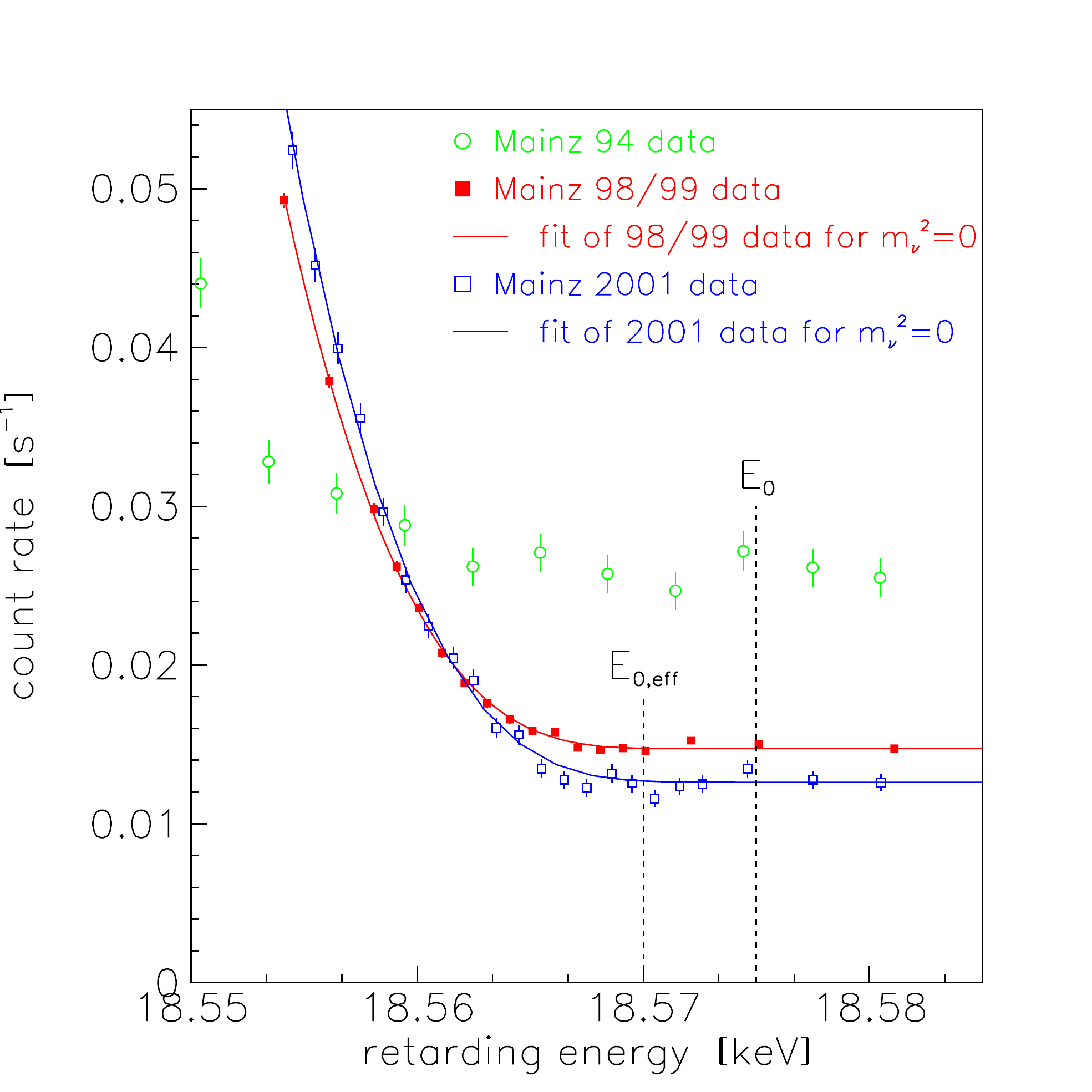}}
\caption{Averaged count rate of the Mainz 1998/1999 data
(filled red squares) with fit for \mnue=0 (red line)
and of the 2001 data (open blue squares) with fit for \mnue=0 (blue line)
in comparison with previous Mainz data
from 1994 (open green circles)
as a function of the retarding energy
 near the endpoint \ezero\  and effective endpoint $E_{0,{\rm eff}}$ (taking into account
the width of the response function of the setup and
the mean rotation-vibration excitation energy of the electronic
ground state of the $\rm ^3HeT^+$ daughter molecule).}
\label{fig-03:mainz_data}
\end{figure}

Figure \ref{fig-03:mainz_data}
shows the endpoint region of the Mainz phase II data (from 1998, 1999 and 2001) in comparison with the former
Mainz 1994 data. An improvement of the signal-to-background ratio
by a factor 10 by the upgrade
as well as a significant enhancement of the
statistical quality of the data by long-term measurements are clearly visible.
The main systematic uncertainties of the Mainz experiment
are the inelastic scattering
of \belec s within the tritium film,
the excitation of neighbour molecules due to sudden change of the nuclear charge
during \bdec ,
and the self-charging of the tritium film as a consequence of its radioactivity.
As a result of detailed investigations in Mainz
\cite{03-aseev00, 03-barth_erice97, 03-bornschein03, 03-kraus05}
-- mostly by dedicated experiments --
the systematic corrections became
much better understood and their uncertainties were reduced
significantly.
The high-statistics Mainz phase II data from 1998-2001 allowed the first
determination of the probability of the neighbour excitation
to occur in $(5.0 \pm 1.6 \pm 2.2)~\%$ of all \bdec s \cite{03-kraus05} in good agreement
with the theoretical expectation \cite{03-kolos88}.

The analysis of the last 70~eV below the endpoint of the phase II data,
resulted in \cite{03-kraus05}
\begin{equation}
\mtwonue = (-0.6 \pm 2.2 \pm 2.1)~ \evtwo,
\end{equation}
which -- using the Feldman-Cousin method  \cite{03-feldman98} -- corresponds to an upper limit  of
\begin{equation}
\mnue < 2.3~ \ev \quad {\rm (95~\%~C.L.)}
\end{equation}
An analysis of the Mainz phase II data with respect to setting a limit on the contribution of a light sterile neutrino is underway \cite{03-kraus12}.

\subsection{The Troitsk Neutrino Mass Experiment}
\label{sec-03:troitsk}

The windowless gaseous tritium source of the Troitsk experiment
\cite{03-lobashev99}  is essentially a tube of 5~cm diameter filled
with \ttwo\ resulting in a column density of $10^{17}$~molecules$\rm /cm^2$.
The source is connected to the
ultrahigh vacuum of the  MAC-E-Filter type spectrometer by a series a differential
pumping stations (see fig. \ref{fig-03:troitsk_setup}).

From their first measurement in 1994 on the Troitsk group had reported for more than a decade
the observation of a small, but significant
 anomaly in its experimental spectra starting a few eV below the
$\beta$-endpoint \ezero . This anomaly appeared as a sharp step of
the count rate \cite{03-belesev95}.
Because of the integrating property of the MAC-E-Filter, this step should correspond to a narrow line in the
primary spectrum with a relative intensity of about $10^{-10}$ of
the total decay rate.
In 1998 the Troitsk group claimed that the
position of this line was oscillating with a frequency of 0.5 years
between 5~eV and 15~eV below \ezero\ \cite{03-lobashev99}. By 2000
the anomaly  no longer followed the 0.5 year periodicity, but
still existed in most data sets.
The reason for such an anomaly with these features were not clear but gave rise to many speculations.
In presence of this problem, the Troitsk experiment  corrected for this
anomaly by fitting an additional line to the \bspec\ run-by-run.

\begin{figure}[t!]
    \centerline{\includegraphics[angle=0,width=0.8\textwidth]{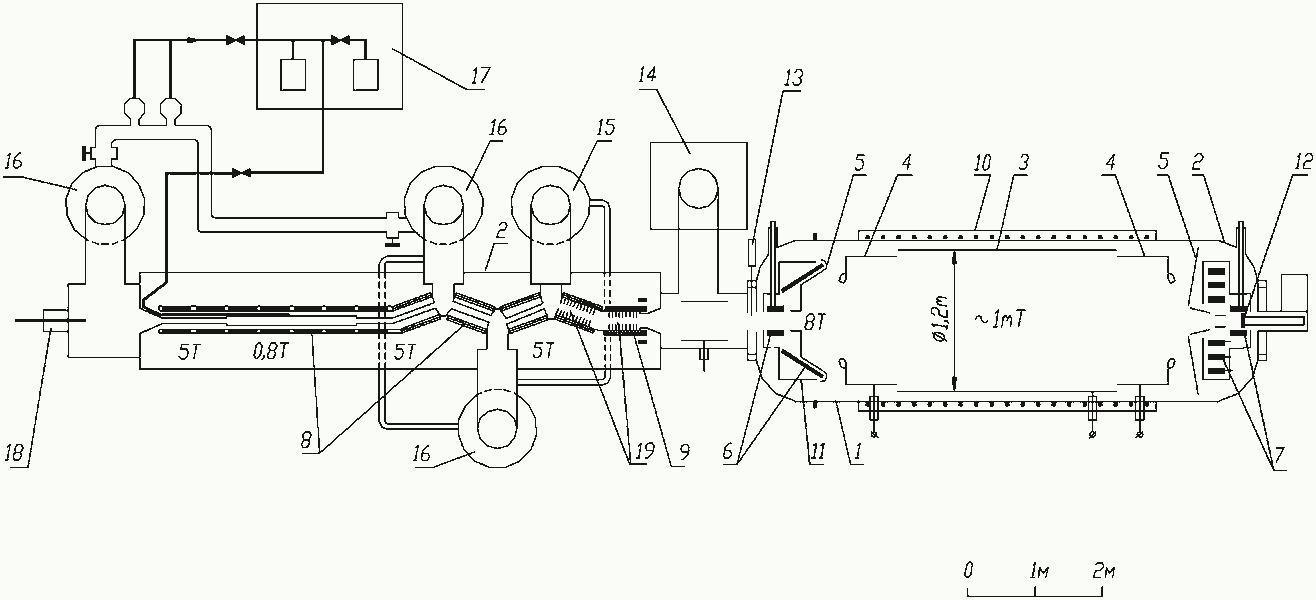}}
  \caption{The Troitsk Neutrino Mass Experiment: Vacuum vessel of MAC-E-Filter (1) and of windowless gaseous tritium source (2), retarding electrodes (3,4), ground electrode (5),  superconducting solenoids (6-9), warm solenoid (10), LN$_2$ shield (11),
  Si(Li) detector (12), emergency valve (13), magneto-discharge pump (14), mercury diffusion pumps (15-16)
  tritium purification system (17), electron gun (18), argon trap (19).
  (Reprinted with permission from~\cite{03-aseev11}, Copyright (2011) by the American Physical Society).}
  \label{fig-03:troitsk_setup}
\end{figure}

In 2011 the Troitsk group repeated the analysis of their data \cite{03-aseev11}.
Special care was taken for calculating the time-dependent
column density of the windowless tritium source and applying these values to the analysis. Secondly the data were very carefully selected with regard to data quality and stability of the experiment.
Thus a time and intensity varying anomaly was not any more needed to describe the Troitsk $\upbeta$-spectra.
Combining all the selected Troitsk data from 1994 to 2004 gave \cite{03-aseev11}
\begin{equation}
\mtwonue = (-0.67 \pm 1.89 \pm 1.68)~ \evtwo
\end{equation}
from which using the Feldman-Cousin method \cite{03-feldman98} an upper limit can be deduced:
\begin{equation}
\mnue < 2.05~ \ev \quad {\rm (95~\%~C.L.)}
\end{equation}

\subsection{Cryo-bolometers}
\label{sec-03:bolometer}

\begin{figure}[t!]
  \centerline{\includegraphics[angle=0,width=0.45\textwidth]{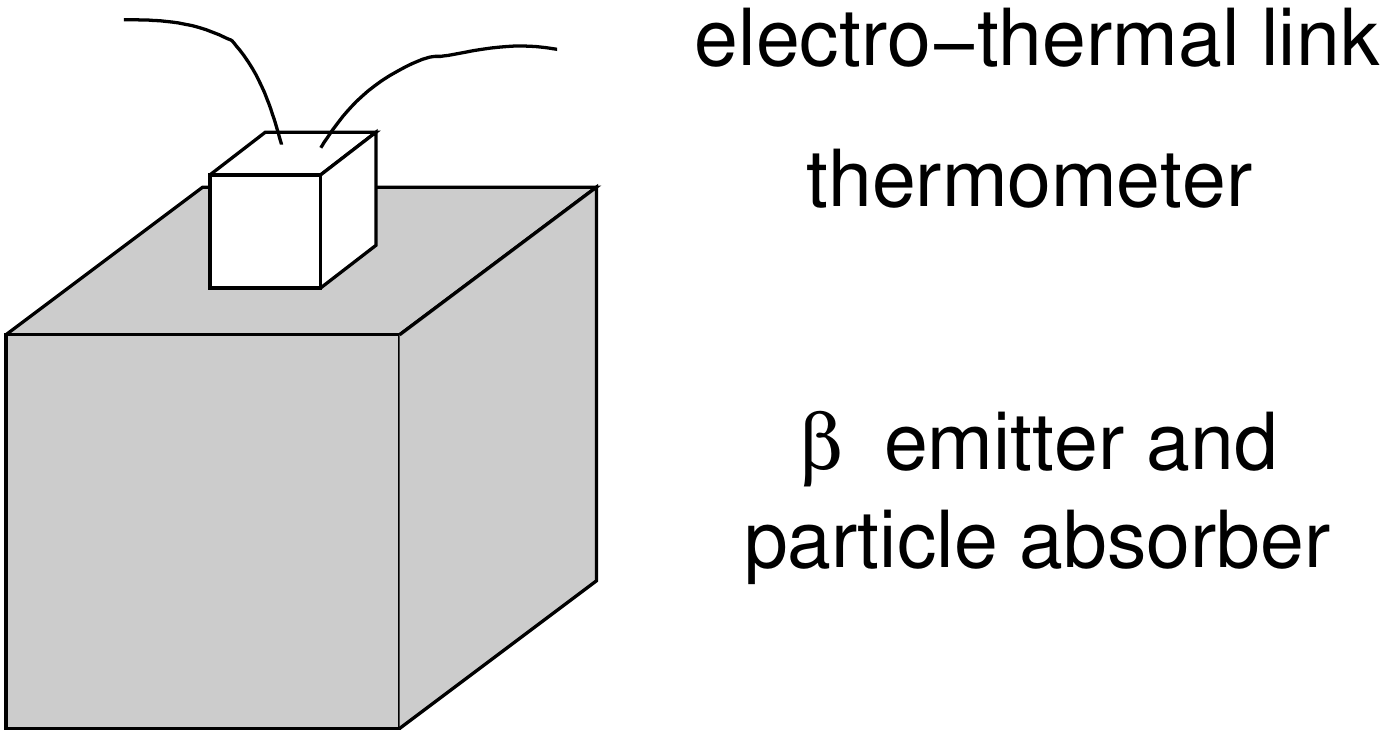}}
  \caption{Principle scheme of a cryo-bolometer for direct
   neutrino mass measurements consisting of a $\upbeta$ emitting crystal, which
   serves at the same time as the particle and energy absorber. The
  electric read-out wires of the thermometer link the whole bolometer
  to a thermal bath.}
  \label{fig-03:cryo_bolometer}
 \end{figure}

Due to the complicated electronic
structure of \rhenium\ and its \bdec\ (compare to subsection
\ref{sec-03:beta_spec}) the advantage of the 7 times lower
endpoint energy \ezero\ of \rhenium\ with respect to tritium can
only be exploited if the $\upbeta$-spectrometer measures  the entire
released energy, except that of the neutrino. This situation
requires a $\upbeta$-spectrometer, which at the same time contains the \belec\ emitting \rhenium .
The advantage of this approach is that many systematic uncertainties -- like the electronic final state spectrum, energy losses by inelastic scattering, etc. -- drop out in first order, since all released energy except that of the neutrino is measured in the same way and summed up automatically\footnote{We assume that all de-excitation processes are faster than the integration time of a detector signal.}.

To reach the required energy resolution such a {\it source=detector} arrangement can be ideally realized by a cryo-bolometer (see figure \ref{fig-03:cryo_bolometer}) \cite{03-giuliani_rev12}.
The energy release $\Delta W$  by the \bdec\ results in  a tempe\-ra\-ture rise $\Delta T$ of the crystal. This temperature increase can be measured by the thermometer, if $\Delta T$ is large enough.
The temperature rise $\Delta T$ depends on the energy release $\Delta W$ and  on the heat capacity $C$:
\begin{equation}
  \Delta T = \frac{\Delta W}{C}
\end{equation}
In order to obtain a large temperature increase $\Delta T$ in the presence of a very small energy release
$\Delta W$ \footnote{It should be noted, that the sensitivity on the neutrino mass requires a very low endpoint energy \ezero .},  the heat capacity $C$ has to be extremely small. The first measure is using tiny cryo-bolometers of typical masses of ${\cal O}(1)$~mg. Secondly, the temperature of the cryo-bolometer has to be as low as possible. The Debye model states that the phonon part of the heat capacitance of a crystal
consisting of $N$ atoms scales with the third power on temperature in units of the Debye temperature $\Theta_D$:
\begin{equation}
  C = \frac{12 \pi^4}{5} \cdot N \cdot k_{\rm B} \cdot \left( \frac{T}{\Theta_D} \right)^3
\end{equation}

There is a second reason why the crystals should be small: The cryo-bolometer is not an integral spectrometer like the MAC-E-Filter but measures always the entire spectrum. Therefore pile-up of two random events may pollute the endpoint region of a \bdec\ on which the neutrino mass is imprinted. The pile-up rate of a detector of random rate $\dot N_\mathrm{tot}$ requiring a minimal time interval to distinguish two events of $\delta t$
amounts to $\dot N_\mathrm{tot}^2 \delta t$.
With a half-life $4.3 \cdot 10^{10}$~y and a natural abundance of 62.6~\%
the specific activity of pure rhenium amounts to about 1~Bq~mg$^{-1}$.
The rise time $\delta t$ typically scales with the mass of the cryo-bolometers. Rise times of cryo-bolometers with temperature read-out of ${\cal O}(1)$~mg mass are typically in the ${\cal O}(100)~\mu$s range\footnote{New cryo-bolometer approaches like Metallic Magnetic Calorimeters (MMC) as discussed in section \ref{sec-03:echo} could be much faster.}.
 Therefore cryo-bolometer detectors with mg masses are
required to suppress pile-up by 4 or more orders of magnitude.
This has to be compared to the fraction of the \bspec\ which contains the information on the neutrino mass.
Even for the lowest known endpoint energy of \rhenium\ with $\ezero = 2.47$~keV, the relative fraction of \rhenium\ \bdec\ events
in the last eV below \ezero\ is of order $10^{-11}$ only (compare
to Figure \ref{fig-03:beta_spec}). Therefore arrays of many bolometers are required to reach a high sensitivity on the neutrino mass.

Another technical challenge is the energy resolution of  the cryo-bolometers. Although cryogenic
bolometers with an energy resolution of a few eV have been produced
with other absorbers, this resolution has  not yet been achieved with rhenium.
\subsection{\texorpdfstring{$^{187}$Re \bdec\ experiments}{187Re beta decay experiments}}
\label{sec-03:rhenium}
Two groups have started the field of \rhenium\ \bdec\ experiments at Milan (MiBeta) and Genoa (MANU):
The MANU experiment used a single metallic rhenium crystal of about 1.6~mg as absorber read out by a neutron transmutation doped (NTD) germanium thermistor~\cite{03-Gal00}. Sensor and absorber were connected using epoxy glue and were suspended by aluminum wires used for readout of the device. These lines also provided the thermal link to a heat bath cooled down to temperatures below 100~mK using a $^3$He-$^4$He dilution refrigerator.
At a detector threshold of 350~eV, an event rate of about 1.1~counts per second was observed with this absorber.
The rise time of this early detector was of the order of 1~ms with a decay time of the signals of several tens of ms. This clearly limits
the amount of activity allowed per absorber crystal, in order to avoid pile-up problems. The energy resolution of the detector was determined to be 96~eV FWHM. While the main objective of the initial measurement was a determination of the endpoint energy and the half life of \rhenium\ \bdec, an interesting side effect was the first observation of the so called $\beta$ environmental fine structure (BEFS) in pure rhenium~\cite{03-Gal00}. This fine structure is caused by an interference of the outgoing direct wave of the beta electron with incoming waves reflected by the neighboring atoms and leads to a modulation of the shape of the \bspec\ that is most pronounced at low electron energies.
\begin{figure}
\centering
\includegraphics[width=0.37\textwidth]{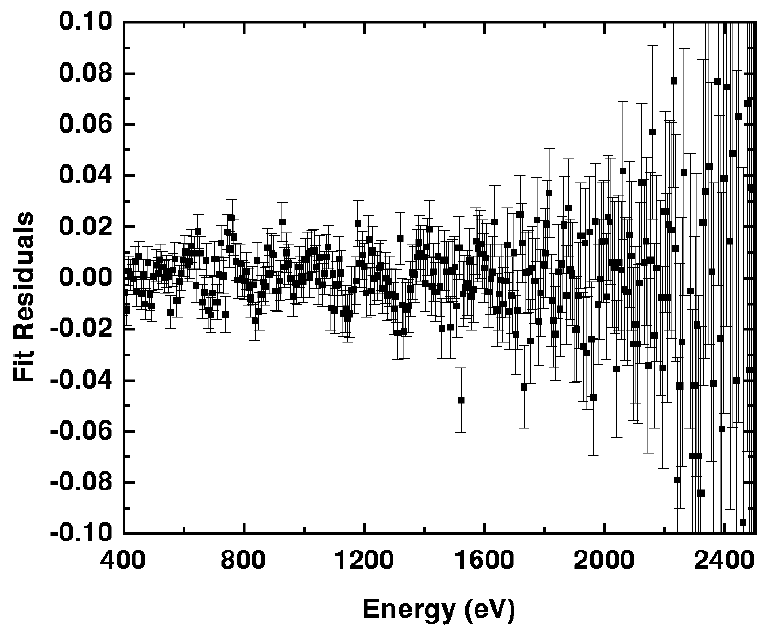}\hspace{10mm}
\includegraphics[width=0.45\textwidth]{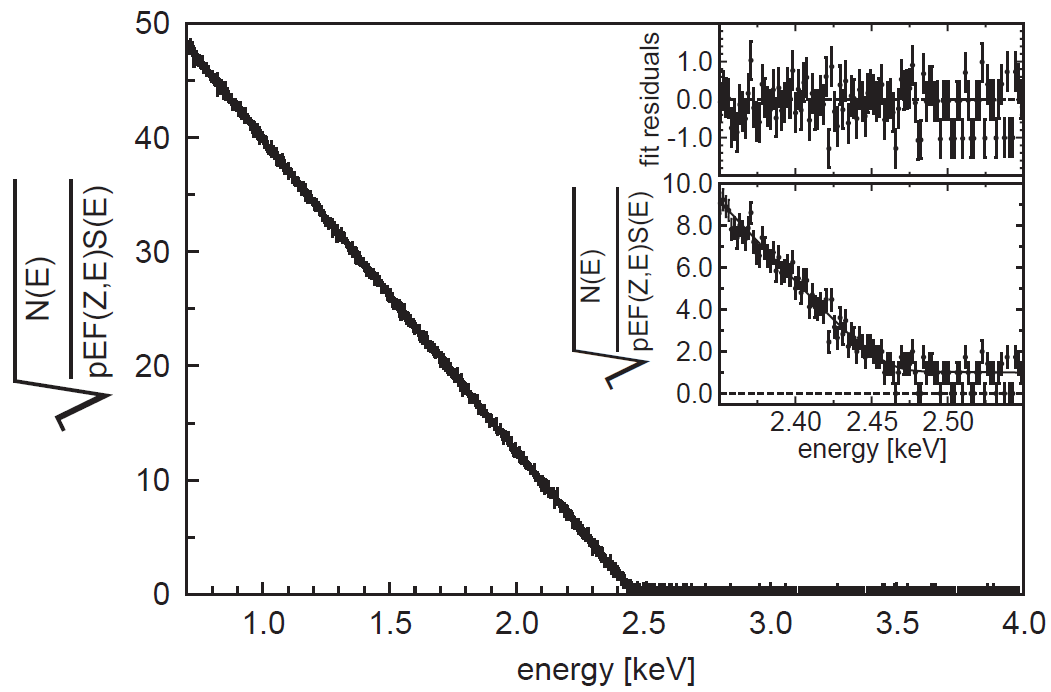}
\caption{Left: The residuals of the theoretically expected \rhenium\ \bspec\ that has been fitted to the data collected by the MANU experiment exhibit effects of a $\beta$ environmental fine structure (BEFS) modulation most clearly visible at low electron energies (Reprinted with permission from~\cite{03-Gal00}, Copyright (2000) by the American Physical Society).
Right: Kurie plot of the experimental \rhenium\ \bspec\ obtained by the MiBeta collaboration (Reprinted with permission from~\cite{03-Arn06}, Copyright (2006) by the American Physical Society).
\label{fig-03:mibeta}}
\end{figure}
(see figure~\ref{fig-03:mibeta}, left). This effect is of concern to future calorimetric neutrino mass experiments, as it produces difficulties to calculate distortions of the \rhenium\ \bspec\ which will become important for experiments aiming at a sensitivities in the sub-eV region~\cite{03-Arn06}. The effect was later also observed in the MiBeta experiment~\cite{03-Sis04}, albeit with a different characteristic due to the use of a different absorber material. In the subsequent analysis the data taken by MANU set an upper limit on the neutrino mass of $\mnue < 26$~eV/c$^2$ at 95\%CL~\cite{03-Gat01}.\\
The second pioneering \rhenium\ \bdec\ experiment was set up by the MiBeta collaboration who were working with AgReO$_4$ absorbers with a mass of about $0.25-0.30$~mg each, read out by silicon implanted thermistors. The group was the first to work with an array of detectors to circumvent the problem of the low maximum activity allowed for the individual crystals. Their setup contained 10 detectors with an average energy resolution of 28.5~eV FWHM~\cite{03-Sis04}. The rise times of the detector signals were of the order 0.5~ms.
Figure~\ref{fig-03:mibeta}, right, shows a Kurie plot of the accumulated spectrum obtained during one year of data taking.
The analysis of the spectrum near the endpoint resulted in an upper limit on the electron neutrino mass of $\mnue < 15$~eV at 90\% CL.

\section{The KATRIN experiment}
The Karlsruhe Tritium Neutrino (KATRIN) experiment is a next-generation direct neutrino mass experiment which is currently under construction by a large international collaboration with groups from Lawrence Berkeley National Laboratory (LBNL), Bonn University, Fulda University of Applied Sciences, Max-Planck-Institute for Nuclear Physics Heidelberg (MPIK), Karlsruhe Institute of Technology (KIT), Mainz University, Massachusetts Institute of Technology (MIT), M\"unster University, University of North Carolina at Chapel Hill, Academy of Sciences of the Czech Republic, University of Santa Barbara, University of Washington (UW) Seattle, Swansea University and Institute for Nuclear Research (INR) Troitsk, with associated groups from Aarhus University, University College London (UCL) and Federal University of Paran$\acute{a}$. The experiment is housed at Tritium Laboratory Karlsruhe (TLK) at KIT's Campus North site. KATRIN has been designed to substantially increase the sensitivity of the Mainz and Troitsk forerunners, while employing the same general spectroscopic principles. It will push the MAC-E-filter technology and tritium process technology, as well as many other methods, to their technological limits to press forward into the sub-eV sensitivity regime of the averaged electron (anti-) neutrino mass \mnue.

\subsection{Introduction}
\label{sec-03:intro}
KATRIN has been designed to improve the \mnue\ sensitivity by a factor of 10 from the present value of 2~eV to 200~meV at 90\%C.L. This increase in sensitivity by one order of magnitude will allow covering almost the entire region of quasi-degenerate neutrino masses (nuclear and particle physics motivation) and to directly probe the neutrino hot dark matter fraction in the universe down to a contribution $\Omega_{\text{HDM}} \approx 0.01$ of the total matter-energy budget of the universe (cosmological motivation) \cite{03-Dre05}. As the experimental observable in \bdec\ is the \emph{square} of the averaged electron (anti-) neutrino mass \mtwonue, this requires an improvement of the experimental precision in $\upbeta$-spectroscopy by two orders of magnitude as compared to Mainz and Troitsk. This in turn requires significant, major improvements to key experimental parameters such as source activity, energy resolution and background rate. In addition, it requires a much better control of systematic effects (by one order of magnitude). The systematic effects mainly arise from parameters related to the tritium source such as fluctuations of the source temperature or the injection pressure or source composition (such as different hydrogen isotopologues or ions), but other systematics such as fluctuations of the retarding potential have to be limited to very small values as well.

\begin{figure}[b!]
  \centering
  \includegraphics[width = \textwidth]{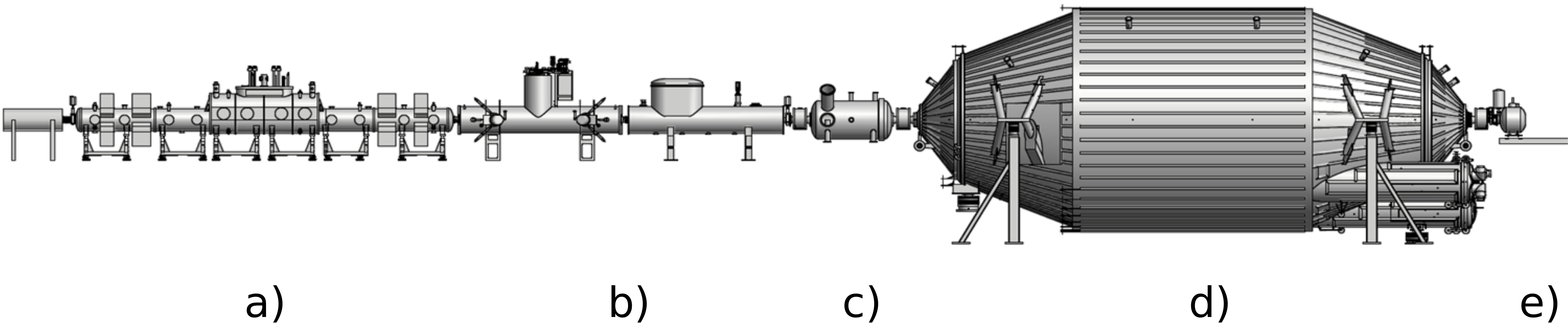}
  \caption{Schematic view of the complete 70~m long KATRIN setup. a) Windowless Gaseous Tritium Source (WGTS).
  b) Transport section consisting of a differential (DPS2-F) and a cryogenic (CPS) pumping section. c) Pre-spectrometer: pre-filter of \bspec . d) Main spectrometer: energy analysis of $\upbeta$-electrons. e) Detector: position-sensitive detection of transmitted electrons~\cite{03-KAT04}}
  \label{fig-03:katrin-beamline}
\end{figure}

Extensive design work by the KATRIN Collaboration reported in \cite{03-KAT04} has revealed that the key statistical parameters to reach a neutrino mass sensitivity of 200 meV are an
increase of the source strength by a factor of 100 and of the measurement time by a factor of 10. The required energy resolution $\Delta E$ is less critical, as the information on the neutrino mass is derived from an energy interval which typically extends to about 30 eV below the endpoint, so that an improvement by about a factor of four to a value of $\Delta E = 0.93$~eV at KATRIN is sufficient. On the other hand, the background rate close to the endpoint has to be limited to a very low value of $<10^{-2}$~cps (counts per second) to achieve the design sensitivity. This is identical to the background value obtained at the Mainz and Troitsk experiments, so that novel background reduction mechanisms have to be developed for the much larger KATRIN spectrometer section.

The 70~m long KATRIN setup has been designed along these general criteria (see figure~\ref{fig-03:katrin-beamline}). The experimental configuration can be grouped into the following major components: a) the Windowless Gaseous Tritium Source (WGTS), where $10^{11}$  electrons are produced per second by the \bdec\ of molecular high-purity tritium gas at a temperature of 30~K, b) an electron transport and tritium elimination section, comprising an active differential pumping (DPS) followed by a passive cryo-pumping section (CPS), where the tritium flow is reduced by more than 14 orders of magnitude c) the electrostatic pre-spectrometer of MAC-E-Filter type, which offers the option to pre-filter the low-energy part of the tritium \bdec\ spectrum, d) the large electrostatic main spectrometer of MAC-E-Filter type which represents the precision energy filter for electrons and which is, with its dimensions of 10 m diameter and 24 m length, the largest UHV recipient in the world (not shown here are its inner electrode system and its outer air coil system) and e) a segmented Si-PIN diode array to count the transmitted electrons. The experiment is completed by a rear section which allows to control and monitor key source-related parameters as well as the monitor spectrometer for redundant monitoring of the retarding potential. Finally, there
are extensive infrastructure facilities.

The unique properties of the gaseous molecular tritium source (high activity and stability) and of the large main spectrometer (ppm precision of the retarding high voltage) allow KATRIN to extent its physics reach from its main goal of measuring the neutrino mass in the sub-eV range to look for contributions by possible sterile neutrinos from the sub-eV up to the multi-keV range. The sensitivity of KATRIN for (sub-) eV sterile neutrinos, as suggested by~\cite{03-Men11, 03-Gal95, 03-Gal06, 03-Agu07} is expected to be very high~\cite{03-Rii11, 03-For11, 03-Esm}. Interestingly, a high precision search of Warm Dark Matter in form of keV sterile neutrinos looks very attractive, as a heavy sterile neutrino would manifest itself as a tiny kink and subsequent spectral distortion deep in the \bspec\, further away from the endpoint. Finally, KATRIN will allow performing stringent tests of other physics issues beyond the Standard Model such as large extra dimensions~\cite{03-Gon}, right handed currents~\cite{03-bonn11} and Lorentz violation~\cite{03-Car00}.

\subsection{KASSIOPEIA: a full monte carlo simulation software for KATRIN}
\label{sec-03:kassiopeia}
The main principles of the MAC-E filter and its application in the KATRIN experiment can be understood analytically via the adiabatic approximation. However, in order to understand the behavior of electrons not only in perfect vacuum and beyond the assumption of adiabatic motion, a precise and fast computational tool is required. Over the past years the KATRIN Collaboration has developed the universal code package \textsc{Kassiopeia}~\cite{03-Fur}, which is based on the ground-laying work described in ~\cite{03-Glu11,03-Glu11b}. The software package includes various particle generators for signal electrons as well as background processes.

\textsc{Kassiopeia} comprises a number of modules for the creation of particles, the calculation of their subsequent trajectories in electromagnetic fields, and the detection of particles in Si-based detectors. To do so, \textsc{Kassiopeia} provides a detailed WGTS model, a number of electric and magnetic field calculation methods and different methods for the calculation of particle trajectories. The particle detection module includes backscattering of electrons on the detector surface and a comprehensive number of physical phenomena of low energy-electrons in silicon~\cite{03-Ren11}.

At KATRIN very large dimensions of the order of tens of meters (main spectrometer) concur with very small dimensions of the order of$~\mu$m (e.g. wire electrode). This fact constitutes the biggest challenge for the electromagnetic field calculation, that can not be handled adequately by commercial programs. A fast and precise field calculation is achieved by a variety of field calculation methods, ranging from very fast axisymmetric field calculations~\cite{03-Glu11,03-Glu11b}, to fully three-dimensional field calculation methods~\cite{03-For12} based on the boundary element methods which is most suitable for large dimensional differences.

The particle trajectory calculations are based on explicit Runge-Kutta methods described in~\cite{03-Ver78, 03-Pri81, 03-Tsi99} and include physical processes like synchrotron radiation and elastic, electronic excitation and ionization collisions of electrons with molecular hydrogen~\cite{03-Hwa96, 03-Tra83, 03-Taw90}.

The detailed tritium source model allows to simulate the actual neutrino mass measurement with high precision. The source model includes, among other things, the final state distribution of tritium \cite{03-doss06,03-doss08}, the Doppler broadening and scattering in the source~\cite{03-Bab, 03-Kae12}.

\textsc{Kassiopeia} was successfully used to study background processes~\cite{03-Wan} and transmission properties of the spectrometer~\cite{03-Pra}, to optimize the electromagnetic design of the spectrometer section and to study systematic effects related to the source section of KATRIN \cite{03-Kae12}.

\subsection{Source and Transport Section}
\label{sec-03:sourcesection}

The tritium source-related parts of KATRIN comprise the Windowless Gaseous Tritium Source (WGTS), the Differential (DPS) and Cryogenic (CPS) Pumping Sections, as well as the Calibration and Monitoring System (CMS) at the rear end. The main tasks of these elements are to (a) control and monitor the tritium column density in the source tube to a precision of better than $10^{-3}$ and (b) reduce the tritium partial pressure from the source inlet to the entrance into the spectrometers by more than 14 orders of magnitude, while at the same time transporting the \bdec\ electrons adiabatically towards the spectrometers.

The extensive tritium infrastructure required for the continuous operation of the KATRIN tritium source is provided by Tritium Laboratory Karlsruhe (TLK), which has developed key technologies for experiments involving large amounts of tritium since the early 1990s. As of today, TLK has an amount 24~g of tritium on-site and holds a license for up to 40~g. Tritium process technology at TLK is based on closed tritium cycles with their central elements of tritium storage, isotope recovery and separation, and tritium retention, among others. The yearly throughput of 10~kg of high purity tritium in the KATRIN source will be equivalent to the ITER operation in the late 2020s~\cite{03-Bor11}. The KATRIN tritium cycle is subdivided into two closed loops, with an Inner Loop being optimized for a highly stabilized injection pressure into the WGTS, and an Outer Loop designed to maintain a high tritium purity by withdrawing a small tritium fraction ($<$1\%) for clean-up and isotope separation (the amount withdrawn is reinjected to the Inner Loop after purification).

\subsubsection{Gas dynamics of the WGTS}
\label{subsec-03:WGTS}

The geometry of the tritium source of KATRIN is defined by a cylindrical stainless steel beam tube of 90~mm diameter and 10~m length, which is housed inside the WGTS cryostat. At the center of the beam tube, high purity molecular tritium gas is injected via capillaries with an inlet pressure of about 10$^{-3}$~mbar. At both ends of the beam tube the injected flow is pumped out by large turbomolecular pumps (TMP). The isotopic composition of the injected hydrogen isotopologues (T$_2$, D$_2$, H$_2$, HT, DT, HD), with T$_2$ dominating by a large factor, is constantly being monitored by Laser Raman spectroscopy. In this windowless geometry the systematic effects for electrons close to the 18.6~keV end point of tritium $\beta$-decay are minimized, allowing to adiabatically transport electrons out of the source by a homogeneous magnetic field of $B=3.6$~T. This field is provided by a system of three large superconducting solenoids surrounding the beam tube. The operating temperature of the beam tube is in the 27-30~K temperature range to minimize the tritium throughput through the beam tube as well as the contribution of thermal Doppler broadening of electron energies due to the molecular motion. Finally, this temperature regime also mitigates systematic effects due to clustering and condensation of hydrogen isotopologues.

Over the past years a detailed 2D/3D model of the gas dynamical characteristics of the tritium source has been developed, which is much more detailed than the initial 1D model described in ~\cite{03-KAT04}, where perfect temperature homogeneity and isotropy of the source was assumed. The new gas dynamical model of the source has been used to derive precise values for the integral WGTS column density and its radial variation and the  pressure distribution along the beam tube~\cite{03-Glu12}. The stability of the column density, which is of the order of $5 \cdot 10^{17}$~T$_2$-molecules/cm$^2$, plays a central role in the sensitivity of KATRIN, as it impacts not only the statistical accuracy of the measurement, but also defines the energy losses of the \bdec\ electrons inside the source due to inelastic scattering~\cite{03-Ass00}. An important aspect in the detailed modelling of the source characteristics is the fact that the gas rarefaction parameter along the tube varies from the strong hydrodynamic regime at the injection point to free molecular flow at the beginning of the differential pumping chambers. Accordingly, a significant part of the beam tube exhibits values in the intermediate regime, where the phenomenological intermediate conductance formula of Knudsen is applied~\cite{03-Dus62}.

\subsubsection{The WGTS cryostat and associated control and monitoring systems}
\label{subsec-03:sourcemonitoring}

Initial estimates in ~\cite{03-KAT04} have revealed that a stabilisation of the column density on the 0.1~\% level is required to obtain a neutrino mass sensitivity of 200~meV. This translates into a stabilisation of the inlet pressure and the beam tube temperature on the same level, which certainly is a major technological challenge, given the macroscopic WGTS beam tube dimensions. A stable gas feed into the WGTS is achieved by injecting high-purity tritium gas from a pressure controlled buffer vessel over a 5 m long tritium capillary with constant conductance, while pumping out the gas at both ends of the beam tube with large TMPs.
To fulfil the requirements on temperature stability, a novel beam tube cooling system based on a 2-phase boiling fluid has been designed and tested successfully (a two phase system, when used as a cooling method, has the well-known distinct advantage that it can absorb heat without changing its temperature). The choice of neon as cooling fluid has the advantage that it coexists in two phases at temperatures around 30~K for moderate pressures in the range from 1\,-\,3~bar.

In combination with the requirements of adiabatic electron transport and safe tritium throughput in a closed loop system, the required stability of the tritium column density on the 10$^{-3}$ level has resulted in a technical design of a complex cryostat (see figure~\ref{fig-03:WGTS}) with a length of 16.1~m and a weight of 27~tons~\cite{03-Gro08} and a cryogenic system  comprising 13~fluid circuits where 6~cryogenic fluids are processed. The cryostat also houses an extensive array of more than 500~sensors (magnetic field, pressure, gas flow, liquid levels, voltage taps) as part of a dedicated measurement and control system. The precision temperature readings of the sensor array will later on be used as input to model minute fluctuations of the actual source characteristics over typical time scales of about 1~min.

\begin{figure}
  \centering
  \begin{minipage}{0.74\textwidth}
    \includegraphics[width = 11cm]{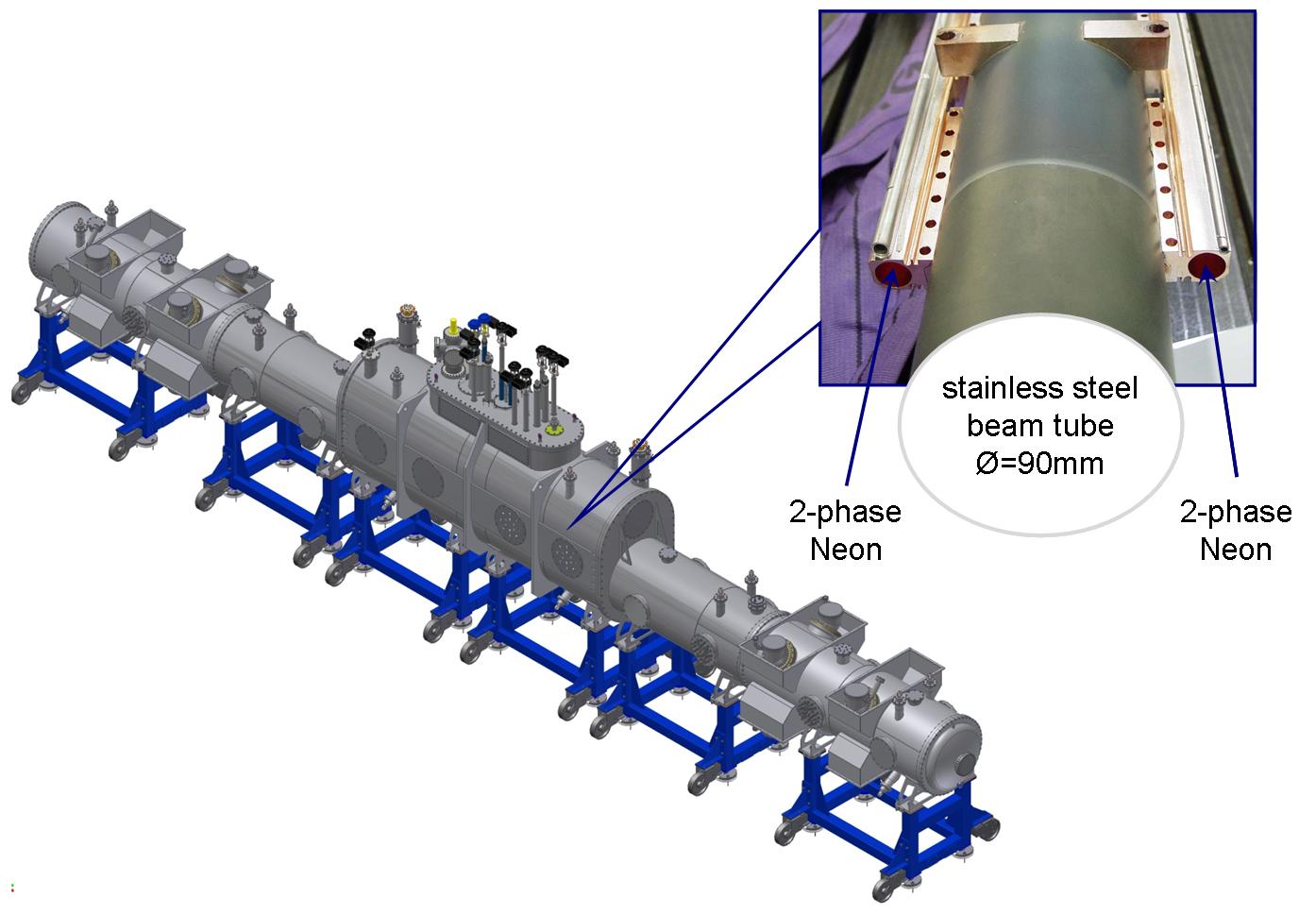}
  \end{minipage}
  \hfill
  \begin{minipage}{0.25\textwidth}
    \caption{Schematic view of the 16~m long WGTS cryostat. The inlet shows the beam tube with the two phase liquid neon cooling system.}
    \label{fig-03:WGTS}
  \end{minipage}
\end{figure}

In order to test the advanced cryogenic technologies required for the cooling of the beam tube, a large-scale test unit, the WGTS demonstrator, was built. This 10 m~long (1:1 scale) unit makes use of many original components which will be used later on in the final WGTS cryostat. In 2011 extensive measurements with the WGTS demonstrator were performed at TLK targeted at a proof-of-principle and optimization of the beam tube cooling system. To obtain a precise measurement of the beam tube temperature, an array of 24 metallic resistance temperature sensors (Pt500) were distributed along the beam tube. The sensors feature a sub-mK resolution and, via calibration with sensors measuring the saturation pressure of neon~\cite{03-Gro11}, an uncertainty for the absolute temperature of 4~mK only. During the measurements, a peak to peak variation of $\Delta T = 3$~mK over a time period of 4~h at $T = 30$~K was obtained, which is one order of magnitude better than required~\cite{03-Gro11, 03-Bab}. This result offers the potential to substantially reduce the systematic errors from the tritium source during the long-term measurements. At present, the WGTS demonstrator is being reassembled to the final WGTS cryostat. These extensive works will be completed by the end of 2014.

A dedicated Laser Raman (LARA) system, based on inelastic Raman scattering~\cite{03-Lon02} of photons from gas molecules, is connected to the inner tritium loop and allows for in-line and near time monitoring of the gas composition.
The different tritiated gas species (T$_2$, DT, HT) have to be distinguished not only because of their different properties in the gas dynamical calculations, but also due to their slightly different endpoint energies as well as rotational and vibrational final states.
Test measurements with a Raman cell suitable for operation with tritium~\cite{03-Lew08, 03-Stu10} show that a precision of $\Delta_{prec}(\epsilon_{T_2})/\epsilon_{T_2} = 0.1\%$
for $T_2$ monitoring can be achieved in measurement intervals of $< 250$~s under KATRIN operating conditions~\cite{03-Sch11, 03-Fis11}. With an optimized setup the acquisition time recently has been reduced to 60~s only.

\subsubsection{Calibration and Monitoring System upstream of the WGTS}
\label{subsec-03:cms}
To monitor the stability of the tritium activity and of the column density of the windowless gaseous tritium source, the rear side of the WGTS is connected to a
calibration and monitoring system. An angular-selective electron gun  is used to measure the energy loss function of the electrons due to inelastic scattering in the WGTS and to monitor the integral column density at different source radii. This novel UV-laser based e-gun is based on the photo-electric effect producing mono-energetic electrons with stable intensity and with a well-defined starting angle $\theta$ w.r.t. the magnetic $\vec{B}$-field~\cite{03-beck12, 03-Val09, 03-hugenberg10,03-Val11}. Secondly this electron source will be used to determine the transmission function of the main spectrometer and the response function of the whole setup with high precision.

The calibration and monitoring system has another very important task related to the source operation.
The \belec s will be guided from the source to the spectrometer and mapped onto the detector within a magnetic flux tube of 191~T~cm$^2$.
The guiding by superconducting solenoids implies a tight transverse confinement by the Lorentz force to all charged particles. This includes
the $10^{11}$ daughter ions per second, which emerge from \bdec\ in the source tube,
as well as the $10^{12}$ electron-ion pairs per second produced therein by the \belec -flux through ionization of \ttwo\ molecules.
The strong magnetic field of 3.6~T within the source is confining this plasma strictly in the transverse direction so that charged particles
cannot diffuse to the conducting wall of the source tube where they would get neutralized \cite{03-Nas05}. The charges within the windowless gaseous tritium source will thus be neutralized by a gold-plated crystalline rear wall which is part of
the calibration and monitoring system, and on which all magnetic field lines from the windowless gaseous tritium source end. Accordingly, the high longitudinal conductance of the plasma
will define and also stabilise the electric potential within the tritium source.

Finally,  the source activity will be monitored by $\upbeta$-induced X-ray spectroscopy (BIXS), where the potential-defining rear wall acts as an X-ray converter.
An encapsulated X-ray detector positioned behind the rear wall will monitor the source activity by recording the X-ray intensity, while the fluorescence lines from the surrounding structures would allow an in-line calibration of the setup.

\subsubsection{Tritium retention techniques}
\label{subsec-03:transportsection}

Since the spectrometer section must be essentially tritium-free for background reasons (see next chapter), the tritium flow along the beam line must be reduced from its initial value at the injection point of $1.8$~$\text{mbar}\cdot\ell$/s to a level of $10^{-14}$~$\text{mbar}\cdot\ell$/s at the entry to the spectrometers. This unprecedented large suppression factor will be achieved by a combination of differential and cyrogenic pumping~\cite{03-Luk11,03-Luo06,03-Gil10,03-Luo08}. While reducing the tritium flux by 14~orders of magnitude these sections also have to maintain adiabatic guidance of $\beta$-decay electrons from the WGTS to the spectrometers over a distance of more than 15~m.

\paragraph{Differential pumping section (DPS2-F)}

Differential pumping by large TMPs is the first tritium retention technique in use at the KATRIN beam line, with the initial stages being performed inside the WGTS cryostat by pump ports at the rear (DPS1-R) and front (DPS1-F) section. Thus, the DPS2-F cryostat is the second pumping unit in the forward direction. Its scientific objectives are the following: (a) active pumping of tritium molecules with TMPs (b) reduction of the tritium flow by a factor larger than 10$^4$ (c) maintaining a stable tritium circulation in the Outer Loop, and (d) diagnostics of the composition and suppression of ion species from the WGTS.

\begin{figure}
  \centering
  \begin{minipage}{0.7\textwidth}
    \includegraphics[width = 10.5cm]{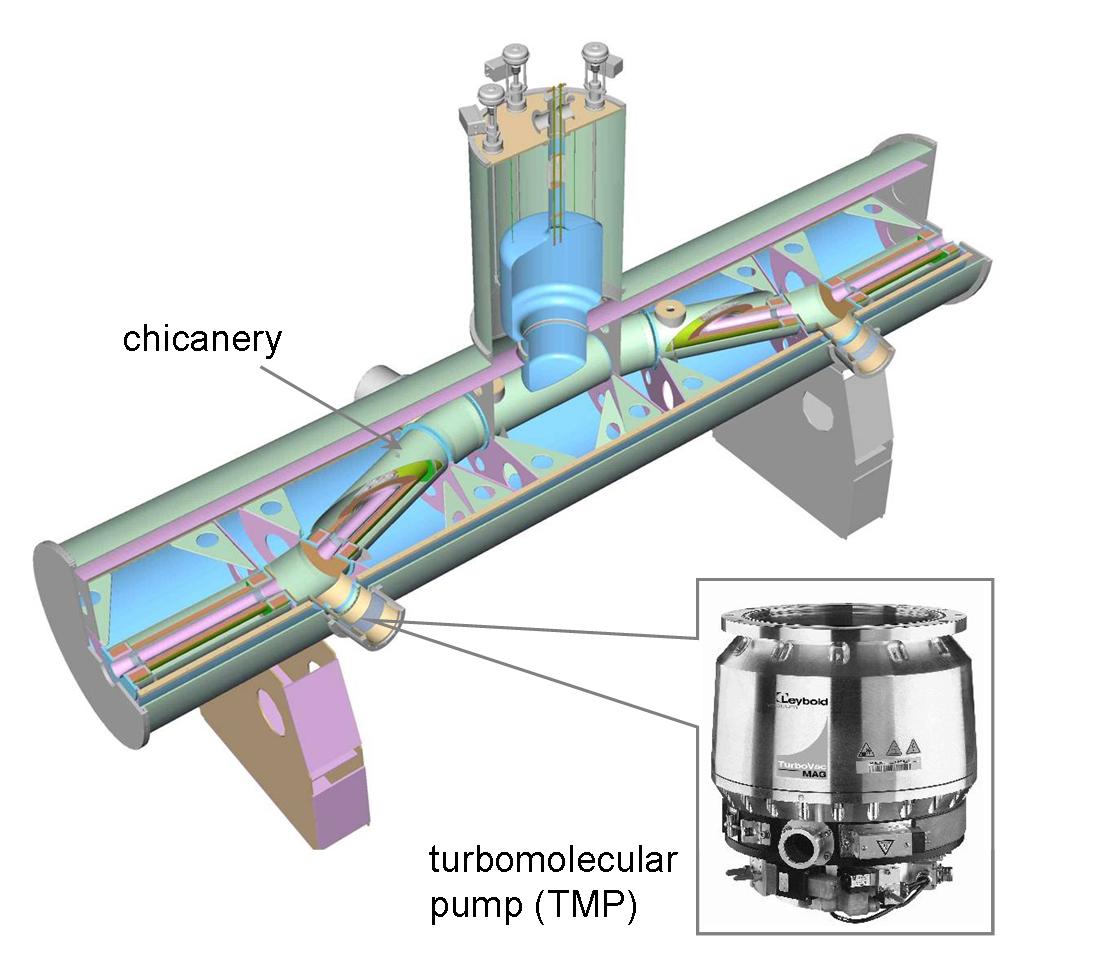}
  \end{minipage}
  \hfill
  \begin{minipage}{0.29\textwidth}
    \caption{Schematic view of the 6.96\,m long DPS2-F cryostat. The beam tube guiding the \bdec\ electrons features a chicane
     to avoid the molecular beaming effect. At the four pump ports large TMPs are installed for differential pumping. The beam tube instrumentation includes FT-ICR units and through-shaped electrodes for ion diagnostics and suppression.}
    \label{fig-03:DPS}
  \end{minipage}
\end{figure}

The DPS2-F has been designed as a single large cryostat of 6.96~m length (see figure~\ref{fig-03:DPS}). It was the first KATRIN source-related component on site at KIT and was successfully commissioned in 2010/11~\cite{03-Kos12}. The unit features a beam line with an $\Omega$-shaped chicane to avoid the {\it molecular beaming} effect, in which case gas particles would not be pumped out due to the alignment between gas particle momenta and beam tube axis. The cascaded differential pumping system consists of four large TMPs mounted on pump ducts, which minimize the effect of thermal radiation from the TMPs operated at room temperature to the beam tube kept at 77~K, while also limiting TMP rotor heating due to induced eddy-currents~\cite{03-Wol11}. During the commissioning measurements a retention factor for tritium (based on measurements for other hydrogen-like and noble gas species) of $\approx$\,2.5\,$\times$\,10$^4$ was obtained, demonstrating the excellent tritium retention characteristics of differential pumping already at room temperature~\cite{03-Luk11}.

The DPS2-F beam tube instrumentation is used for diagnostics and suppression of ions which result from ionization processes of primary \bdec\,-\,electrons with the neutral gas molecules of the source. As ions are guided by the field lines (and thus are not being pumped out) they remain trapped between a ring electrode at the end of the DPS2-F, where they are reflected electrostatically, and the WGTS, where they are reflected back by the gas pressure. The number density and composition of ions will be measured on a near-time basis by two Fourier Transform-Ion Cyclotron Resonance (FT-ICR) devices \cite{03-Ubi09}, which discriminate Penning-trapped ions by detecting differences in Ion Cyclotron frequencies in the 5.6 T B-field at the center of the DPS2-F solenoids. The ions stored between
the WGTS and the end of the DPS2-F are eliminated by a system of three electrostatic dipoles by an induced $\vec{E}\times\vec{B}$-drift, irrespective of their mass, charge sign or motion with respect to the electric field, within a time period of 20~ms or less, which is much shorter than the allowed upper time limit of 10~s for ion storage.

The magnetic flux tube of 191~T~cm$^2$ of \bdec\ electrons which is used for the neutrino mass analysis is guided through the beam tube chicane with a minimum margin of 10~mm from the walls of the inner vacuum system. The system of 15 superconducting solenoids is cooled by liquid helium supplied from an external Linde TCF~50 refrigerator, with a measured DPS2-F cryogenic stand-alone time of about eight hours~\cite{03-Gil11}. Unfortunately, due to a malfunction of one of the protective diodes of the magnet system during the commissioning measurements, further tests with the DPS2-F had to be stopped in mid-2011. At present different options to ensure a future fail-safe operation of the differential pumping system during long-term tritium running are being studied.

\paragraph{Cryogenic pumping section (CPS)}

The Cryogenic Pumping Section (CPS) is the final tritium retention unit in the KATRIN beam line, featuring a large-area cryosorption pump based on a thin layer of argon snow frosted onto gold-plated steel as adsorbent (see figure~\ref{fig-03:CPS}) \cite{03-kaz07}). It is designed for a tritium flow reduction factor of $\geq$ 3\,$\times$\,10$^7$.

\begin{figure}
  \centering
  \begin{minipage}{0.7\textwidth}
    \includegraphics[width = 10cm]{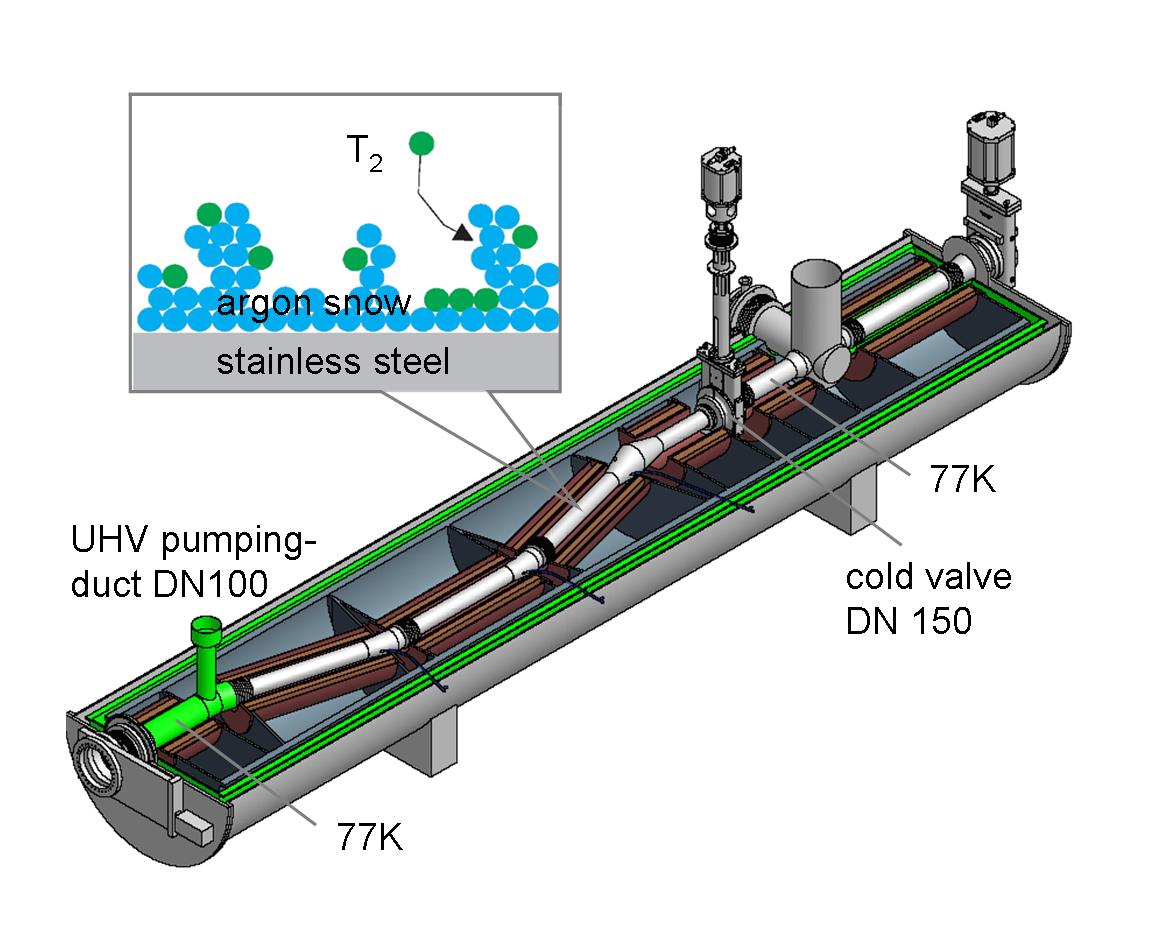}
  \end{minipage}
  \hfill
  \begin{minipage}{0.29\textwidth}
    \caption{Schematic view of the seven beam tube elements of the CPS. The inner surface of the cryopump elements is covered with argon snow to capture the remaining tritium molecules.}
    \label{fig-03:CPS}
  \end{minipage}
\end{figure}

The 6.73~m long CPS cryostat features an even more complex beam tube design than DPS2-F with a beam tube segmented into 7~sections, as shown in figure~\ref{fig-03:CPS}. To prevent residual tritium absorption onto the walls, the inner surfaces of the beam tube elements are gold-plated. The beam tubes of the cryopump section are tilted by an angle of 15$^{\circ}$ with respect to each other and are covered by a thin argon frost layer to enlarge the inner surface for sorption. The CPS design allows the insertion of a Condensed $^{83m}$Kr Source (CKrS)~\cite{03-Ost09} for calibration purposes, as well as a small Si-diode for source intensity monitoring. The CPS is currently being manufactured by an industrial partner in Genoa (Italy) and will be shipped to KIT at the end of 2013. After commissioning and initial bake-out, the CPS will regularly cycle through several modes of operation. In a first step the beam tube elements will be cooled to their respective operational temperature, which is followed by the preparation of an argon frost layer. This requires a temperature in the range of 6~K to favour the growth of small to medium sized crystallites. After argon frost preparation, the cryopump temperature will be lowered again to 3~K. After a data collection cycle of about 60~days the CPS has to be regenerated to remove the adsorbed tritium activity of about 2~Ci by purging 100~K gaseous helium through the cryopumping segments.

By combining differential pumping by TMPs in the DPS with cryosorption on a large-area argon frost pump in the CPS, the number of migrating HT-molecules to the pre- and main spectrometer is kept at a minimum. When including all relevant tritium retention techniques, i.e. not only DPS and CPS, but also tritium interactions with the spectrometer walls and tritium pumping by the large non-evaporable getter (NEG) strips in the spectrometers, a tritium partial pressure in the main spectrometer of about 10$^{-22}$ mbar can be expected~\cite{03-Kos12}. This ultra-low level should result in a tritium-related background rate of only 10$^{-4}$ cps, which is two orders of magnitude below the design specification. It is only by this huge tritium reduction factor that the KATRIN experiment can combine an ultra-luminous windowless gaseous tritium source of 10$^{11}$ $\beta$-decay electrons with a high-resolution electrostatic spectrometer operated with an extremely low background level of 0.01 cps~\cite{03-Mer12}.

\subsection{Spectrometer and detector section}
\label{sec-03:sds}
In the following the non-tritium related parts of KATRIN will be described. First the tandem spectrometer system performing the energy analysis of the $\upbeta$-electrons is addressed and then the detector and data acquisition system are described.

\subsubsection{Overview: pre-filtering and precision scanning}
\label{subsec-03:spectrometer}
The spectrometer section consists of two electrostatic retarding filters of the MAC-E-Filter type: the \prespectrometer{} and the main spectrometer. The \prespectrometer{} offers the option to act as a pre-filter, reflecting all electrons with energies up to 300~eV below the endpoint, while transmitting the interesting part of the spectrum undisturbed. All electrons transmitted through this first stage are guided to the main spectrometer for precise energy analysis close to the endpoint. A third spectrometer, the monitor spectrometer, is used to monitor precisely the high voltage of the main spectrometer.

\paragraph{Pre-spectrometer}
The \prespectrometer{} has a length of 3.4~m and a diameter of 1.7~m. At both ends a superconducting magnet is installed providing a magnetic field of 4.5~T at the center of the magnet and 15.6~mT in the center of the spectrometer. As a novel electromagnetic design feature, if compared to the Mainz and Troitsk set-ups, the tank itself is set on high voltage. An inner electrode system consisting of two conical full electrodes and a central cylindrical wire electrode can be set to a different potential as the vessel. Figure \ref{fig-03:prespec} shows a schematic view of the \prespectrometer{}.

One of the \prespectrometer{}'s major tasks has been to serve as a prototype for advanced technologies and experimental methods later applied to the much larger main spectrometer. Many basic concepts, such as the design of the ultra high vacuum (UHV) system and high voltage stabilization were successfully tested at the \prespectrometer{}. Especially important was the detailed investigation of background processes in MAC-E-filters. Two new classes of background were identified and studied at the \prespectrometer{}: background due to small Penning traps and radon induced background~\cite{03-Fra10}. The knowledge gained during the extensive measurement period at the \prespectrometer{} has proven to be extremely valuable for the design and operation of the main spectrometer.

\begin{figure}
  \centering
  \begin{minipage}{0.7\textwidth}
    \includegraphics[width = \textwidth]{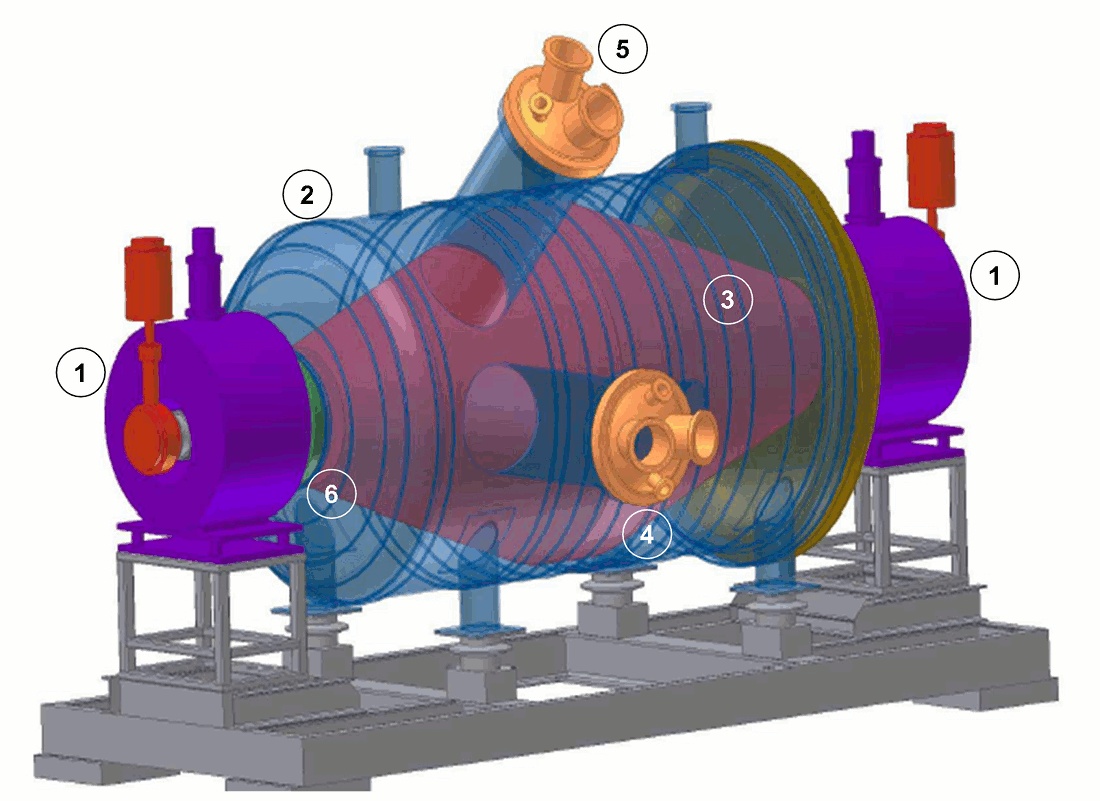}
  \end{minipage}
  \hfill
  \begin{minipage}{0.29\textwidth}
    \caption{Schematic view of \prespectrometer{}. 1: superconducting solenoids, 2: \prespectrometer{} vessel, 3: inner electrode system, 4: $90^{\circ}$ pump port, 5: $45^{\circ}$ pump port, 6: insulator.}
    \label{fig-03:prespec}
  \end{minipage}
\end{figure}

\paragraph{Main spectrometer}
The exceedingly large dimensions of the main spectrometer are essential to operate it as an extremely precise high-energy filter
with an energy resolution of $\Delta E = 0.93$~eV (0\%-100\% transmission). This sharp resolution requires that the stability of the retarding high voltage is in the ppm range. The scanning voltage is varied in steps of $\Delta U = 0.5 - 1$~V around a narrow region close to the endpoint at $qU = E_0 = 18.6$ keV. The highest electrostatic potential is located in the central plane of the spectrometer, perpendicular to the beam axis, the so called ``analyzing plane''.

\begin{figure}[tb]
  \centering
  \begin{minipage}{0.5\textwidth}
    \includegraphics[width = 0.9\textwidth]{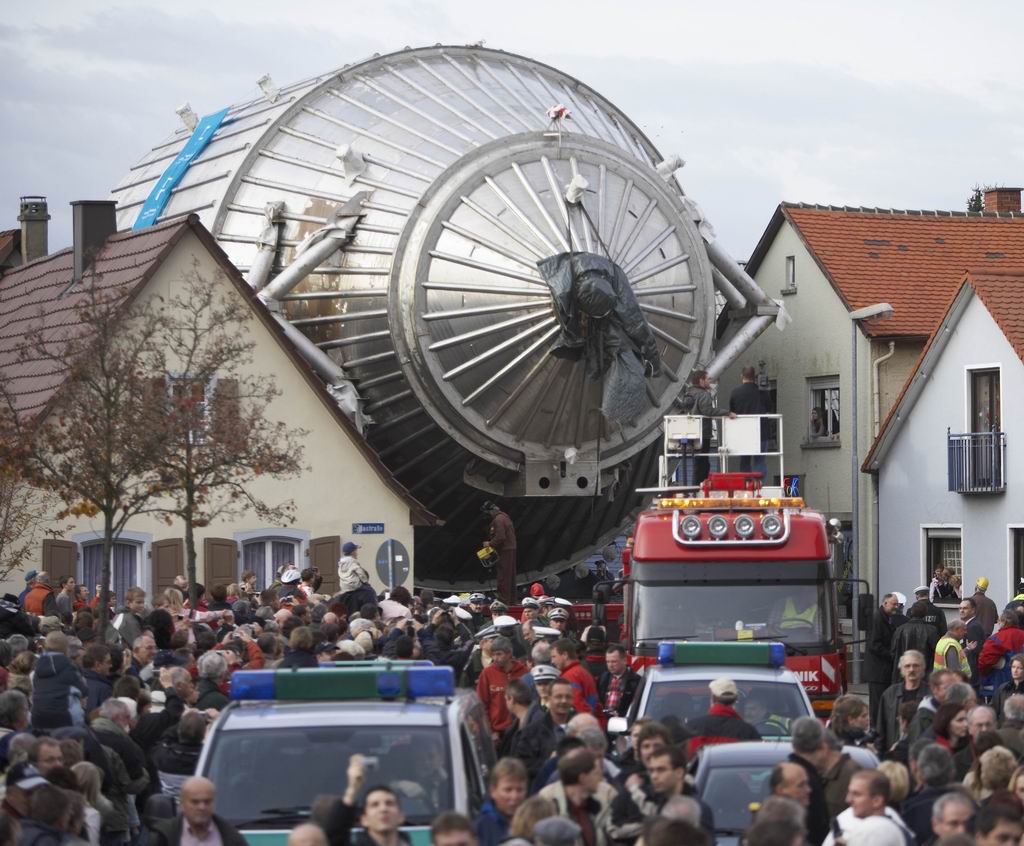}
  \end{minipage}
  \hfill
  \begin{minipage}{0.49\textwidth}
    \caption{Photograph of the spectrometer transport during the final careful maneuvering through the nearby village of Leopoldshafen in November 2006. Due its size it could not be transported to Karlsruhe on motorways, but had to travel nearly 9000~km through the Danube River, the Black Sea, the Mediterranean Sea, the Atlantic Ocean, the North Sea and the River Rhine~\cite{03-KAT}.}
    \label{fig-03:KATRIN}
  \end{minipage}
\end{figure}

To achieve a very high energy resolution, the cyclotron motion of the electrons, being isotropically created in the WGTS, needs to be almost fully transformed into longitudinal motion parallel to the magnetic field lines, since only the latter is analyzed by the retarding potential (see section~\ref{sec-03:mace}). To obtain adequate parallelization, the magnetic field drops by four orders of magnitude from the entrance to the center of the main spectrometer. Since the magnetic flux $\Phi$ is conserved, the cross section of the flux tube in the center plane has to be four orders of magnitude larger than at the entrance. This scaling explains the huge size of the main spectrometer (length $\text{L} = 23.8~\text{m}$, diameter $\text{d} = 9.8~\text{m}$, surface area $\text{A} = 650~\text{m}^2$ and volume $\text{V} = 1450~\text{m}^3$) (see figure~\ref{fig-03:KATRIN}).

To compensate for the distorting earth magnetic field and to fine-tune the magnetic field inside the main spectrometer, the vessel is surrounded by a large external air coil system (see figure~\ref{fig-03:AirCoil}). It consists of 10~horizontal current loops and 16~vertical ones to compensate the earth magnetic field (EMCS). Additionally, a system of 15~Helmholtz-like coils with individually adjustable currents allows for precise adjusting of the gradients and the overall strength of the magnetic field (LFCS). Two arrays of magnetic field sensors, one at fixed positions, and the other a mobile one attached to robots \cite{03-osipowicz12}, allow to measure and monitor precisely the magnetic field.

\begin{figure}[tb]
  \centering
  \begin{minipage}{0.6\textwidth}
    \includegraphics[width = 0.9\textwidth]{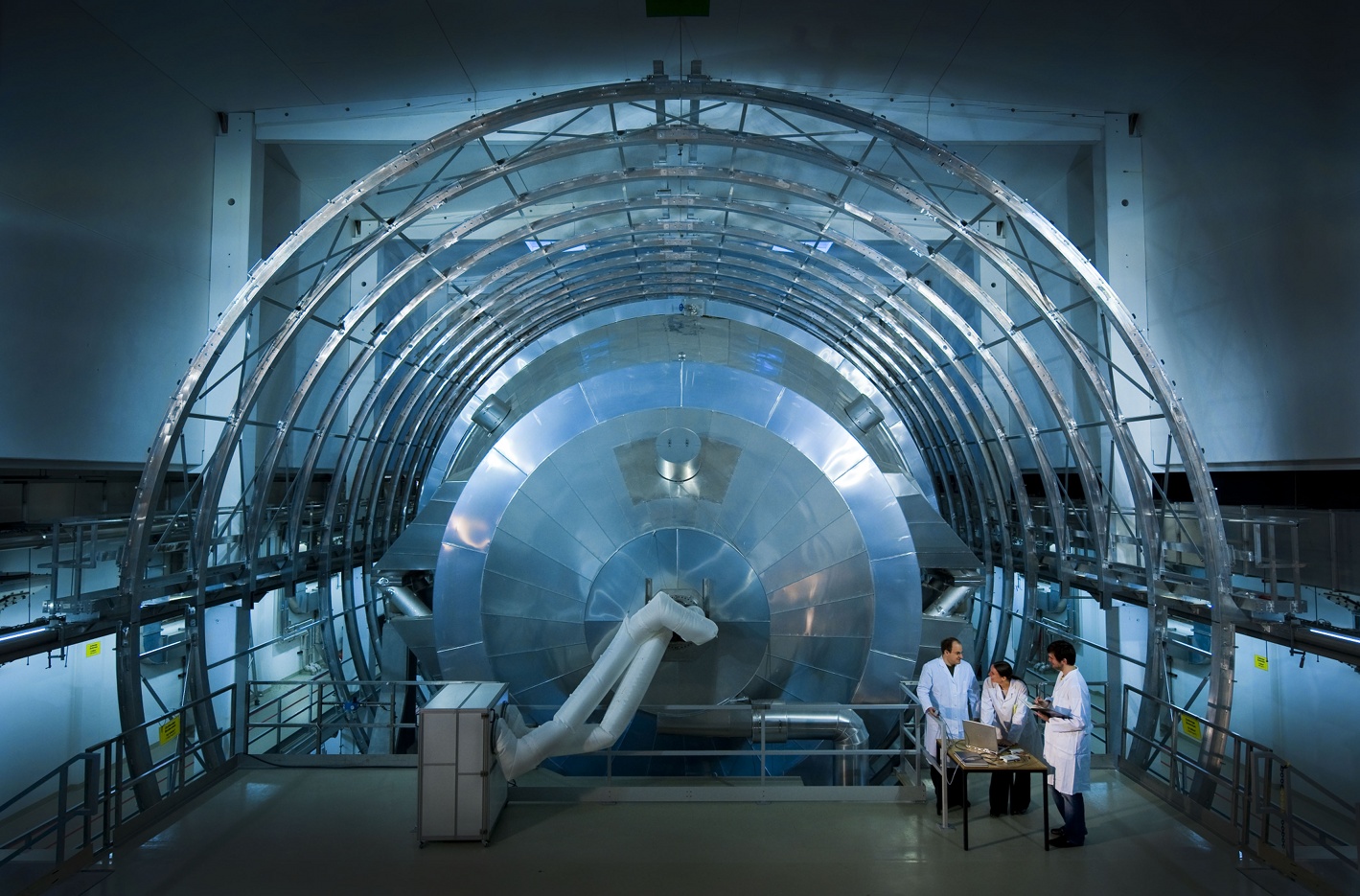}
  \end{minipage}
  \caption{Photograph of air coil system surrounds the main spectrometer vessel. Its purpose is to compensate the earth magnetic field and to fine tune the magnetic field inside the main spectrometer.}
  \label{fig-03:AirCoil}
\end{figure}

\begin{figure}[b]
  \centering
  \begin{minipage}{0.4\textwidth}
    \includegraphics[width = 0.9\textwidth]{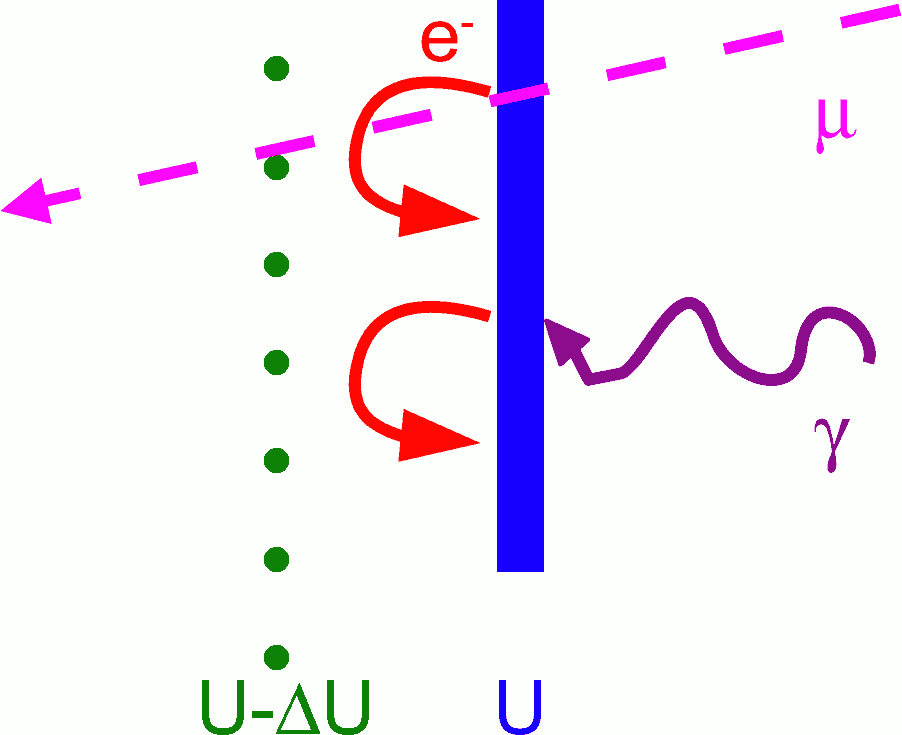}
  \end{minipage}
  \hfill
  \begin{minipage}{0.59\textwidth}
    \caption{Principle of electrical screening of a surface at high voltage $U$ by a nearly massless wire electrode at a slightly more negative voltage $U-\Delta U$. Secondary electrons, which are emitted from the surface at potential $U$ -- {\it e.g.} by the interaction of cosmic muons or $\gamma$ radiation by environmental radioactivity  -- will be reflected by the wire potential back
to the wall, if the kinetic energy of the secondary electrons is not sufficient to pass the electrical potential barrier $U-\Delta U$. If the wire electrode has a sufficiently small cross section, the probability that unscreened secondary electrons emitted from the wires is small.}
       \label{fig-03:screening_principle}
  \end{minipage}
\end{figure}

The main background source of the Mainz experiment was secondary electrons from the electrodes
at high potential. These electrons are created by cosmic muons or by environmental radioactivity. The shielding effect of the magnetic field will prohibit most of these electrons to enter the magnetic flux tube and thus to reach the detector. However, measurements at Mainz \cite{03-dipl_schall01} and Troitsk yielded a transmission rate of $10^{-5} - 10^{-7}$ through the magnetic shielding of a MAC-E-filter. A large shielding factor is important considering the large surface of 650~m$^2$ of the KATRIN main spectrometer. To suppress this kind of background a new idea of additional background suppression has been developed and successfully tested at the Mainz spectrometer \cite{03-dipl_ostrick02, 03-phd_flatt05} and optimized at the pre-spectrometer. This technique will also be applied to the KATRIN main spectrometer:
The vessel walls at high potential will be covered by a system of nearly massless wire electrodes, which are put to a slightly more negative potential (see fig. \ref{fig-03:screening_principle}).
Secondary electrons ejected by cosmic rays or environmental radioactivity from the vessel wall will thus be repelled
and prohibited from entering the magnetic flux tube.
To achieve a high suppression factor of 10, as demonstrated in the Mainz experiment, the KATRIN main spectrometer has been instrumented by a two-layer wire electrode system \cite{03-Val06, 03-Val10, 03-Pra12b} (see figure \ref{fig-03:wire_electrode}).

This system consists of 248 modules of a typical size of $2 \times 1.5$~m$^2$. The wires of the inner (outer) layer have a diameter of 200~$\mu m$ (300~$\mu m$) and  a distance of 22~cm (15~cm) from the surface of the spectrometer vessel. Essentially the inner (outer) wire layer will be on a potential of -200~V (-100~V) with respect to the main spectrometer vessel. The total number of wires amounts to 23440. Stringent demands are required for this electrode system: The two wire layers in the central part have to be mounted with a precision of 200~$\mu m$ to avoid distortions of the electrical potential. The wires have to withstand baking at $350~^\circ$C. The outgassing of the wire electrode system has to be extremely low to allow a final residual gas pressure of $10^{-11}$~mbar. The mounting of the wire electrode system within the huge KATRIN main spectrometer under clean-room conditions was a big challenge on its own (see figure \ref{fig-03:wire_electrode}). The installation of the wire electrode system was successfully completed in January 2012.

\begin{figure}
  \centerline{\includegraphics[width = 0.49\textwidth]{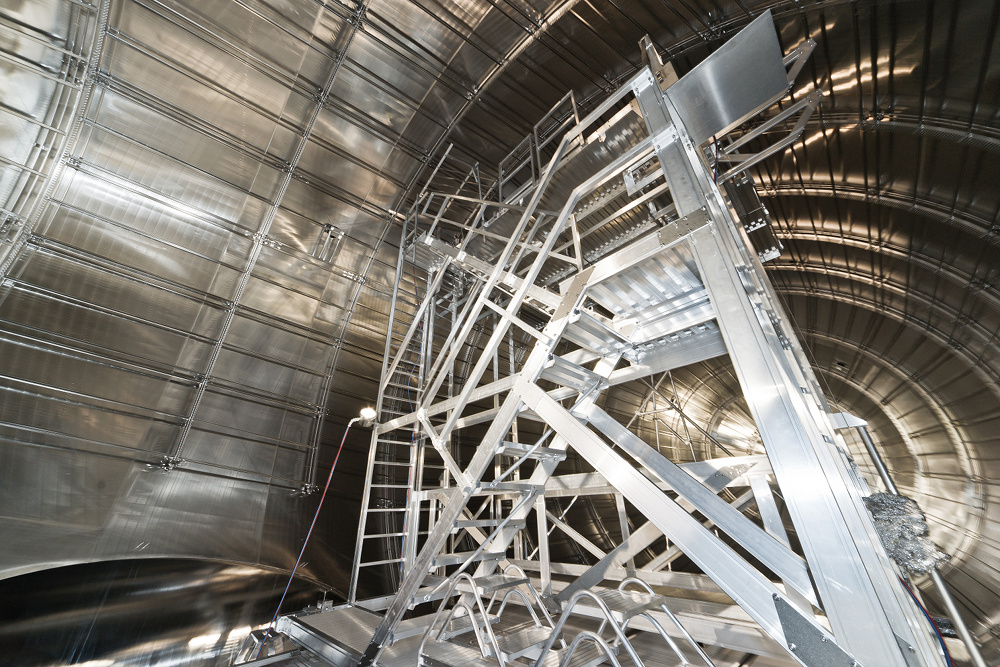}
  \hfill
  \includegraphics[width = 0.49\textwidth]{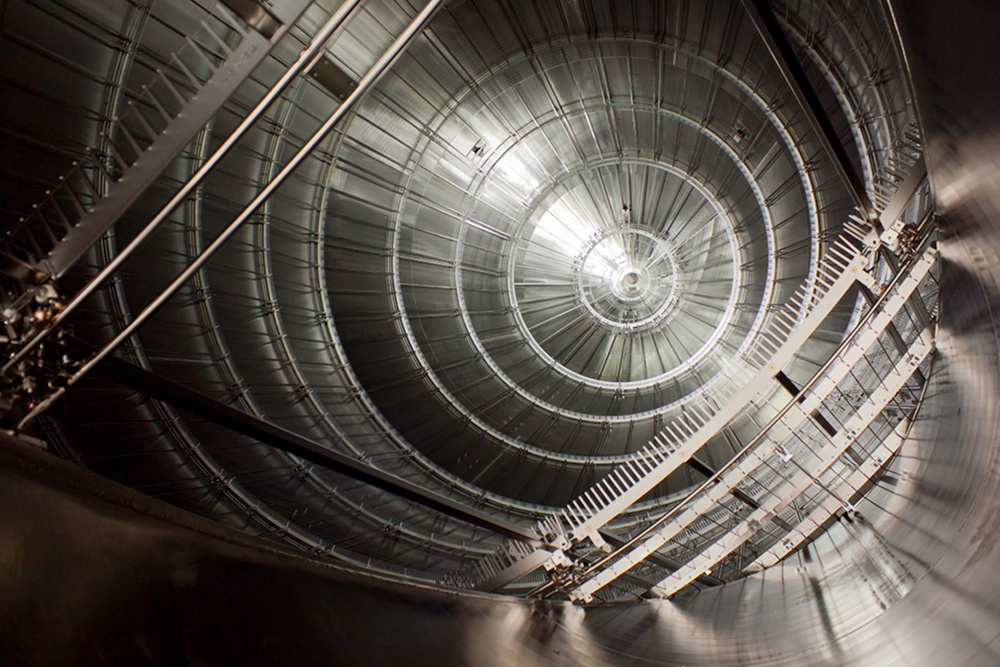}}
   \caption{Photographs of the oil- and dust-free scaffold to install the wire electrode system in the main spectrometer (left) and view into the main spectrometer before the last two wire electrode modules at one of the pump ports were installed
   to finish the wire electrode installation (right). Photographs: Michael Zacher.}
       \label{fig-03:wire_electrode}
\end{figure}

In addition to the background reduction the wire electrode serves three other purposes:
Fine-shaping of the electric retarding potential is achieved by applying slightly different voltages on the different axial rings of the wire modules to guarantee full adiabatic motion of the electrons. Reducing fluctuations of the electric retarding potential is done by decoupling the low-noise high-precision retarding voltage
of the inner-most wire layer from the high-voltage of the vessel, on which electrical devices such as TMPs are mounted.  The multi-layer system acts as a Faraday cage.
Finally, by splitting the wire module arrangement into an eastern and a western half an electric dipole voltage can be applied to eject low-energy stored particles by an induced
$\vec E \times \vec B$ drift (see. equation (\ref{eq-03:drift})).

\paragraph{High voltage and monitoring layout}
The energy analysis of \bdec\ electrons based on the MAC-E filter principle  \cite{03-pic92a,03-Lob85} relies primarily on the stability of the electrostatic filter potential \cite{03-Ott06,03-Kas04}.
The latter has been identified as one of the five main contributions to the KATRIN uncertainty budget \cite{03-KAT04}.
Its contribution to the systematic error of the energy filter potential has to be restricted to $\Delta m^2 < 0.0075$\evtwo\ in order to meet the desired level of sensitivity.

We showed earlier in equation (\ref{eq-03:sigma_mtwonue}) that any unknown fluctuation of the retarding voltage $U$ with variance $\sigma_U^2$ leads to a shift of the squared neutrino rest mass $\Delta \mtwonue \approx - 2 \cdot q^2 \sigma_U^2$ (The electron charge $q$ takes care of the conversion from voltage to energy.).
This relation and the systematic error limit given above have been taken into account in order to define the stability requirements for the KATRIN high voltage monitoring system, which is $\sigma_U < 60$\,mV \cite{03-KAT04}.
At the tritium endpoint energy $E_0 = qU = -18.6$\,keV this corresponds to a relative voltage stability of $\frac{\Delta U}{U} < 3.3 \cdot 10^{-6}$, which has to be maintained for the whole KATRIN measurement time.

Therefore KATRIN will apply two redundant monitoring methods, one being based on two high voltage dividers with highest possible precision, i.e.\ $10^{-6}$ relative precision and long-term stability for up to 35~kV \cite{03-Thu09}, and the other one being based on a monitoring beam-line which observes a mono-energetic electron source \cite{03-KAT04}. To do so in both cases, the low-noise retarding high voltage applied to the inner layer of the central wire electrode modules is distributed to the precision high voltage dividers and to the monitor spectrometer.

The monitor spectrometer consists of the refurbished and upgraded MAC-E filter spectrometer of the Mainz experiment  (see subsection \ref{sec-03:mainz}) and observes an implanted \rbkr\ source \cite{03-ven10, 03-zboril11}. For the calibration source \rb\ is produced either at the ISOLDE facility at CERN or at the cyclotrons at Nuclear Research Institute (Rez) in the Czech Republic or at Bonn University in Germany \cite{03-ven05, 03-rasulbaev08}.
The \kr\ decay provides mono-energetic conversion electrons of the K shell with an energy close to the tritium endpoint at $E_{\rm K-32} = 17.8$~keV \cite{03-ven06}.
While applying the MAC-E filter principle as well as the main spectrometer retarding potential, the \kr\ source has to be elevated by about 800~eV in order to monitor the K conversion electrons.
This method relies on atomic and nuclear physics standards and provides monitoring with a stability at the 50~meV level as required by the systematic uncertainty limit.

Based on a self-compensating principle~\cite{03-marx01}
two precision high voltage dividers have been developed in cooperation with PTB Braunschweig. These HV dividers rank among the most stable ones in the world.
For voltages of up to 35~kV they provide a long-term stability of $< 10^{-6}$ per month as well as negligible temperature and voltage dependencies \cite{03-Thu09}. The two dividers have served recently to calibrate the ISOLDE facility at CERN \cite{03-krieger11}.
The permanent voltage monitoring with the precision voltage divider and frequent \kr\ calibration measurements at the monitor spectrometer will provide reliable and stable calibration and monitoring for the retarding potential of the main spectrometer over long periods of time.

\subsubsection{Focal plane detector}
\label{subsec-03:detector}
\begin{figure}[bt]
  \centering
  \begin{minipage}{0.63\textwidth}
    \includegraphics[width = 0.95\textwidth]{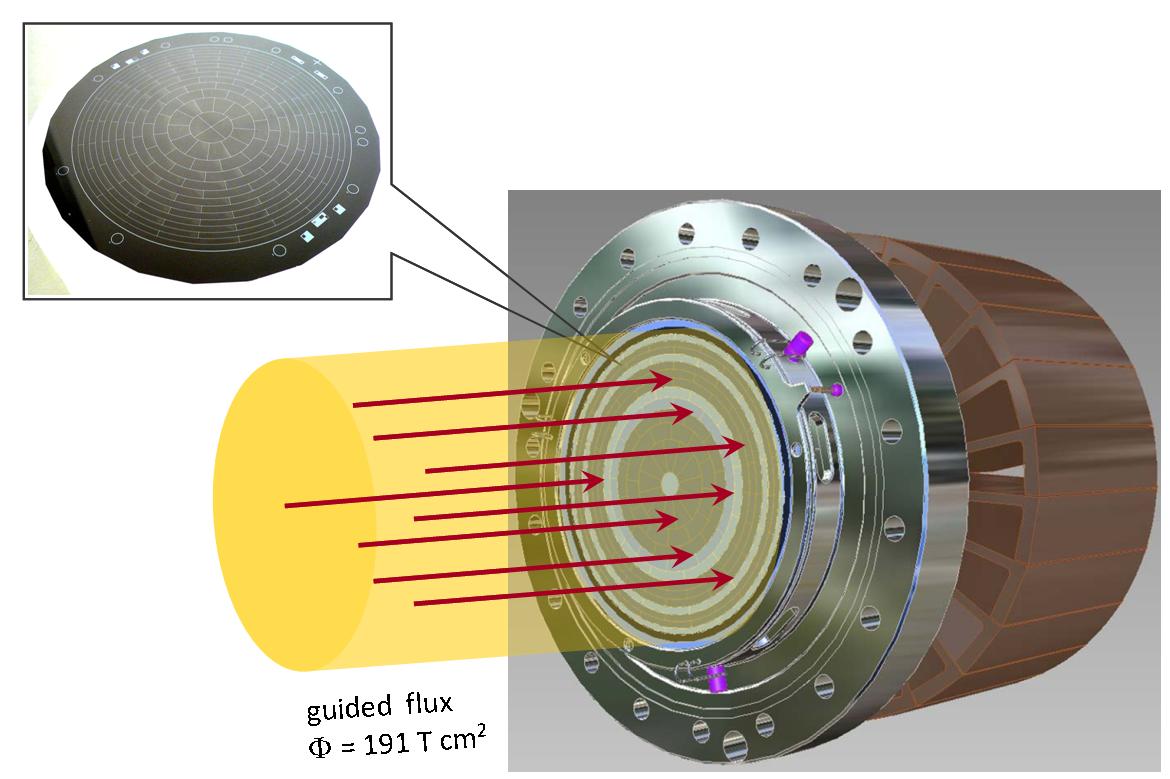}
  \end{minipage}
  \hfill
  \begin{minipage}{0.36\textwidth}
    \caption{Schematic view of the detector. The electrons are guided along magnetic field lines to the sensitive area of the detector, which has a diameter of 10~cm. The inset shows a photograph of the detector, in which the 148~pixels are visible.}
    \label{fig-03:Detector}
  \end{minipage}
\end{figure}

The focal plane detector is a semi-conductor based silicon PIN diode \cite{03-vanDevender12}. Its main goal is to detect transmitted electrons with a detection efficiency of $>90\%$.

The detector is subdivided into 148~pixels to achieve good spatial resolution (see figure \ref{fig-03:Detector}). This is important, as electrons passing the analyzing plane at different radii will experience slightly different retarding potentials, due to the radial inhomogeneity of the electric  potential. To account for this effect the detector consists of 12~concentric rings subdivided azimuthally into 12~pixels each. The innermost part is composed of 4 quarter circle pixels. This pixel arrangement allows for a precise mapping of inhomogeneities of the retarding potential as well as of the magnetic field. Each detector pixel measures an independent tritium \bspec\, which has to be corrected for the actual retarding potential and magnetic field value at the analyzing plane.

The signal of each detector pixel is connected by pogo-pins to a charge-sensitive pre-amplifier. The amplified signals are
digitized by FADC- and FPGA-boards \cite{03-auger10} developed for the Pierre Auger Observatory. The Data Acquisition
System is based on the software package ORCA\footnote{Object-oriented Real-time Control and Acquisition}, which was originally developed at the University of Washington as a general purpose, highly modular, object-oriented, acquisition and control system. Since 2008 further development continues at the University of North Carolina at Chapel Hill \cite{03-How}.

The detector is cooled by LN$_2$. To suppress external background it is surrounded by low radioactivity lead shield. Additionally an active muon veto system consisting of a cylindrical plastic scintillator read out by wavelength shifting fibers is installed.
There is an option to use post-acceleration to increase the kinetic energy of the $\upbeta$-electrons to about 30~keV or above before they hit the detector. This could help in discriminating signal from fluorescence background. Additionally, the impact angle of the electrons relative to the detector surface would be increased, which decreases the probability of backscattering from the detector surface~\cite{03-Ren11}.

The detector is situated in a superconducting magnet providing a magnetic field of $B_{\text{det}} = 3-6$ T. The detector magnet is adjacent to the so-called pinch magnet which provides the maximal magnetic field of $B_{\text{\text{max}}} = 6$ T of the entire KATRIN setup. All electrons that started in the source with an angle larger than $51^{\circ}$ will be reflected by the pinch magnet before they reach the detector (see equation (\ref{eq-03:pinch})). This is advantageous, as electrons emitted under a large angle perform a lot of cyclotron motion, which in turn increases their total path length and therefore their scattering probability and also their synchrotron losses. To exclude those electrons the magnetic field of the source is smaller than the maximal field.

\subsection{Signal and sensitivity}
\label{sec-03:sensitivity}
By counting $\upbeta$-electrons at different retarding potentials KATRIN measures the integral \bspec\
(\ref{eq-03:betaspec})
\begin{equation}
\label{eq-03:Ntheo}
 N(qU, E_0, \mtwonue) = N_{\text{tot}} t_U \int_0^{E_0} \dot N(E,E_0, \mtwonue ) R(E, qU) dE,
\end{equation}
which is a convolution of the response function $R(E, qU)$, incorporating the energy loss in the WGTS and the spectrometer transmission function, with the differential energy spectrum $\dot N(E,E_0,\mtwonue)$, describing the number of decays per second, nucleus and energy bin. In contrast to equation (\ref{eq-03:betaspec}) here we have explicitly
stated the dependence of $\dot N$ on the endpoint energy \ezero\ and on the squared neutrino mass \mtwonue .
The parameters $N_{\text{tot}}$ and $t_U$ denote the total number of tritium nuclei and the measurement time at a certain retarding potential, respectively. Assuming a constant background rate $\dot N_{\text{b}}$ the fit function to the measured spectrum is given by
\begin{equation}
\label{eq-03:NtheoB}
 N_{\text{th}}(qU, R_{\text{s}}, R_{\text{b}}, E_0, \mtwonue) = R_{\text{s}} \cdot N(qU, E_0, \mtwonue ) + R_{\text{b}} \cdot \dot N_{\text{b}} \cdot t_U,
\end{equation}
where $R_{\text{s}}$ and $R_{\text{b}}$ are the relative fraction of signal and background. In the fit $R_{\text{s}}$, $R_{\text{b}}$, $E_0$ and \mtwonue\ are free parameters.

\subsubsection{Sources of systematic errors}
Sources of systematic uncertainties arise from:
\begin{itemize}
 \item unconsidered corrections to the \bspec\ $\dot N(E, E_0, \mtwonue)$
 \item unaccounted variations of experimental parameters, e.g. retarding potential $U$ and number of decays $N_{\text{tot}}$
 \item an imprecise knowledge of the response function $R(E,qU)$
 \item and a non-constant background $\dot N_{\text{b}}$ in time and energy
\end{itemize}

Most of the systematic effects are related to the tritium source. They include energy losses of $\upbeta$-electrons, primarily due to inelastic scattering in the source, fluctuations of the column density, potential charging effects due to remaining ions, and the accuracy of the quantum-chemical calculations of the final state distribution of molecular tritium (see section \ref{sec-03:sourcesection}). The main systematic effect related to the spectrometer section is the high voltage stability (see section \ref{sec-03:sds}).

New physics effects such as right handed currents~\cite{03-bonn11}, sterile neutrinos~\cite{03-Veg, 03-For11, 03-Esm, 03-Rii11}, extra dimensions~\cite{03-Gon} etc.\ also affect the shape of the tritium spectrum (see equation (\ref{eq-03:betaspec_nusterile})). However, the influence of these phenomena typically becomes important only further away from the endpoint. Another new physics contribution that might affect the spectrum also close to the endpoint is violation of Lorentz invariance~\cite{03-Car00}.

\subsubsection{Background sources}
Only a small fraction of $10^{-13}$ of the $10^{11}$ tritium decays per second in the WGTS will produce $\upbeta$-electrons in the interesting energy region close to the endpoint (1~eV below the endpoint). This leads to a generic rather low count rate of only $10^{-2}$~cps in this energy region. To achieve a sensitivity of 200~meV (90\% C.L.), the background must be of the same order of magnitude, or smaller. Of major concern in this context are backgrounds which are not constant in time or energy.

The background originates mainly from the spectrometer section and partly from the detector.
With a detector energy resolution of the order of 1~keV all non-signal electrons detected in the region-of-interest (i.e.\ from approximately 15~keV to 20~keV) will contribute to the background level. Possible detector background sources are electrons produced by cosmic muons (and subsequent neutrons and gammas), high energetic gammas of environmental radioactivity (mainly from the thorium and uranium decay chains in the surrounding area) and decays of radio-nuclei in the detector material.

To reduce the backgrounds, the materials used in the construction of the detector system were radio-assayed very thoroughly. In addition, the detector is surrounded by a muon veto system. Finally, there is the option to use post acceleration to increase the energy of signal electrons up to 30~keV, which would allow for better signal--background discrimination. The total expected detector background is about $10^{-3}$~cps, which has been verified in the on-going commissioning measurements of the focal plane detector system.

The spectrometer-related background is especially challenging since all low-energy electrons created in the spectrometer are accelerated to the retarding potential before hitting the detector. Consequently, they cannot by distinguished from signal electrons.
In section \ref{sec-03:sds} we have already discussed that secondary electrons produced at the spectrometer walls are prevented from entering the sensitive magnetic flux tube by the dominant inherent magnetic shielding of the MAC-E filter, and by sub-dominant electrostatic shielding of the wire electrode system. However, secondary electrons can also be generated inside the sensitive flux tube as a result of ionisation processes of residual gas. Of particular concern are electrons that are stored, either in Penning traps, or by the inherent magnetic bottle property of the MAC-E-configuration.
These processes have been studied in the \prespectrometer{} set-up in an extended measurement program.

In~\cite{03-Fra10, 03-Mer12} is was shown that small Penning traps of the size of $\text{V} < 100~\text{cm}^3$ can lead to background rates easily exceeding 10$^3$ cps. Electrons being stored in Penning traps produce background via messenger particles, such as positive ions or photons, which can leave the trap and subsequently can create secondaries in the sensitive volume. This background has been tackled by a very precise and careful electromagnetic design~\cite{03-Zacher12,03-Mer10} that avoids the creation of Penning traps ab initio.

\begin{figure}[b!]
\begin{center}
\begin{minipage}{0.495\textwidth}
\includegraphics[width = \textwidth]{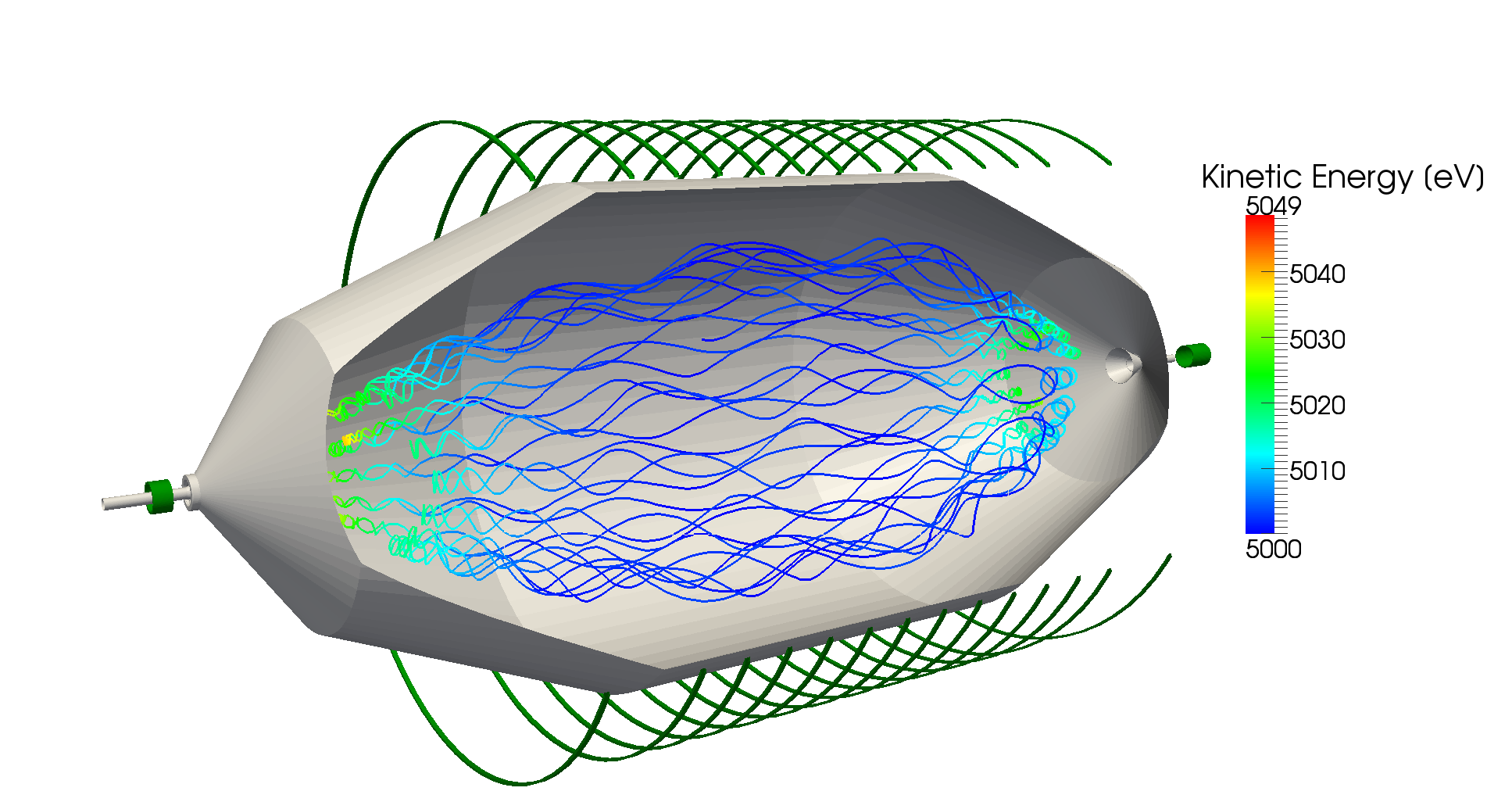}
\end{minipage}
\hfill
\begin{minipage}{0.495\textwidth}
\includegraphics[width = \textwidth]{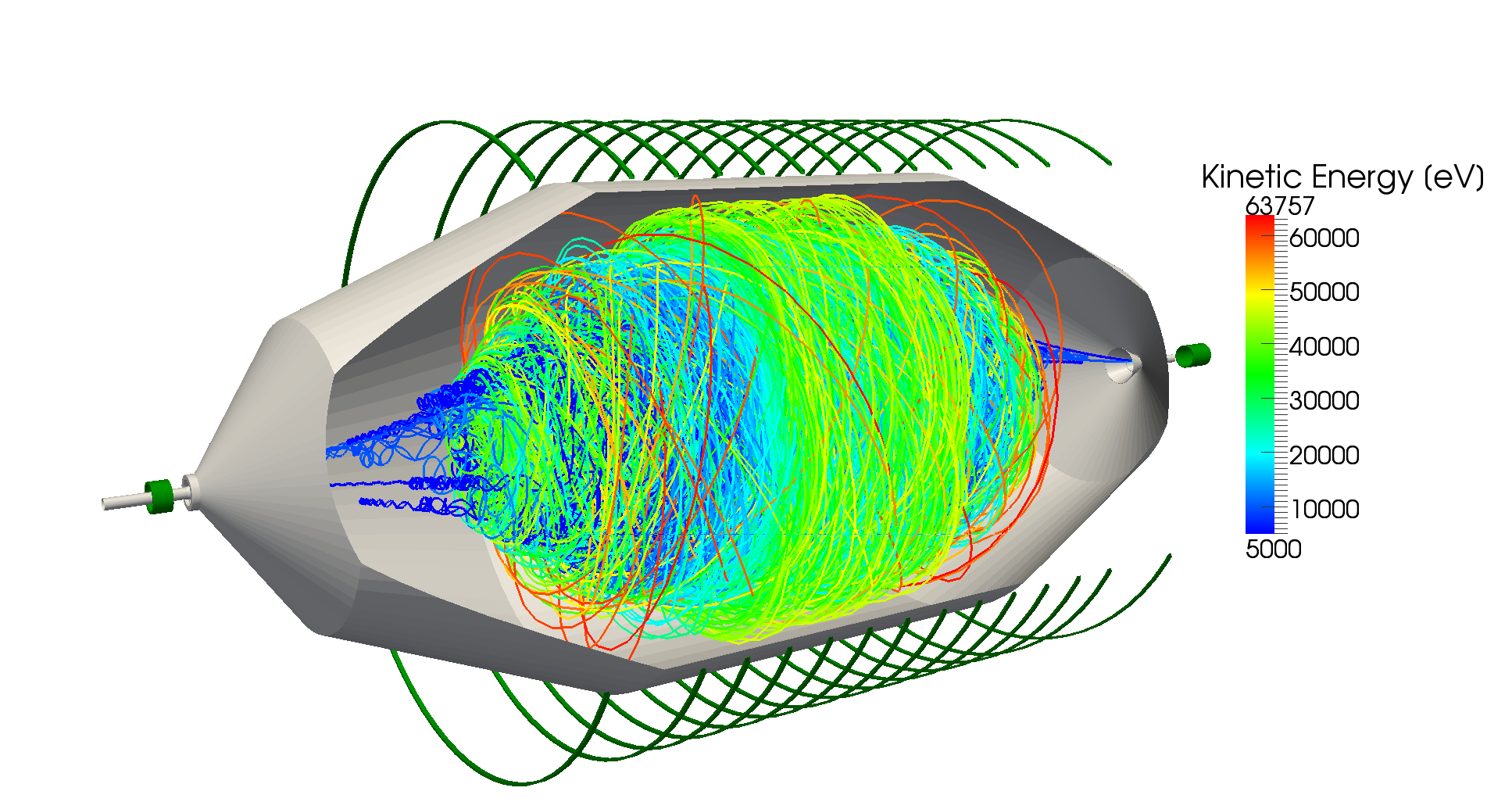}
\end{minipage}
\caption{The left picture shows the stable storage condition without external RF field, where the electron trajectory is a superposition of a fast cyclotron motion, axial oscillation,  and much slower magnetron drift. The right picture shows an electron trajectory in presence of an RF field with a frequency which is tuned to the cyclotron frequency of the electron in the central part of the spectrometer. The stored electron is stochastically heated up every time it passes the resonance region. After less than 5~ms its cyclotron radius is so large that the electron hits the wall and is absorbed there.}
\label{fig-03:effect_of_ECR}
\end{center}
\end{figure}

The simultaneous operation of two electrostatic retarding spectrometers (pre-spectrometer and main spectrometer) next to each other in the beam line, and with ground potential between them, would create a large Penning trap for electrons \cite{03-Bec10}. Signal electrons that experience no energy loss on their way through the spectrometers are not affected. However, all electrons that are created between the spectrometers with less kinetic energy than the \prespectrometer{} retarding energy remain trapped.
A stored electron will fill the trap in an avalanche effect by continually ionizing residual gas molecules. Simulations have shown that a primary stored electron together with all its secondaries may create up to $10^{8}$ positive ions, which are not trapped and can freely propagate into the main spectrometer. There, they can ionize residual hydrogen molecules thereby producing further electrons that can reach the detector and contribute to the background level.

A possible countermeasure would be the installation of a wire scanner between the spectrometers which regularly wipes through the trap, thereby removing the stored electrons and suppressing the background production~\cite{03-Bec10, 03-Hillen11}. Another most promising option would be to operate the \prespectrometer{} at reduced filter energy~\cite{03-Pra,03-Ren11}, even down to zero potential. This mode would avoid the creation of the Penning trap. Test measurements and simulations have proven the effectiveness of both approaches.

Another major background source results from magnetically stored electrons~\cite{03-Fra11,03-Mer12b}. At the main spectrometer this will cause all electrons with (transversal) energies larger than 1~eV (i.e. larger than the width of the transmission function) to be stored. Owing to the excellent UHV conditions in the KATRIN main spectrometer ($\text{p} = 10^{-11}$~mbar) the storage times of electrons in the multi-keV-range can reach several hours. During its cooling time, a stored electron produces several hundred secondary electrons, which eventually leave the spectrometer and hit the detector. The main source of primary electrons originate from $^{219}$Rn, $^{220}$Rn~\cite{03-Fra11} and tritium decays inside the spectrometer volume. This background source features large time fluctuations and thereby a strong non-Poissonian characteristic. This fact has a large impact on the neutrino mass sensitivity~\cite{03-Mer12b}.

As a counter measure liquid nitrogen cooled baffles will passively shield the spectrometer volume from ${}^{219}$Rn, which is mainly emanating from the large NEG pumps~\cite{03-Fra11}. The tritium decay rate inside the main spectrometer volume can be reduced by the option to use the \prespectrometer{} as an additional tritium pump. Finally, a number of active background reduction techniques, which are targeted at  removing the stored electrons, have been successfully investigated experimentally and via extensive MC simulations. A method based on stochastic heating with an RF field by the well known technique of Electron Cyclotron Resonance (ECR)~\cite{03-Mer12c} and a method based on nulling the magnetic field by magnetic pulsing~\cite{03-Hillen11} seem most promising. Figure~\ref{fig-03:effect_of_ECR} shows the strong background-reducing effect of a very short RF-pulse fed into the main spectrometer on magnetically stored electrons as calculated by KASSIOPEIA
(even for very moderate RF-amplitudes). These short RF-pulses would empty all stored particles with negligible duty-cycle.

\subsubsection{Neutrino mass sensitivity of KATRIN}
For the reference setup of the KATRIN experiment, the quadratic sum of all known systematic uncertainties is expected to be $\sigma_{\text{sys,tot}} \le 0.017$~$\text{eV}^2$. After three ``full beam'' years of measurement time the statistical error is about as small as  the systematic error. This yields a total error of $\sigma_{\text{tot}} = 0.025$~$\text{eV}^2$. Figure~\ref{fig-03:Sensitivity} shows the discovery potential of KATRIN as a function of beam time for different neutrino masses. Accordingly, a neutrino mass of $m_{\nu} = 350$~meV would be seen with 5~$\sigma$ significance. The figure also shows the 90\% C.L. upper limit as a function of measurement time in case that no neutrino mass signal is seen. In this case, an upper limit of $m_{\nu} \leq 200$~meV can be stated at 90\%~C.L after three full beam years of measurements.

\begin{figure}
  \centering
  \begin{minipage}{0.49\textwidth}
    \includegraphics[width = \textwidth]{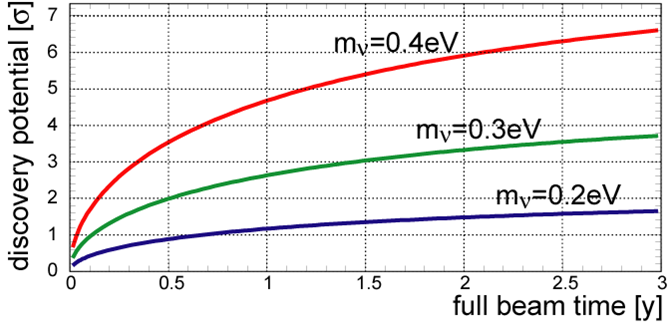}
  \end{minipage}
  \hfill
  \begin{minipage}{0.49\textwidth}
	\includegraphics[width = \textwidth]{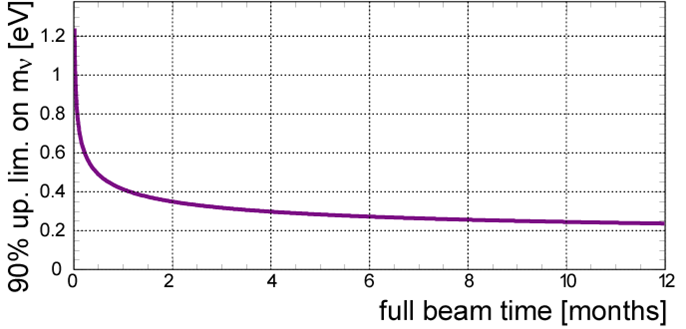}
  \end{minipage}
  \caption{Left: Discovery potential of KATRIN as function of time for different neutrino masses. Right: Upper limit on neutrino mass at 90\% C.L. as a function of time.}
  \label{fig-03:Sensitivity}
\end{figure}

\subsection{Status and Outlook}
\label{sec-03:status}

The commissioning of the KATRIN beam line elements is progressing well. On one side, the multi-faceted technical hurdles are challenging,
causing the commissioning of the overall setup to take longer than originally planned. On the other side, the excellent results of the
large-scale test units have opened the possibility to substantially reduce the systematic effects during the long-term
measurements with the final configuration, thus improving the neutrino mass sensitivity of the experiment.

Beginning at the source-related components, an important breakthrough was the verification of the novel beam tube cooling system.
In a dedicated setup, consisting largely of original components, the WGTS demonstrator, a temperature stabilization
of the 10~m long beam tube of $\Delta T/T$ of $\approx$ 10$^{-4}$ was achieved by using two-phase neon fluid as cooling agent. This is
one order of magnitude better than specified. Together with the achieved pressure stabilisation of the inner loop mock-up of $\Delta p/p$ of $\approx$ 10$^{-4}$
this opens up the possibility of reduced systematic errors from column density fluctuations (this is one of the largest overall systematic errors). At present the
WGTS demonstrator is being reassembled to the final WGTS cryostat. The final WGTS assembly at KIT is expected to be completed by the end of 2014. Further
progress has been made with regard to tritium analytics (LARA setup), as well as the design of the rear section which will include extensive control
and monitoring units.

Major progress has also been achieved in the field of large-scale tritium retention. After the successful
commissioning of the DPS2-F cryostat, first tritium retention measurements with a beam tube at room temperature have yielded
experimental flow suppression factors which are in very good agreement with corresponding MC simulations. Due to the
malfunction of a protective diode of the superconducting magnet system of DPS2-F, a new magnet safety concept for all s.c. solenoids has been designed.
This concept is currently being implemented for WGTS and CPS, as well as a fail-safe differential pumping section. The manufacture of the cryopump CPS
is well under way with assembly works expected to be finished by the end of 2013.

In the spectrometer section, the extensive measurement program with the \prespectrometer{} facility has given important insights into background reduction
techniques, precision electromagnetic layout, vacuum technologies and high voltage stability. At present the \prespectrometer{} is ready for beam line integration.

The main spectrometer together with its external air coil system and its inner electrode system, which was completed at the beginning of 2012, is currently being
prepared for test measurements. These measurements will be focused first on extensive background studies, with the objective to remove any remaining small-scale Penning traps, to quantify the contribution of cosmic muon induced background and to study its signature by making use of external muon detectors. An important aspect of the background studies will be the identification
of background due to stored electrons following nuclear decays, and the optimisation of active and passive background reduction techniques to limit the spectrometer
background to a level of $< 10^{-2}$~cps. Another important task will be to map the transmission properties of the spectrometer with an angular-selective electron gun.
In all these investigations the recently commissioned focal plane detector system with its excellent properties will be of vital importance. Finally, the extensive software
developments for simulation and analysis tools are in an advanced state and the software packages are continually being refined and extended.

After integration of all source-related and spectrometer-related components, the first runs in the final KATRIN configuration are expected in the second half of 2015.

\section{New Approaches}
While spectrometer experiments based on the MAC-E filter principle~\cite{03-pic92a} currently provide the highest sensitivities in direct neutrino mass experiments, there are alternative approaches that aim for comparable performance and better scalability in the study of weak decays.\\
A very recent development is promoted by the Project~8 team (see section~\ref{sec-03:project8}) where tritium technology from the KATRIN experiment is used in conjunction with microwave antennas to detect coherent cyclotron radiation emitted by individual decay electrons in a magnetic field. The aim is to extract a \bdec\ spectrum without the need for a large electrostatic spectrometer.\\
Most of the work on alternative experimental methods is, however, focused on using microcalorimeters to study rhenium \bdec s (see section~\ref{sec-03:mare}) or holmium electron capture decays (see sections~\ref{sec-03:ec} and \ref{sec-03:echo}). The main advantage of using microcalorimeters lies in the {\it source = detector} principle that allows to measure the complete decay energy (excluding the energy carried away by the emitted neutrino) as opposed to only measuring the kinetic energy of the decay electrons. On the other hand the comparably slow signals produced by calorimetric detectors bring the challenge of using large arrays of small detectors to avoid pile-up in the individual calorimeter crystals.\\
Another approach has been proposed by Jerkins and co-workers~\cite{03-Jer10} who suggest to perform a full kinematic reconstruction of the decays of trapped tritium atoms. This idea, however, seems to be hampered by conceptual difficulties~\cite{03-Ott11}.
\subsection{Project 8}
\label{sec-03:project8}
Project~8 is a new effort to measure the neutrino mass lead by groups from MIT, University of California, Santa Barbara and University of Washington. The idea is to use technology from KATRIN's gaseous tritium source combined with a sensitive array of microwave antennae to extract energy spectra of tritium decay electrons from the coherent cyclotron radiation emitted by individual electrons in a magnetic field (see
\begin{figure}[h]
\centering
\includegraphics[width=0.7\textwidth]{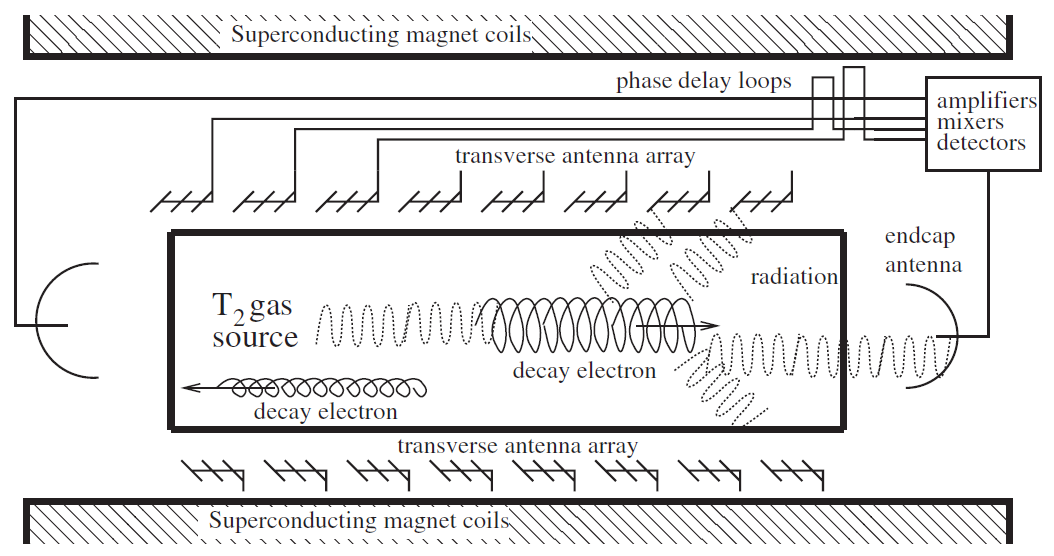}
\caption{Schematic view of the experimental layout to measure coherent cyclotron radiation from tritium decay electrons (Reprinted with permission from~\cite{03-Mon09}, Copyright (2009) by the American Physical Society).
\label{fig-03:project8}}
\end{figure}
figure~\ref{fig-03:project8}). In such a setup electrons will follow a circular or spiral path with a cyclotron frequency $\omega$ of
\begin{equation}
 \omega = \frac{\omega_0}{\gamma} = \frac{qB}{m + E} \; ,
\end{equation}
where $E$ is the kinetic energy of the electron, $B$ is the magnetic field of the source, $q$ is the electron charge, $m$ is the electron mass, $\omega_0$ is the unshifted cyclotron frequency and $\gamma$ is the Lorentz factor.
The total power emitted by each electron depends on its relative velocity $\beta$ and the pitch angle $\theta$ between the direction of movement and the magnetic field vector
\begin{equation}
 P(\beta, \theta) = \frac{1}{4\pi\epsilon_0} \frac{2 q^2 w_0^2}{3c} \frac{\beta^2 sin^2(\theta)}{1-\beta^2} \; .
\label{eq-03:p8_power}
\end{equation}
With a suitable set of source parameters, the power of the cyclotron radiation emitted will be large enough to be detected, but not large enough to rapidly change the electron's momentum and will therefore allow to reconstruct its kinetic energy.
The energy resolution $\Delta E$ of the method depends on the relative uncertainty with which the cyclotron frequency can be determined, which in turn depends on the observation time for an individual decay electron. \\
Monreal and Formaggio~\cite{03-Mon09} discuss a reference design with a 1~T magnetic field strength. In this design $f_0 = \omega_0/2\pi \approx 27$~GHz for decay electrons near the endpoint energy $E_0 = 18575$~eV.
For a resolution of $\Delta E = 1$~eV we therefore require a relative uncertainty $\Delta f / f = \Delta E/m = 2 \cdot 10^{-6}$ and, via the Nyquist theorem, a minimum observation time of $t_{min} = 2/\Delta f \approx 38~\mu$s.
This observation time determines the necessary mean free path of decay electrons in the source before they undergo inelastic scattering from T$_2$ molecules and therefore places constraints on the density of the source. In the reference design the required free path length is given by $t_{min} \, \beta \, c \approx 3000$~m corresponding to a maximum T$_2$ density of $\rho_{max} = (t_{min} \, \beta \, c \, \sigma_i)^{-1} \approx 1.1\cdot 10^{12}/\mbox{cm}^3$ with a total cross section $\sigma_i \approx 3 \cdot 10^{-18}\,\mbox{cm}^2$ for inelastic scattering of electrons near the endpoint energy from molecular tritium.\\
Given the long pathlength required for the electrons to achieve the necessary observation time, it becomes clear that the decay electrons need to be trapped within an instrumented volume to realize the actual experiment. This can be achieved using the magnetic mirror effect, where the $B$-field of the source volume is increased by some amount at both ends of the apparatus. The ratio between the field strength in the source volume to that at the edges then determines the minimum pitch angle $\theta_{min}$ of the stored electrons. Only storing electrons with pitch angles above $\theta_{min}$ has the added benefit that they are also emitting higher signal powers via the dependence of equation~\ref{eq-03:p8_power} on $\theta$. For stored electrons the emitted signal powers are in the $10^{-15}$~W region while the authors of reference~\cite{03-Mon09} estimate the thermal noise contribution of suitable amplifiers for the signal detection to be of the order $10^{-17}$~W. Incoherent noise from non-endpoint and/or low-pitch beta electrons is expected to contribute about $10^{-24}$~W/Bq in the signal region which would allow a tritium activity of the source of the order $10^8-10^9$~Bq. Of course these estimates also depend on how efficient an actual antenna configuration could collect the signal power emitted by the electrons.\\
The correct reconstruction of electron energies from the detected cyclotron radiation is complicated by the Doppler shift the signals experience due to the motion of the electrons parallel to the magnetic field lines. Two different antenna configurations are proposed to remedy this problem. In the first case, two antennae are placed on axis at both ends of the source volume (see endcap antennae on figure~\ref{fig-03:project8}). While one of the antennae will detect a blue-shifted signal due to the parallel velocity component of the electron towards it, the other will detect the red-shifted version of the same signal. Combining these signals, the unshifted electron energy can then be reconstructed. \\
The second configuration discussed consists of arrays of evenly spaced microwave antennae oriented transverse to the magnetic field. Passing electrons will induce signals that sweep from blueshift to redshift in the individual antennae. When these signals are summed up with an appropriate choice of delay lines, the unshifted cyclotron frequency should sum coherently, while the contributions from Doppler shifts sum incoherently and should thus be suppressed. The resulting output will then appear at the unshifted frequency $\omega$ with an amplitude modulated due to the periodically varying response function of the antennae along the array. This modulation gives rise to two sideband peaks in the signal, that can help to discriminate a real high energy electron signal from background signals caused e.g. by sidebands from lower energy signals.\\
A simulation performed by Monreal and Formaggio~\cite{03-Mon09} for $10^5$ tritium decays observed within a $30\,\mu$s interval is shown in
\begin{figure}
\centering
\includegraphics[width=0.6\textwidth]{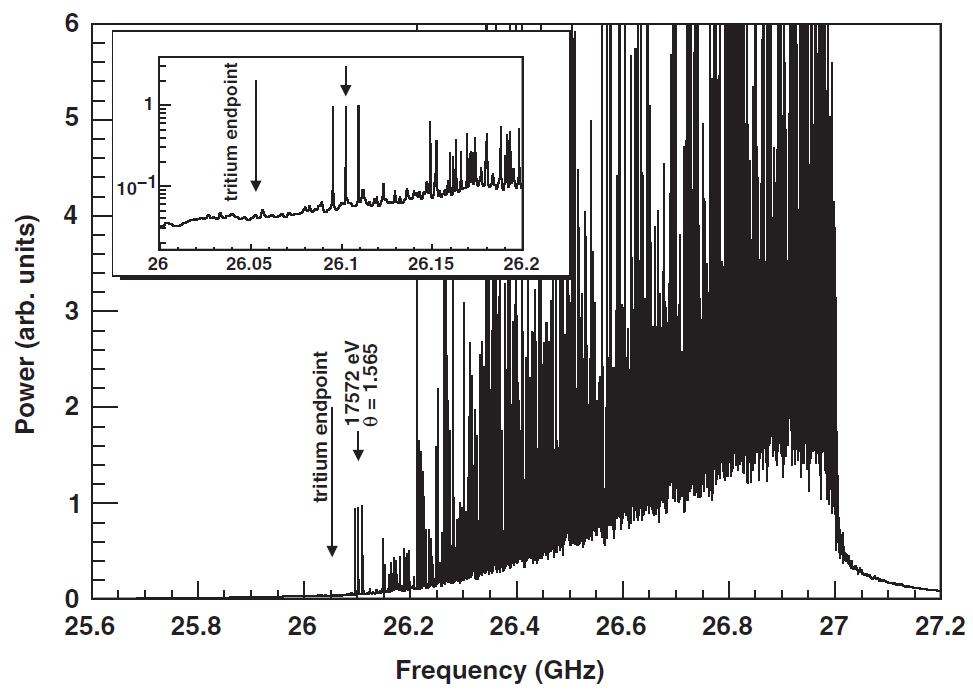}
\caption{Simulated frequency spectrum of $10^5$ tritium decays observed within a $30\,\mu$s interval in a 10~m long source with a field of 1~T. The inset shows the triplet of lines produced by a high energy electron (Reprinted with permission from~\cite{03-Mon09}, Copyright (2009) by the American Physical Society).
\label{fig-03:p8-spectrum}}
\end{figure}
figure~\ref{fig-03:p8-spectrum}. As discussed above, signals of high energy electrons show up as triplets of lines in the lower part of the frequency spectrum. Because this region is also populated by sideband signals from lower energy electrons, the coincident detection of at least two of the three lines is required to confidently identify an endpoint electron. Extending this requirement to all three lines for the positive identification of a high energy electron, a total source strength of ca. $10^9$~Bq can be allowed to keep background from accidental coincidences on an acceptable level.\\
The proposed technique presents very different systematic errors than those present in MAC-E filter experiments like KATRIN.
Advantages are the ability to correct for fluctuations of the source density, as the complete spectrum is monitored by the experiment.
The absence of an electrostatic spectrometer should enable a more scalable experiment as the energy resolution is not limited by the size of a MAC-E filter, but by the observation time for individual electrons and ultimately by the irreducible energy spread due to the excitation of rotational and vibrational final states of the daughter molecule emerging from the tritium decay.
On the other hand, care has to be taken to eliminate the effects of Doppler shifts that alter the frequency picked up by the microwave antennae. Inhomogeneities in the B-field are a source of line broadening, while drifts of the magnetic field will shift the overall spectrum. Electron - T$_2$ scattering and signal pile-up will limit the allowed source density and therefore influence the achievable signal rate in the important endpoint region of the spectrum.\\
To verify the ability to detect the cyclotron emission of individual electrons and to investigate the achievable resolution, a test experiment~\cite{03-For11a} is currently being set up at the University of Washington. In the setup a small magnetic bottle within a 1~T superconduction magnet is used to capture 17.8~keV conversion electrons from a $^{83m}$Kr source. Different antenna designs will be tested in an attempt to observe the cyclotron emission from these electrons.
\subsection{MARE}
\label{sec-03:mare}
The ``Microcalorimeter Arrays for a Rhenium Experiment'' (MARE) collaboration is working to further the development of sensitive micro-calorimeters to investigate \rhenium\ \bdec .
The current activities in the MARE collaboration are organized in two phases~\cite{03-Nuc08}: in MARE-1 several groups are working on alternative micro-calorimeter concepts which will be tested by setting up neutrino mass experiments with sensitivities in the order of a few eV. Besides the selection of the most sensitive detector technology, this phase will also be used to investigate the use of the EC decay of $^{163}$Ho as an alternative to the study of rhenium \bdec\ to determine the neutrino mass~\cite{03-Gal12} (see also section~\ref{sec-03:ec}). The consideration of $^{163}$Ho was triggered by persisting difficulties with superconducting metallic rhenium absorbers coupled to the sensors~\cite{03-Fer12}. Thermalization of the energy deposited in \bdec s seems to be hindered by the excitation of long lived states in the rhenium absorber. The nature of these states can presently only be speculated upon~\cite{03-Ran12}.
After selecting the most successful technique, a full scale experiment with sub-eV sensitivity to the neutrino mass will then be set up in MARE phase 2.\\
The technical developments generally aim at two main goals: first to improve the energy resolution of the detectors and secondly to shorten the response time of the signals in order to reduce pile-up problems. The various groups in the MARE collaboration work on different techniques to achieve these goals.\\
Groups from University Milano-Bicocca, NASA/GSFC and Wisconsin are working together to develop arrays of silicon implanted thermistors coupled to AgReO$_4$ absorbers~\cite{03-Kra08} (see figure~\ref{fig-03:mare-xrs}).
\begin{figure}
\centering
\includegraphics[width=0.35\textwidth]{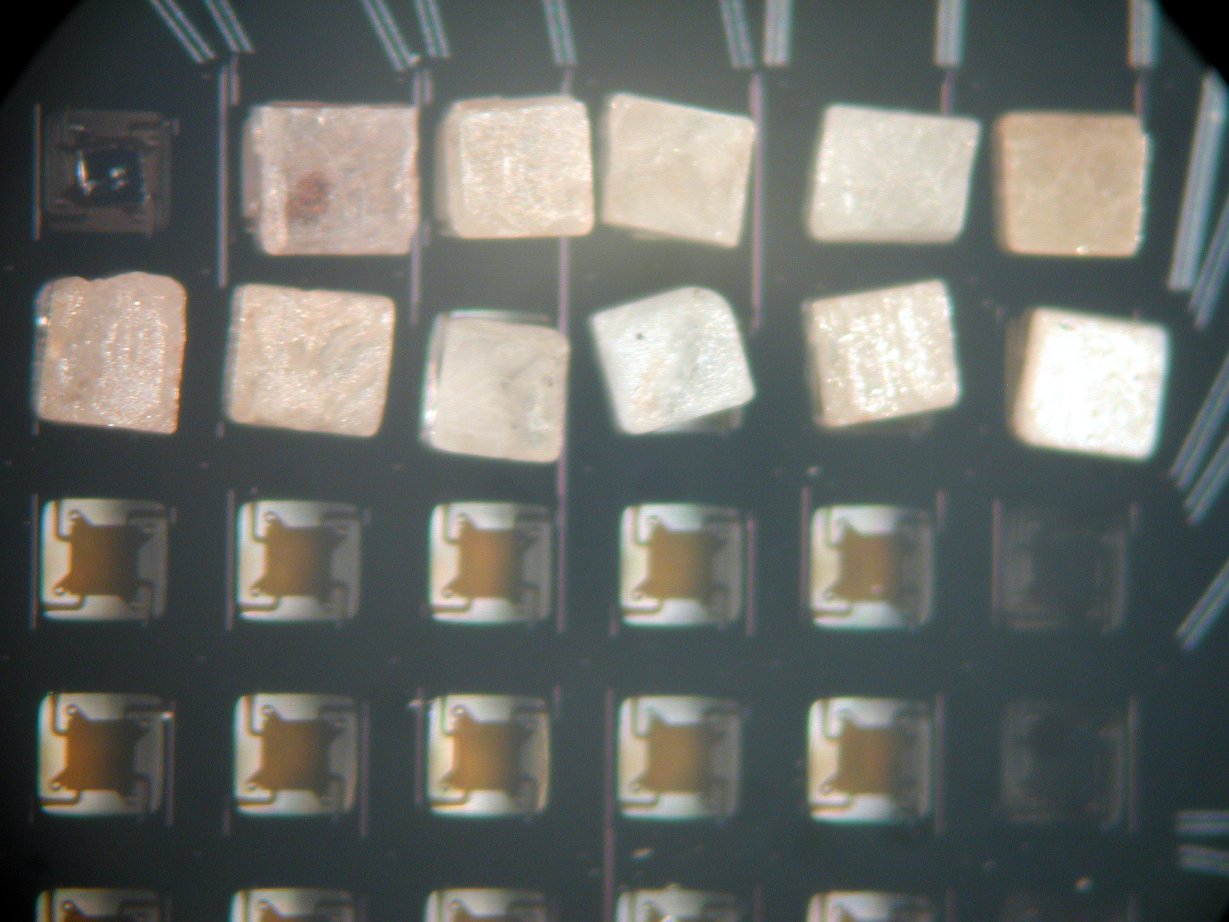}\hspace{10mm}
\includegraphics[width=0.4\textwidth]{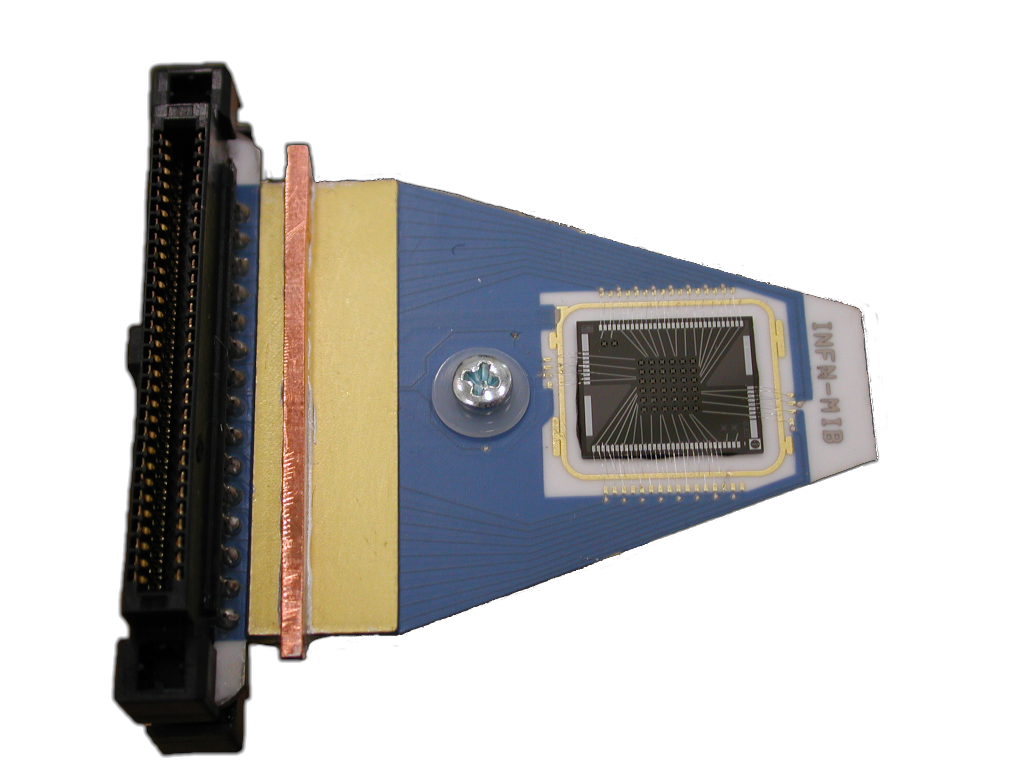}
\caption{Left: AgReO$_4$ crystals glued to a $6\times 6$ pixel XRS thermistor array.
Right: View of the $6\times 6$ array with readout lines (images courtesy of A.~Nucciotti~\cite{03-Nuc11}).
\label{fig-03:mare-xrs}}
\end{figure}
The experiment that is currently being set up can accommodate up to eight 36 pixel arrays with 0.5~mg AgReO$_4$ absorbers of 0.27~Bq activity each. The absorber crystals used are cut in a regular shape of $600\times 600 \times 250\, \mu$m$^3$ from large single crystals and are glued to the thermistors with an intermediate layer of thin pure silicon spacers. For readout of the thermistors a cold buffer stage using JFETs at 120~K is installed as close as possible to the sensors, followed by a main amplifier stage at room temperature. With this system it is possible
to achieve an energy and time resolutions of 25~eV and 250~$\mu$s, respectively~\cite{03-Fer12}. After a first test run with 10 AgReO$_4$ crystals on one array and two Sn absorbers on a second array to investigate the environmental background near the rhenium endpoint, a \rhenium\ \bdec\ measurement with 72 channels should start and provide a sensitivity of 4.5~eV at 90\% CL within three years running time~\cite{03-Fer12}. With all 8 arrays instrumented a sensitivity of 3.3~eV at 90\% CL would be reached with $7\cdot 10^{9}$ rhenium decays.\\
Besides their work on thermistors with AgReO$_4$ absorbers, the group from Milano-Bicocca is also investigating the use of Microwave Kinetic Inductance Detectors (MKIDs) for measurements of $^{163}$Ho EC decays. These devices are superconducting resonators in the 1-10~GHz region that exploit the temperature dependence of superconducting films. An advantage with these sensors is the easy readout of a few thousand detectors using frequency multiplexing techniques~\cite{03-Fav12}.\\
The University of Genoa is working together with groups from University of Miami and University Lisbon/ITN on the development of Transition Edge Sensors (TES) coupled to metallic rhenium absorber crystals~\cite{03-Vac08}. TES sensors exploit the sharp rise of resistance with temperature of a superconductor operated at the phase transition from normal to superconducting behaviour. This allows for a very sensitive detector compared to conventional semiconductor thermistors, but has the drawback of being less stable in operation and having a lower saturation energy. The sensors are based on a Ir-Au bi-layer configured in a double S-shape.
%
\begin{figure}
\centering
\includegraphics[width=0.4\textwidth]{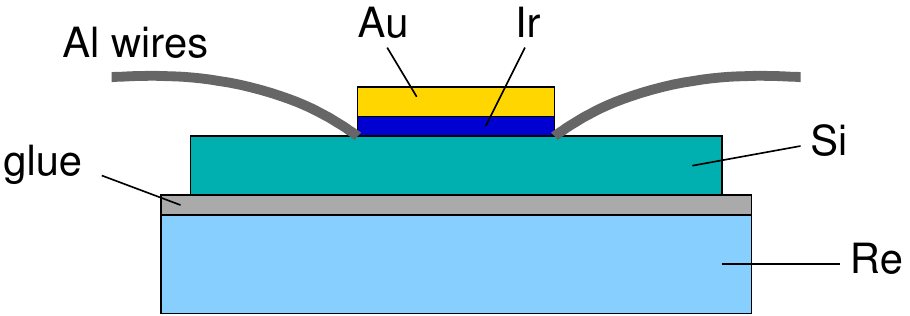}
\caption{Schematic drawing of the Ir-Au/Re TES microcalorimeter developed at Genoa.
\label{fig-03:mare-tes}}
\end{figure}
The normal resistance of the circuit is $2\,\Omega$ and by adjusting the layer thickness, the transition temperature is brought to ca. 80~mK.
The superconducting layer is deposited on a silicon substrate onto which, on the other side, a superconducting rhenium absorber of 200-300~mg is glued (see figure~\ref{fig-03:mare-tes}). Two Al-Si wires are used for readout of the TES and for connection to a heat bath cooled by a dilution refrigerator. With this design an energy resolution of 11~eV FWHM and a rise time of 160~$\mu$s have been achieved.~\cite{03-Vac08}.
Their planned phase 1 experiment will accommodate 300 TES detectors with about 0.25~Bq activity per element, which would allow a sensitivity on the neutrino mass of 2~eV with three years worth of data taking.
As an alternative to rhenium the group at Genoa is also working on the development of TES sensors with $^{163}$Ho loaded absorbers.
\\
At the University of Heidelberg work is ongoing on the development of so-called Metallic Magnetic Calorimeters (MMC)~\cite{03-Gas09}. In contrast to TES sensors or thermistors that exploit the change of resistivity of the sensors with temperature, these detectors are measuring the change in magnetization of a paramagnetic material. Promising results have been obtained with gold absorbers, where resolutions down to 2~eV FWHM and signal rise times of about 90~ns have been achieved for soft x-rays around 6~keV energy~\cite{03-Pie12}. The resolution with rhenium absorbers, however, was found to be 44~eV FWHM at signal rise times below 10~$\mu$s. The use of MMC detectors with holmium implanted absorbers produced more encouraging results and led to the formation of the ECHO collaboration to further investigate this technique (see section~\ref{sec-03:echo}).\\
All these activities should finally lead to a selection of the most suitable technique for a rhenium or holmium based large scale neutrino mass experiment with sub-eV sensitivity. According to reference~\cite{03-Nuc08} a staged approach where a total of five $10^4$ detector arrays is deployed one per year, would, after 10 years running time, enable a statistical sensitivity better than 0.25~eV. One has to keep in mind however that effects like the beta environmental fine structure (see section~\ref{sec-03:rhenium}) are difficult to estimate and have to be investigated in parallel to the technical developments within the next years.
\subsection{\texorpdfstring{Electron capture on $^{163}$Ho}{Electron capture on 163Ho}}
\label{sec-03:ec}
A promising alternative to \bdec\ measurements is the study of electron capture (EC) decays of $^{163}$Ho to measure the neutrino mass. The decay process considered is:
\begin{equation}
 ^{163}{\rm Ho}^+ + e^- \rightarrow ^{163}{\rm Dy}^*_i + \nu_e \rightarrow ^{163}{\rm Dy} + E_i + \nu_e \; .
\end{equation}
The de-excitation spectrum of the intermediate state $^{163}$Dy$^*_i$ is given by a series of lines at energies $E_i$ which correspond to the dissipated binding energy of the electron hole in the final atom.
The $Q$-value of the reaction is given by the mass difference of mother and daughter nucleus in the ground state.
%
%
Like the electron energy spectrum in \bdec s this spectrum depends on the square of the neutrino mass~\cite{03-Lus11}:
\begin{equation}
 \dot N(E_C) = \frac{d\lambda_{EC}}{dE_C} = \frac{G_F^2 \cos^2\theta_C}{4\pi^2} (Q-E_C) \sqrt{(Q-E_C)^2 - m^2_{\nu_e}} \cdot
  \sum_i n_i C_i \beta_i^2 B_i \frac{\Gamma_i}{2\pi}\frac{1}{(E_C - E_i)^2 + \Gamma_i^2/4} \; .
\label{eq-03:ec_spec}
\end{equation}
The measured calorimetric energy $E_C$ when the source is embedded in a micro-calorimeter contains the complete de-excitation energy of the daughter atom that is dispersed in the form of electrons and x-rays. The atomic levels involved are described by Breit-Wigner resonances with finite widths $\Gamma_i$.
Additionally, $n_i$ is the fraction of occupancy of the $i$-th atomic shell, $C_i$ is the nuclear shape factor, $\beta_i$ is the squared Coulomb amplitude of the electron radial wave function at the origin of the electron radial wave function and $B_i$ is an atomic correction for electron exchange and overlap. The use of $^{163}$Ho is favored due to its very low Q-Value in the range of 2.3~keV to 2.8~keV~\cite{03-Lus11}.
Due to the low $Q$-value and selection rules only electrons from the $M_1$, $M_2$, $N_1$, $N_2$, $O_1$, $O_2$ and $P_1$ shells can be captured and the spectrum is expected to have the shape shown in figure~\ref{fig-03:ho_spectr}, left plot, with the influence of the
\begin{figure}
\centering
\includegraphics[width=0.4\textwidth]{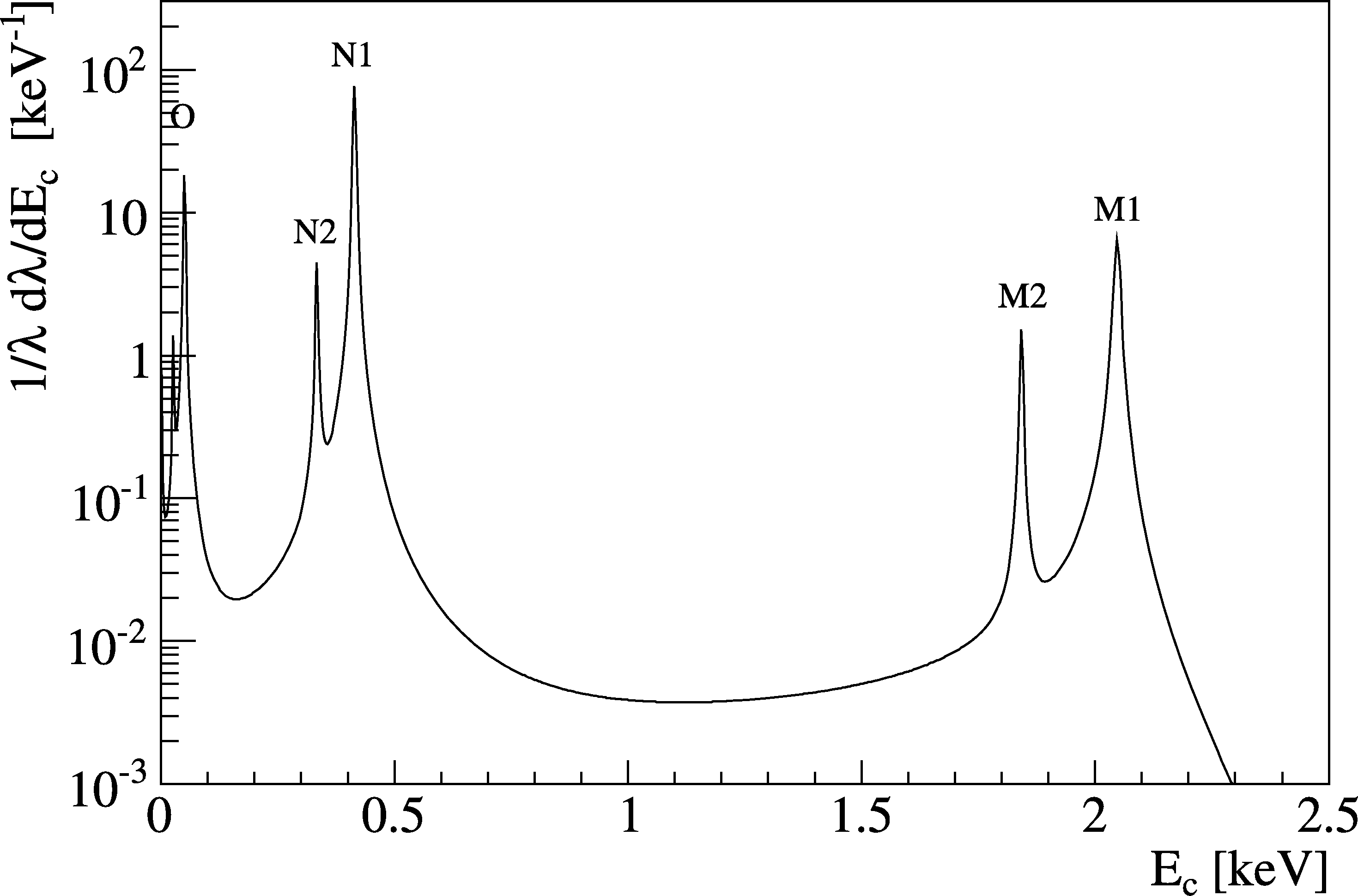}\hspace{10mm}
\includegraphics[width=0.4\textwidth]{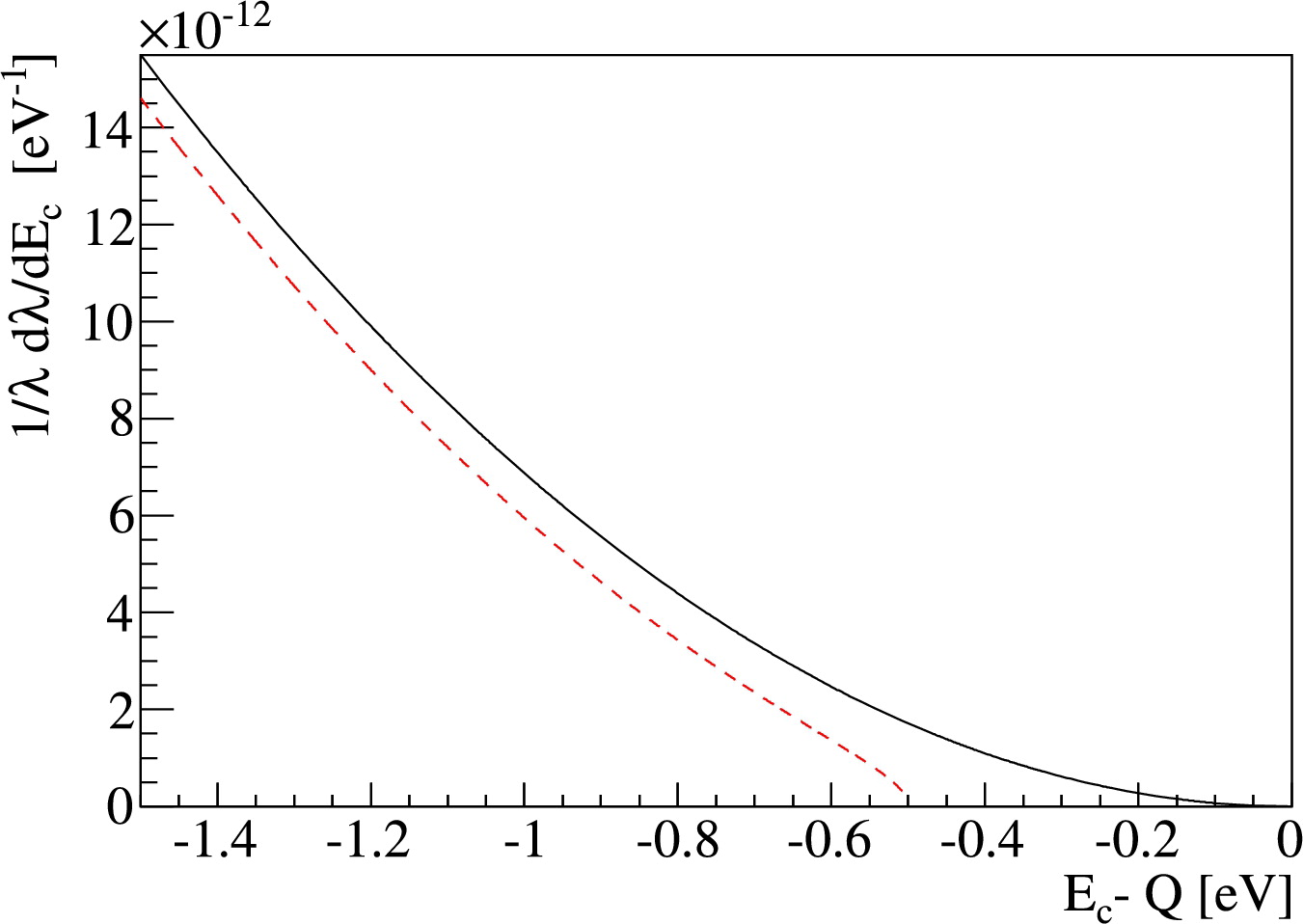}
\caption{De-excitation spectrum of $^{163}$Ho for $Q = 2.5$~keV (left). The right plot shows a zoom into the endpoint region of the spectrum with the effect of a 0.5~eV neutrino mass indicated by the red dashed line (Reprinted from~\cite{03-Lus11}, Copyright 2011, with permission from Elsevier).
\label{fig-03:ho_spectr}}
\end{figure}
neutrino mass most pronounced near the endpoint as shown in figure~\ref{fig-03:ho_spectr}, right plot.
The count rate in the endpoint region strongly drops with increasing distance between the closest atomic level and the Q-value of the reaction. At the same time the amount of activity that can be allowed in a single calorimetric detector has to be limited in order to reduce the unresolvable pile-up of signals which otherwise distorts the measured spectrum. In order to gather the required statistics to reach sub-eV sensitivity with this method it is therefore necessary to operate large numbers of small detectors in parallel.\\
In contrast to the calorimetric method described above, the first measurements applying $^{163}$Ho EC decay to constrain the neutrino mass actually made use either of X-rays emitted after the decay~\cite{03-Ben81} or of inner Bremsstrahlung photons created in the process of radiative electron capture~\cite{03-Ruj81}.
Studying X-rays emitted after the EC decay allows one to determine the ratios of capture rates from the $M$ and $N$ shells as well as the absolute $M$ capture rates, from which the neutrino mass can be reconstructed. Applying this method, Yasumi \etal\ obtained an upper limit on the electron neutrino mass of $\mnue < 490$~eV at 68\% CL~\cite{03-Yas94}. On the other hand, Springer \etal\ made use of the inner Bremsstrahlung method to obtain an upper limit of $\mnue < 225$~eV at 95\% CL~\cite{03-Spr87}. \\
First measurements of the calorimetric $^{163}$Ho EC spectrum~\cite{03-Rav84, 03-Har92, 03-Gat97} have not yet achieved sufficient sensitivity to improve on the above mentioned limits due to limitations in energy resolution and statistics of the measurements. A very recent measurement of the calorimetric de-excitation spectrum using Metallic Magnetic Calorimeters (see section~\ref{sec-03:echo}) reached a high energy resolution of 12~eV FWHM~\cite{03-Ran12}, but was hampered by a background of EC decays from $^{144}$Pm that contaminated the detector during the $^{163}$Ho implantation process.
\subsection{ECHO}
\label{sec-03:echo}
Groups from Heidelberg University and MPIK, the Saha institute, ISOLDE/CERN and the Petersburg Nuclear Physics institute are working on the development of the Electron Capture $^{163}$Ho experiment ECHO~\cite{03-Ran12}, that is based on micro-structured MMC detectors~\cite{03-Bur08}.
In these detectors, the temperature change following an energy absorption is measured by the change of magnetization of a paramagnetic sensor material (Au:Er) sitting in an external magnetic field.\\
The MMC detectors developed by the Heidelberg group consist of pairs of superconducting meander shaped Nb pickup coils covered with planar Au:Er sensors
\begin{figure}
\centering
\includegraphics[width=0.3\textwidth]{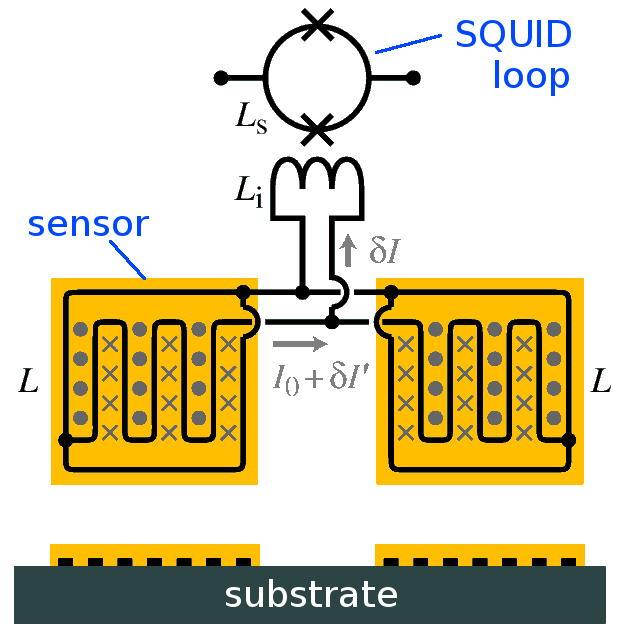}\hspace{10mm}
\includegraphics[width=0.35\textwidth]{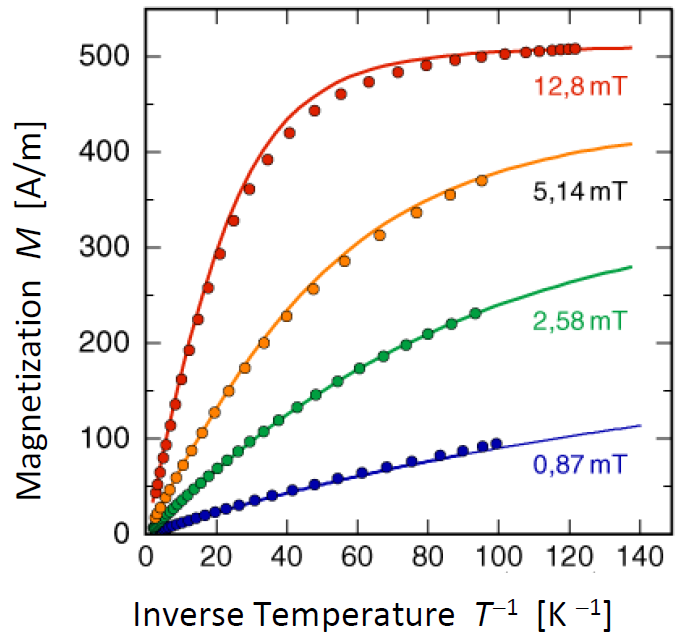}
\caption{Left: Schematic view of a micro-structured MMC type detector.
Right: Dependence of the sensor magnetization on the inverse sensor temperature (figures courtesy of
L.~Gastaldo~\cite{03-Gas11}).
\label{fig-03:mmc-squid}}
\end{figure}
(see figure~\ref{fig-03:mmc-squid}, left). To produce a magnetic field in the sensor material, a persistent current is injected into the
superconducting loop formed by the two meanders. A temperature rise in one of the sensors leads to a decrease in magnetization of the material (see figure~\ref{fig-03:mmc-squid}, right) and consequently to a small current $\delta I$ through the input coil of a SQUID circuit connected for readout in parallel to the meander coils. The sensors, which are operated at temperatures below 100~mK, are coupled via thermalization leads to the heat-sink areas on the chip that act as a thermal bath. Absorber materials containing radioisotopes can be coupled to the Au:Er sensor areas. Energy deposited in the absorber due to a radioactive decay is thermalized and causes a temperature pulse in the sensor.\\
For Holmium EC decay measurements the relatively short half-life of the isotope of about 4570~years makes it possible to work with $^{163}$Ho implanted gold absorbers that are deposited onto the Au:Er sensors. In a measurement with a prototype detector~\cite{03-Ran12} this technique enabled an energy resolution of 12~eV (FWHM) and fast rise times of the signals of about 90~ns.
Figure~\ref{fig-03:ho_spectrum} shows the calorimetric spectrum obtained in that measurement
\begin{figure}
\centering
\includegraphics[width=0.4\textwidth]{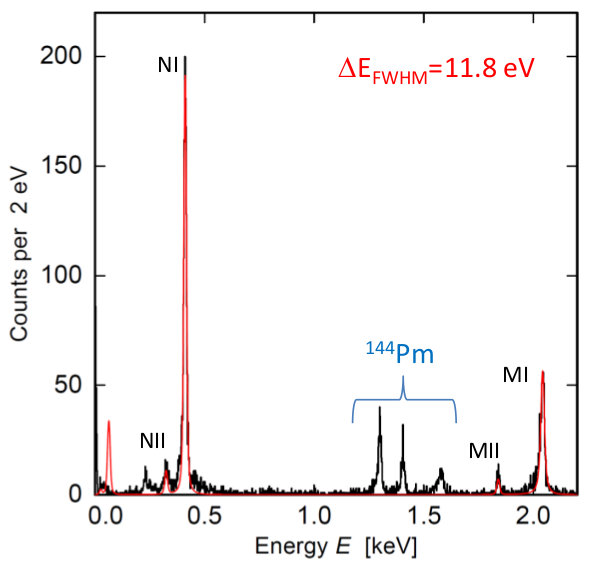}
\caption{Calorimetric energy spectrum for electron capture decays of $^{163}$Ho. The peaks around 1.5~keV originate from $^{144}$Pm decays. $^{144}$Pm was present as an admixture in the implantation beam (figure courtesy of L.~Gastaldo~\cite{03-Gas11}).
\label{fig-03:ho_spectrum}}
\end{figure}
from which a $Q$-value of $(2.80 \pm 0.08)$~keV has been extracted. This is not compatible with the recommended value from atomic mass measurements of $Q = (2.555 \pm 0.016)$~keV~\cite{03-Aud03}, but agrees with other calorimetric measurements~\cite{03-Gat97}.
The reasons for this discrepancy are suspected to be uncertainties in the theoretical parameters required to extract the Q-value (namely the square of the electron wavefunction at the nucleus and exchange and overlap corrections as well as  the width of the peaks $\Gamma_i$). Clearly more investigations are needed to resolve this discrepancy, as the Q-value is crucial for the interpretation of future neutrino mass experiments based on Holmium EC decay.\\
Besides the single pixel performance, where work is ongoing to further improve the energy resolution into the $\Delta E<3$~eV range, the use of MMC detectors in a neutrino mass experiment requires parallel operation and readout of a large number of detector pixels in order to gather the necessary statistics. The ECHO team proposes to read out arrays of MMC detector pixels using frequency domain multiplexing techniques in the microwave region. For this purpose the SQUIDs of the individual detector pixels are coupled to superconducting resonators, each operating at a characteristic frequency between 4 and 8~GHz and connected to a single readout line.\\
In order to set up a competitive neutrino mass experiment based on the EC decay of $^{163}$Ho, efforts are made to improve on the experimental and theoretical aspects of the method. Precision mass measurements using high resolution Penning traps~\cite{03-Bla10} should enable a determination of the Q-value of the decay with eV accuracy. Alternative methods are tested to produce a high purity $^{163}$Ho source, as EC decays of $^{144}$Pm, that was implanted together with the Holmium ions, contributed the largest background to the present measurement.
An improved description of the atomic physics aspects of the decay will be worked on to obtain a more accurate shape of the $^{163}$Ho calorimetric spectrum.
The aim of the these efforts of the ECHO collaboration is to set up a first neutrino mass experiment with sufficiently large detector arrays ($\leq 1000$ sensors) to reach a sensitivity on the electron neutrino mass in the few eV range. If this is successful, a large scale experiment with up to $10^5$ detectors can then be deployed to reach sub-eV sensitivity.
\section{Conclusion}
Direct neutrino mass measurements only rely on kinematic variables in $\beta$-decay
(as well as energy-momentum conservation) to deduce the average electron neutrino
mass \mtwonue\ in a model-independent way. The experimental observable  \mtwonue\
in $\beta$-decay (or EC) is formed by the incoherent sum of the neutrino mass
eigenstates \mnui\,, resulting in a tiny spectral modification in a narrow region close
to \ezero\,, where the emitted neutrino is still non-relativistic. Experimental
challenges in $\beta$-spectroscopy are thus related to obtaining excellent statistics close
to the $\beta$-decay endpoint, favoring $\beta$-emitters with a very short half-life
such as tritium and a low endpoint energy such as \rhenium , tritium and the EC isotope $^{163}$Ho, as well as to maintaining a very small background rate at \ezero\,.
The latter is a non-trivial issue, given that direct neutrino mass experiments are
performed at the surface of the earth, where they are exposed to the full flux
of cosmic rays. Finally, an excellent energy resolution with precisely
known characteristics as well as an excellent control of systematic effects is mandatory.
These requirements have resulted in the development
of two generic experimental techniques.

One the one side, there is the calorimetric approach, where the $\beta$-emitter is embedded into
or identical to the detector, usually operated as a microcalorimeter (absorber materials include AgReO$_4$ crystals or
$^{163}$Ho implanted gold absorbers). This method allows to measure the entire
decay energy, however, the entire $\beta$-spectrum has to be recorded. This
calls for the operation of large arrays of microcalorimeters to circumvent
potential pulse pile-up effects due to the rather slow signal read-out of bolometers.
The main focus in this field is thus targeted at developing new read-out
schemes to improve the energy resolution (down to a few eV) while at the same time improving signal
read-out times (down to a few $\mu$s).

Over the past years substantial progress has been made by several groups (MARE, ECHO) with regard
to improved read-out methods, which now include silicon implanted thermistors,
transition edge sensors (TES) and microwave kinetic inductance detectors (MKID). These
methods are complemented by metallic magnetic calorimeters (MMC), which measure the change
in magnetization. The field is characterized by a rapid progress in this area, so that the most
important decision with regard to a future microcalorimeter array with sub-eV sensitivity will
be to select the most suitable read-out technique for high-precision spectroscopy of
the $\beta$-decay of $^{187}$Re or the EC-process of $^{163}$Ho. The big advantage here is the
possibility to follow a staged approach by continuously enlarging and upgrading the microcalorimeter
array in operation.

On the other side there is the spectrometer approach, where the $\beta$-emitter and the
energy analysis of $\beta$-decay electrons close to the endpoint by a spectrometer
are separated. This approach has been refined by a long list of
tritium $\beta$-experiments. Over the past two decades the Mainz and Troitsk experiments
have pioneered the so-called MAC-E-filter technique, where $\beta$-decay electrons from a gaseous or
quench-condensed source are adiabatically guided to an electrostatic retarding
spectrometer for energy analysis in an integral mode. This technique allows to
combine a source of high intensity with a spectrometer of high energy resolution to
perform superior $\beta$-spectroscopy. This technique has improved
the neutrino mass sensitivity to the present value of 2~eV.

The successor to the Mainz and Troitsk experiments is the large-scale KATRIN project, which is
currently being assembled by an international collaboration at KIT.
The experiment will combine a gaseous molecular tritium
source of highest intensity and stability with a very large electrostatic retarding
spectrometer of unprecedented energy resolution to improve the experimental
sensitivity by one order of magnitude to 200 meV (90\%~C.L.). This sensitivity
is the benchmark for the entire field and will allow investigating almost the
entire parameter space of quasi-degenerated neutrino masses.

Over the past years extensive R\&D work and test measurements by KATRIN groups have resulted in substantial
improvements of the performance of key components (e.g. with regard to source stability and HV stability).
Moreover, a variety of novel background reduction techniques in the electrostatic
spectrometers have been implemented successfully, offering the potential of
measurements almost free of background. An important tool in doing so has been the
KASSIOPEIA code, which allows field calculations and particle tracking with
unprecedented precision and speed. The long-term scanning of the tritium spectrum
will prospectively start at the end of 2015 with first KATRIN sub-eV results shortly
thereafter. At present the experiment is investigating its physics reach in the
search for sterile neutrinos from the sub-eV up to the multi-keV mass regime,
as well as for other physics beyond the Standard Model.

A novel ansatz in $\beta$-spectroscopy is finally pursued by the Project8 collaboration
by developing methods to detect the coherent cyclotron radiation emitted by individual
$\beta$-decay electrons from a gaseous tritium source by a sensitive array of microwave antennae.
The project is also developing other methods to detect this radiation and thus still in the
early stages of R\&D work.

The challenges in further improving the precision in $\beta$-spectroscopy to ultimately push
the neutrino mass sensitivity to the lowest possible value are formidable indeed,
but major advances have already been made in diverse areas such as tritium process control,
cryo-technology, ultra high vacuum methods, precision high voltage, precision electron spectroscopy as well as bolometer read-out technology,
and detector and electronics technology in general. Over the next 5-10 years we can thus
expect high-quality and high-precision neutrino mass results from the large-scale KATRIN experiment,
as well as from other promising techniques, which are exploited in the framework of the MARE, ECHO and Project8 collaborations.

In concluding we would like to emphasize that it is only by comparing high-precision
results from direct neutrino mass measurements with searches for neutrinoless double
$\beta$-decay and cosmological studies that we can obtain the complete picture of neutrino
masses to fully assess the unique r\^{o}le of neutrinos in particle physics and in cosmology.

\section*{Acknowledgments}
This work has been supported in part by the Bundesministerium f\"ur Bildung
und Forschung (BMBF) under contract numbers 05A11PM2 and 05A08VK2, and the
Deutsche Forschungsgemeinschaft (DFG) via Transregio 27 "Neutrinos and beyond"
and Open Access Publishing Fund of KIT. One of us (S.M.) would like to thank
Karlsruhe House of Young Scientists (KHYS) for support.

We would like to express our gratitude and thanks to our colleagues in the ECHO, KATRIN,
MARE, Mainz, Troitsk and Project8 Collaborations, who have given us generous
support and information when preparing this manuscript. Our special thanks go to Ernst Otten,
Markus Steidl and Thomas Th\"ummler.

Finally, we would like to dedicate this article to two of our colleagues,
who died within the last two years and to whom we are very much indebted:
Our colleague Academician Vladimir M. Lobashev of the INR Troitsk, who has pioneered
many techniques described in this review article as head of the Troitsk experiment, and who has
contributed significantly to the progress in this field and others, and our colleague Jochen Bonn from Mainz,
the kind and genius motor in the laboratory of the Mainz experiment,
and one of the most imaginative and smartest colleagues in designing and constructing the KATRIN experiment.


\begin{thebibliography}{99}
%
%
\bibitem{03-atmos} T. Kajita,
  \emph{``Atmospheric Neutrinos''},
  {this issue}

\bibitem{03-solar} V. Antonelli \etal , 
  \emph{``Solar Neutrinos''},
  {this issue}

\bibitem{03-SBL} J. Conrad and M. Shaevitz,
  \emph{``Short baseline neutrino oscillation experiments''},
  {this issue}

\bibitem{03-LBL} J. J. Hartnell, T. Kobayashi and G. Feldman,
  \emph{``Long baseline neutrino oscillation experiments''},
  {this issue}

\bibitem{03-reactor} T. Lasserre, Y. Wang and S.-B. Kim,
  \emph{``Reactor Neutrinos''},
  {this issue}

\bibitem{03-cosmos}  J. Lesgourgues and S. Pastor,
  \emph{``Neutrino mass from Cosmology''},
  {this issue}

\bibitem{03-nature_neutrinos} S. Petcov,
  \emph{``The nature of Massive Neutrinos''}
  {this issue}

\bibitem{03-origin_mass} N.N.,
  \emph{``Origin of neutrino masses and mixings''},
  {this issue}

\bibitem{03-wmap} E. Komatsu \etal ,
  \emph{``Seven-Year Wilkinson Microwave Anisotropy Probe (WMAP) Observations: Cosmological Interpretation''},
  \href{http://dx.doi.org/10.1088/0067-0049/192/2/18}{ApJS 192 (2011) 18}

\bibitem{03-sdss} H. Aihara \etal ,
  \emph{``The Eighth Data Release of the Sloan Digital Sky Survey: First Data from SDSS-III''},
  \href{http://dx.doi.org/10.1088/0067-0049/193/2/29}{ApJS 193 (2011) 29; Erratum ApJS 195 (2011) 26}

\bibitem{03-hannestad12} K.N. Abazajian \etal ,
  \emph{``Cosmological and Astrophysical Neutrino Mass Measurements''},
  \href{http://dx.doi.org/10.1016/j.astropartphys.2011.07.002}{Astropart. Phys. 35 (2011) 177}

\bibitem{03-dbd} A. Giuliani and A. Poves,
  \emph{``Neutrinoless Double Beta Decay''},
  {this issue}

\bibitem{03-klapdor06} H.V. Klapdor-Kleingrothaus and I.V. Krivosheina
  \emph{``The Evidence for the Observation of 0$\nu \beta \beta$ Decay: the Identification of 0$\nu \beta \beta$ Events from the Full Spectra''},
  \href{http://dx.doi.org/10.1142/S0217732306020937}{Mod. Phys. Lett. A 21 (2006) 1547}

\bibitem{03-exo200_2012} M. Auger \etal ,
  \emph{``Search for Neutrinoless Double-Beta Decay in $^{136}$Xe with EXO-200''},
  \href{http://arxiv.org/abs/1205.5608v1}{arXiv:1205.5608v1}

  \bibitem{03-loredo02} T.J. Loredo and D.Q. Lamb,
  \emph{``Bayesian analysis of neutrinos observed from supernova SN 1987A''},
  \href{http://dx.doi.org/0.1103/PhysRevD.65.063002}{Phys. Rev. D65 (2002)  063002}

\bibitem{03-pagliaroli10} G. Pagliarolia, F. Rossi-Torresa and F. Vissania,
  \emph{``Neutrino mass bound in the standard scenario for supernova electronic antineutrino emission''},
  \href{http://dx.doi.org/10.1016/j.astropartphys.2010.02.007}{Astropart. Phys. 33 (2010) 287}

\bibitem{03-robertson_rev88} R.G.H. Robertson and D.A. Knapp,
  \emph{``Direct Measurement of Neutrino Mass''},
  \href{http://dx.doi.org/10.1146/annurev.ns.38.120188.001153}{Ann. Rev. Nucl. Sci. 38 (1988) 185}

\bibitem{03-holzschuh_rev92} E. Holzschuh,
  \emph{``Electron antineutrino mass from beta decay''},
  \href{http://dx.doi.org/doi:10.1088/0034-4885/55/7/004}{Rep. Prog. Phys. 55 (1992) 1035}

\bibitem{03-wilkerson_rev01} J.F. Wilkerson and R.G.H. Robertson,
  \emph{``Direct Measurement of Neutrino Mass''},
  {in ``Current Aspects Of Neutrino Physics,'', ed. D.~O.~Caldwell Springer, Berlin, Heidelberg, 2001, ISBN 978-3-540-41002-7, p. 39}

\bibitem{03-weinheimer_rev03} C. Weinheimer,
  \emph{``Laboratory limits on neutrino masses''},
  {in  ``Massive Neutrinos'', ed. G.~Altarelli and K.~Winter, Springer Tracts in Modern Physics, Springer, 2003, ISBN 3-540-40328-0, p. 25}

\bibitem{03-Otten08} E.W. Otten and C. Weinheimer,
  \emph{``Neutrino mass limit from tritium beta decay''},
  \href{http://dx.doi.org/10.1088/0034-4885/71/8/086201}{Rep. Prog. Phys. 71 (2008) 086201}

\bibitem{03-mount09} B. J. Mount \etal ,
\emph{``Q Value of $^{115}\mathrm{In} \rightarrow ^{115}\mathrm{Sn}(3/2^+)$: The Lowest Known Energy $\beta$ Decay''},
\href{http://dx.doi.org/10.1103/PhysRevLett.103.122502}{Phys. Rev. Lett. 103 (2009) 122502}

\bibitem{03-wieslander09} J. S. E. Wieslander \etal ,
\emph{``Smallest Known Q Value of Any Nuclear Decay: The Rare $\beta^-$ Decay of $^{115} \mathrm{In} (9/2^+) \rightarrow ^{115} \mathrm{Sn} (3/2^+) $''},
\href{http://dx.doi.org/10.1103/PhysRevLett.103.122501}{Phys. Rev. Lett. 103 (2009) 122501}

\bibitem{03-giuliani_rev12} A. Giuliani,
  \emph{``Neutrino Physics with Low-Temperature Detectors''},
  \href{http://dx.doi.org/10.1007/s10909-012-0576-9}{J. Low Temp. Phys. 167 (2012) 991}

\bibitem{03-fogli12} G.L. Fogli \etal ,
  \emph{``Global analysis of neutrino masses, mixings and phases: entering the era of leptonic CP violation searches''},
  \href{http://arxiv.org/abs/1205.5254v3}{arXiv:1205.5254v3}

\bibitem{03-severijns06} N. Severijns \etal ,
  \emph{``Tests of the standard electroweak model in nuclear beta decay''},
  \href{http://dx.doi.org/10.1103/RevModPhys.78.991}{Rev. Mod. Phys. 78 (2006) 991}


\bibitem{03-blaum2010} K. Blaum, Novikov Yu. N. and G. Werth,
\emph{``Penning traps as a versatile tool for precise experiments in fundamental physics''},
\href{http://dx.doi.org/10.1080/00107510903387652}{Contemp. Phys., 51: 2 (2010) 149 — 175}

\bibitem{03-weinheimer93} C. Weinheimer \etal ,
  \emph{``Improved limit on the electron-antineutrino rest mass from tritium $\beta$-decay''},
  \href{http://dx.doi.org/10.1016/0370-2693(93)90355-L}{Phys. Lett. B 300 (1993) 210}

\bibitem{03-masood07}  S.S. Masood  \etal ,
  \emph{``Exact relativistic beta decay endpoint spectrum''},
  \href{http://dx.doi.org/10.1103/PhysRevC.76.045501}{Phys. Rev. C 76 (2007) 045501}

\bibitem{03-repco83} W.W~Repco and C.E.~Wu,
  \emph{``Radiative corrections to the end point of the tritium β decay spectrum''},
  \href{http://dx.doi.org/10.1103/PhysRevC.28.2433}{Phys. Rev. C 28 (1983) 2433}

\bibitem{03-gardner04} S.~Gardner, V.~Bernard and U.G.~Meissner,
  \emph{``Radiative tritium $\beta$-decay and the neutrino mass''},
  \href{http://dx.doi.org/10.1016/j.physletb.2004.08.006}{Phys. Lett. B 598 (2004) 188}

\bibitem{03-stephenson98} G.J.~Stephenson and T.~Goldman,
  \emph{``A possible solution to the tritium endpoint problem''},
  \href{http://dx.doi.org/10.1016/S0370-2693(98)01092-2}{Phys. Lett. B 440 (1998) 89}

\bibitem{03-ignatiev06} A.Yu. Ignatiev and B.H.J. McKellar,
  \emph{``Possible new interactions of neutrino and the KATRIN experiment''},
  \href{http://dx.doi.org/10.1016/j.physletb.2005.11.050}{Phys. Lett. B 633 (2006) 89}

\bibitem{03-simkovic10} R. Dvornicky, F. Simkovic and A. Faessler,
  \emph{``Beyond the Standard Model interactions in $\beta$-decay of tritium''},
  \href{http://dx.doi.org/10.1016/j.ppnp.2009.12.036}{Prog. Part. Nucl. Phys. 64 (2010) 303}

\bibitem{03-bonn11} J. Bonn \etal , 
  \emph{``The KATRIN sensitivity to the neutrino mass and to right-handed currents in beta decay''},
  \href{http://dx.doi.org/10.1016/j.physletb.2011.08.005}{Phys. Lett. B 703 (2011) 310}

\bibitem{03-doss06} N. Doss \etal , 
  \emph{``Molecular effects in investigations of tritium molecule β decay endpoint experiments''},
  \href{http://dx.doi.org/10.1103/PhysRevC.73.025502}{Phys. Rev. C 73 (2006) 025502}

\bibitem{03-doss08} N. Doss and J. Tennyson,
  \emph{``Excitations to the electronic continuum of $^3$HeT$^+$ in investigations of T$_2$ $\beta$-decay experiments''},
  \href{http://dx.doi.org/10.1088/095}{J. Phys. B 41 (2008) 125701}

\bibitem{03-saenz00} A.~Saenz, S.~Jonsell and P.~Froehlich,
  \emph{``Improved Molecular Final-State Distribution of HeT$^+$ for the $\beta$-Decay Process of T$_2$''},
  \href{http://dx.doi.org/10.1103/PhysRevLett.84.242}{Phys. Rev. Lett. 84 (2000) 242}

\bibitem{03-simkovic11} R. Dvornicky \etal , 
  \emph{``The absolute mass of neutrino and the first unique forbidden $\beta$-decay of $^{187}$Re''},
  \href{http://dx.doi.org/10.1103/PhysRevC.83.045502}{Phys. Rev. C 83 (2011) 045502}

\bibitem{03-hirata88} K.S. Hirata \etal ,
  \emph{``Observation in the Kamiokande-II detector of the neutrino burst from supernova SN1987A''},
  \href{http://dx.doi.org/10.1103/PhysRevD.38.448}{Phys. Rev. D 38 (1988) 448}

\bibitem{03-bratton88} C.B. Bratton \etal ,
  \emph{``Angular distribution of events from SN1987A''},
  \href{http://dx.doi.org/10.1103/PhysRevD.37.3361}{Phys. Rev. D37 (1988) 3361}

\bibitem{03-alexeyev87} E.N. Alekseev \etal ,
  \emph{``Possible detection of a neutrino signal on 23 February 1987 at the Baksan underground scintillation telescope of the Institute of Nuclear Research''},
  \href{http://www.jetpletters.ac.ru/ps/1245/article_18825.pdf}{JETP Lett. 45 (1987) 589}

\bibitem{03-raffelt99} G. Raffelt,
  \emph{``Particle Physics from Stars''},
  \href{http://dx.doi.org/10.1146/annurev.nucl.49.1.163}{Ann. Rev. Nucl. Part. Sci. 49 (1999) 163}

\bibitem{03-snews} Webpage of the SuperNova Early Warning System,
  \href{http://snews.bnl.gov}{http://snews.bnl.gov}


\bibitem{03-beacom00} J.F. Beacom, R.N. Boyd and A. Mezzacappa,
  \emph{``Technique for Direct eV-Scale Measurements of the Mu and Tau Neutrino Masses Using Supernova Neutrinos''},
  \href{http://dx.doi.org/10.1103/PhysRevLett.85.3568}{Phys. Rev. Lett. 85 (2000) 3568}

\bibitem{03-Ruj81} A De Rujula,
  \emph{A new way to measure neutrino masses},
  \href{http://dx.doi.org/10.1016/0550-3213(81)90002-X}{Nucl. Phys. B 188 (1981) 414}

\bibitem{03-Spr87} P.T. Springer \etal ,
  \emph{Measurement of the neutrino mass using the inner bremsstrahlung emitted in the electron-capture decay of $^{163}$Ho},
  \href{http://dx.doi.org/10.1103/PhysRevA.35.679}{Phys. Rev. A 35 (1987) 679–689}


\bibitem{03-Yas94} S. Yasumi \etal ,
  \emph{The mass of the electron neutrino from electron capture in $^{163}$Ho},
  \href{http://dx.doi.org/10.1016/0370-2693(94)90616-5}{Phys. Lett. B. 334 (1994) 229}



\bibitem{03-boundstate_bdec92} M.~Jung \etal ,
  \emph{``First observation of bound state Beta- decay''},
  \href{http://dx.doi.org/10.1103/PhysRevLett.69.2164}{Phys. Rev. Lett. 69 (1992) 2164}

\bibitem{03-cur48} S.C. Curran \etal ,
  \emph{``Investigation of Soft Radiations - II. The Beta Spectrum of Tritium''},
  \href{http://dx.doi.org/10.1080/14786444908561210}{Phil. Mag. 40 (1949) 53}

\bibitem{03-itep80} V.A.~Lubimov \etal ,
  \emph{``An estimate of the $\nu_e$ mass from the $\beta$-spectrum of tritium in the valine molecule''},
  \href{http://dx.doi.org/10.1016/0370-2693(80)90873-4}{Phys. Lett. B 94 (1980) 266}

\bibitem{03-itep87} S.~Boris \etal ,
  \emph{``Neutrino Mass from the Beta Spectrum in the Decay of Tritium''},
  \href{http://dx.doi.org/10.1103/PhysRevLett.58.2019}{Phys. Rev. Lett. 58 (1987) 2019}

\bibitem{03-Zuerich86} M.~Fritschi \etal ,
  \emph{``An upper limit for the mass of $\bar \nu_e$ from tritium $\beta$-decay''},
  \href{http://dx.doi.org/10.1016/0370-2693(86)90420-X}{Phys. Lett. B 173 (1986) 485}

\bibitem{03-LANL87}  J.~F.~Wilkerson \etal ,
  \emph{``Limit on anti-$\nu_e$ Mass from Free-Molecular-Tritium Beta Decay''},
  \href{http://dx.doi.org/10.1103/PhysRevLett.58.2023}{Phys. Rev. Lett.  58 (1987) 2023}

\bibitem{03-litmaa85} E.~T.~Lippmaa \etal ,
  \emph{Mass difference of the T--3He doublet and the problem of the rest mass of the electron antineutrino},
  Sov. Phys. Dokl. 30 (1985) 393

\bibitem{03-vanDyck93} R.S. Van Dyck \etal ,
  \emph{``Tritium-–helium-3 mass difference using the Penning trap mass spectroscopy''},
  \href{http://dx.doi.org/10.1103/PhysRevLett.70.2888}{Phys. Rev. Lett. 70 (1993) 2888}

\bibitem{03-nagy06} Sz. Nagy \etal ,
  \emph{``On the Q-value of the tritium $\beta$-decay''},
  \href{http://dx.doi.org/10.1209/epl/i2005-10559-2}{Europhys. Lett. 74 (2006) 404}

\bibitem{03-fleischmann00} L. Fleischmann \etal ,
  \emph{``On dewetting dynamics of solid films of hydrogen isotopes and its influence on tritium $\beta$spectroscopy''},
  \href{http://dx.doi.org/10.1007/s100510070212}{Eur. Phys. J. B 16 (2000) 521}

\bibitem{03-LANL91}     R.G.H. Robertson \etal ,
  \emph{``Limit on anti-$\nu_e$ mass from observation of the $\beta$ decay of molecular tritium''},
  \href{http://dx.doi.org/10.1103/PhysRevLett.67.957}{Phys. Rev. Lett. 67 (1991) 957}

\bibitem{03-Zuerich92}  E. Holzschuh \etal ,
  \emph{``Measurement of the electron neutrino mass from tritium $\beta$-decay''},
  \href{http://dx.doi.org/10.1016/0370-2693(92)91000-Y}{Phys. Lett. B 287 (1992) 381}

\bibitem{03-Tokyo91} H. Kawakami \etal ,
  \emph{``New upper bound on the electron anti-neutrino mass''},
  \href{http://dx.doi.org/10.1016/0370-2693(91)90226-G}{Phys. Lett. B 256 (1991) 105}

\bibitem{03-Bejing95} C.R. Ching \etal ,
  \emph{``A possible explanation of the negative values of $m_{\nu_e}^2$ obtained from the $\beta$ spectrum shape analyses''},
  \href{http://dx.doi.org/10.1142/S0217751X95001340}{Int. J. Mod. Phys. A 10 (1995) 2841}

\bibitem{03-LLNL95} W. Stoeffl and D.J. Decman,
  \emph{``Anomalous Structure in the Beta Decay of Gaseous Molecular Tritium''},
  \href{http://dx.doi.org/10.1103/PhysRevLett.75.3237}{Phys. Rev. Lett. 75 (1995) 3237}

\bibitem{03-weinheimer99} C. Weinheimer \etal ,
  \emph{``High precision measurement of the tritium $\beta$ spectrum near its endpoint and upper limit on the neutrino mass''},
  \href{http://dx.doi.org/10.1016/S0370-2693(99)00780-7}{Phys. Lett. B 460 (1999) 219}

\bibitem{03-kraus05} C. Kraus \etal ,
  \emph{``Final results from phase II of the Mainz neutrino mass search in tritium $\beta$ decay''},
  \href{http://dx.doi.org/10.1140/epjc/s2005-02139-7}{Eur. Phys. Jour. C 40 (2005) 447-468}

\bibitem{03-belesev95} A.I. Belesev \etal ,
  \emph{``Results of the troitsk experiment on the search for the electron antineutrino rest mass in tritium beta-decay''},
  \href{http://dx.doi.org/10.1016/0370-2693(95)00335-I}{Phys. Lett. B 350 (1995) 263}

\bibitem{03-lobashev99}   V.M. Lobashev \etal ,
  \emph{``Direct search for mass of neutrino and anomaly in the tritium beta-spectrum''},
  \href{http://dx.doi.org/10.1016/S0370-2693(99)00781-9}{Phys. Lett. B 460 (1999) 227-235}

\bibitem{03-aseev11} V.N. Aseev \etal ,
 \emph{``Upper limit on the electron antineutrino mass from the Troitsk experiment''},
 \href{http://dx.doi.org/10.1103/PhysRevD.84.112003}{Phys. Rev. D 84 (2011) 112003}

\bibitem{03-kruit83} P. Kruit and F.H. Read,
  \emph{``Magnetic field paralleliser for 2$\pi$ electron-spectrometer and electron-image magnifier''},
  \href{http://dx.doi.org/10.1088/0022-3735/16/4/016}{J. Phys. E 16 (1983) 313}

\bibitem{03-pic92a} A. Picard \etal ,
  \emph{``A solenoid retarding spectrometer with high resolution and transmission for keV electrons''},
  \href{http://dx.doi.org/10.1016/0168-583X(92)95119-C}{Nucl. Instrum. Meth. B 63 (1992) 345}

\bibitem{03-Lob85}  V.M. Lobashev,
  \emph{``The search for the neutrino mass by direct method in the tritium beta-decay and perspectives of study it in the project KATRIN''},
  \href{http://dx.doi.org/10.1016/S0375-9474(03)00985-0}{Nucl. Instr. Meth. A 240 (1985) 305}

\bibitem{03-mueller03} B. M\"uller \etal ,
  \emph{Particle Storage in MAC-E-Filter},
  \href{http://dx.doi.org/10.1016/S0920-5632(03)01371-9}{Nucl. Phys. B (Proc. Suppl.) 118 (2003) 481}

\bibitem{03-KAT04} J. Angrik \etal , (KATRIN Collaboration),
   \emph{``KATRIN Design Report 2004''},
   \href{http://bibliothek.fzk.de/zb/berichte/FZKA7090.pdf}{Wissenschaftliche Berichte, FZ Karlsruhe 7090}

\bibitem{03-bonn99} J. Bonn \etal ,
\emph{A high resolution electrostatic time-of-flight spectrometer with adiabatic magnetic collimation},
\href{http://dx.doi.org/10.1016/S0168-9002(98)01263-7}{Nucl. Instrum. Meth. A 421 (1999) 256}

\bibitem{03-steinbrink12} N. Steinbrink ,
\emph{Simulation of Electron Neutrino Mass Measurements by Time-of-Flight with KATRIN},
Dipl. thesis, University of M\"unster, 2012


\bibitem{03-aseev00} V.N. Aseev \etal ,
 \emph{``Energy Loss of 18~keV Electrons in Gaseous T$_2$ and Quench Condensed D$_2$ Films''},
 \href{http://dx.doi.org/10.1007/s100530050525}{Eur. Phys. J. D 10 (2000) 39}

\bibitem{03-barth_erice97} H. Barth \etal ,
  \emph{``Status and perspectives of the Mainz neutrino mass experiment''},
  \href{http://dx.doi.org/10.1016/S0146-6410(98)00045-3}{Prog. Part. Nucl. Phys. 40 (1998) 353}

\bibitem{03-bornschein03} B. Bornschein \etal ,
  \emph{``Self-Charging of Quench Condensed Tritium Films''},
  \href{http://dx.doi.org/10.1023/A:1022805313162}{J. Low Temp. Phys. 131 (2003) 69}

\bibitem{03-kolos88} W. Kolos \etal ,
  \emph{``Molecular Effects in Tritium Beta Decay. IV. Effect of Crystal Excitations on Neutrino Mass Determination''},
  \href{http://dx.doi.org/10.1103/PhysRevA.37.2297}{Phys. Rev. A 37 (1988) 2297}

\bibitem{03-feldman98} G. J. Feldman and R. D. Cousins,
  \emph{``Unified approach to the classical statistical analysis of small signals''},
  \href{http://dx.doi.org/10.1103/PhysRevD.57.3873}{Phys. Rev. D 57 (1998) 3873}

\bibitem{03-kraus12} C. Kraus \etal , 
  \emph{``Limit on sterile neutrino contribution from the Mainz Neutrino Mass Experiment''}, 
  \href{http://dx.doi.org/10.1140/epjc/s10052-013-2323-z}{Eur. Phys. J. C 73 (2013) 2323}







\bibitem{03-Gal00} M. Galeazzi \etal ,
  \emph{``End-point energy and half-life of the 187Re beta decay''},
  \href{http://dx.doi.org/10.1103/PhysRevC.63.014302}{Phys. Rev. C 63 (2000) 014302}

\bibitem{03-Arn06} C. Arnaboldi \etal ,
  \emph{``Measurement of the p to s Wave Branching Ratio of 187Re beta-Decay from Beta Environmental Fine Structure''},
  \href{http://dx.doi.org/10.1103/PhysRevLett.96.042503}{Phys. Rev. Lett. 96 (2006) 042503}

\bibitem{03-Sis04} M. Sisti \etal ,
  \emph{``New limits from the Milano neutrino mass experiment with thermal microcalorimeters''},
  \href{http://dx.doi.org/10.1016/j.nima.2003.11.273}{Nucl. Inst. and Meth. A 520 (2004) 125-131}

\bibitem{03-Gat01} F. Gatti,
  \emph{``Microcalorimeter measurements''},
  \href{http://dx.doi.org/10.1016/S0920-5632(00)00954-3}{Nucl. Phys. B - Proc. Supp. 91 (2001) 293}

\bibitem{03-Dre05} G. Drexlin \etal ,
  \emph{``KATRIN: Direct Measurement of a sub-eV Neutrino Mass''},
  \href{http://dx.doi.org/10.1016/j.nuclphysbps.2005.04.019}{Nucl. Phys. B (Proc. Suppl.) 145 (2005) 263--267}

\bibitem{03-Men11} G. Mention \etal ,
  \emph{``Reactor antineutrino anomaly''},
  \href{http://dx.doi.org/10.1103/PhysRevD.83.073006}{Phys. Rev. D 83 (2011) 073006}

\bibitem{03-Gal95} GALLEX Collaboration,
  \emph{``GALLEX solar neutrino observations: complete results for GALLEX II''},
  \href{http://dx.doi.org/10.1016/0370-2693(95)00897-T}{Phys.~Lett.~B 357 (1–2) (1995) 237--247}

\bibitem{03-Gal06} J. N. Abdurashitov \etal ,
  \emph{``Measurement of the response of a Ga solar neutrino experiment to neutrinos from a 37Ar source''},
  \href{http://link.aps.org/doi/10.1103/PhysRevC.73.045805}{Phys. Rev. C 73 (2006) 045805}

\bibitem{03-Agu07} A. A. Aguilar-Arevalo \etal ,
  \emph{``Search for Electron Neutrino Appearance at the $\Delta m^2\approx1~eV^2$ Scale''},
  \href{http://link.aps.org/doi/10.1103/PhysRevLett.98.231801}{Phys. Rev. Lett. 98 (2007) 231801}

\bibitem{03-Rii11} A.~Sejersen-Riis and S.~Hannestad,
  \emph{``Detecting sterile neutrinos with KATRIN like experiments''},
  \href{http://iopscience.iop.org/1475-7516/2011/02/011}{JCAP (2011) 1475}

\bibitem{03-For11} J. A. Formaggio and J. Barrett,
  \emph{``Resolving the Reactor Neutrino Anomaly with the KATRIN Neutrino Experiment''},
  \href{http://dx.doi.org/10.1016/j.physletb.2011.10.069}{Phys. Lett. B 706 1 (2011) 68--71}

\bibitem{03-Esm} A.~Esmaili and Orlando L. G. Peres,
  \emph{``KATRIN Sensitivity to Sterile Neutrino Mass in the Shadow of Lightest Neutrino Mass''},
  \href{http://arxiv.org/abs/1203.2632}{arXiv:1203.2632 [hep-ph]}

\bibitem{03-Gon} V.~S.~Basto-Gonzalez \etal , 
  \emph{``Kinematical Test of Large Extra Dimension in Beta Decay Experiments''},
  \href{http://arxiv.org/abs/1205.6212}{arXiv:1205.6212 [hep-ph]}

\bibitem{03-Car00} J.~Carmona and J.~Cortés,
  \emph{``Testing Lorentz invariance violations in the tritium $\beta$-decay anomaly''},
  \href{http://dx.doi.org/10.1016/S0370-2693(00)01182-5}{Phys. Lett. B 494 (2000) 75--80}

\bibitem{03-Fur} D.~Furse \etal ,
  \emph{``KASSIOPEIA - the simulation package for the KATRIN experiment}, to be publ.

\bibitem{03-Glu11} F.~Gl{\"u}ck,
  \emph{``Axisymmetric electric field calculation with zonal harmonic expansion''},
  \href{http://dx.doi.org/10.2528/PIERB11042106}{Prog. In Electromagn. Res. B 32 (2011) 319--350}

\bibitem{03-Glu11b} F.~Gl{\"u}ck,
  \emph{``Axisymmetric magnetic field calculation with zonal harmonic expansion''},
  \href{http://dx.doi.org/10.2528/PIERB11042108}{Prog. In Electromagn. Res. B 32 (2011) 351--388}

\bibitem{03-Ren11} P.~Renschler,
  \emph{``KESS - A new Monte Carlo simulation code for low-energy electron interactions in silicon detectors''},
  \href{http://digbib.ubka.uni-karlsruhe.de/volltexte/1000024959}{PhD thesis, KIT, 2011}

\bibitem{03-For12} J.~Formaggio \etal ,
  \emph{``Solving for micro- and macro-scale electrostatic configuration using the Robin Hood Algorithm''},
  \href{http://www.jpier.org/pierb/pier.php?paper=11112106}{Prog. In Electromagn. Res. B 39 (2012) 1--37}

\bibitem{03-Ver78} J.~H. Verner,
  \emph{``Explicit runge--kutta methods with estimates of the local truncation error''},
  \href{http://dx.doi.org/10.1137/0715051}{SIAM J. on Num. Anal. 15~(4) (1978) 772--790}

\bibitem{03-Pri81} P.~Prince and J.~Dormand,
  \emph{``High order embedded Runge-Kutta formulae''},
  \href{http://dx.doi.org/10.1016/0771-050X(81)90010-3}{J. of Comp. and Appl. Math. 7~(1) (1981) 67--75}

\bibitem{03-Tsi99} C.~Tsitouras and S.~N. Papakostas,
  \emph{``Cheap error estimation for runge--kutta methods''},
  \href{http://dx.doi.org/10.1137/S1064827596302230}{SIAM Journal on Scientific Computing 20~(6) (1999) 2067--2088}



\bibitem{03-Hwa96} W.~Hwang, Y.-K. Kim and M.~E. Rudd,
  \emph{``New model for electron-impact ionization cross sections of molecules''},
  \href{http://dx.doi.org/10.1063/1.471116}{J. of Chem. Phys. 104 (1996) 2956--2966}

\bibitem{03-Tra83} S.~Trajmar, D.~Register and A.~Chutjian,
  \emph{``Electron scattering by molecules ii. experimental methods and data''},
  \href{http://dx.doi.org/10.1016/0370-1573(83)90071-6}{Phys. Rep. 97 (1983) 219--356}

\bibitem{03-Taw90} H.~Tawara \etal , 
  \emph{``Cross sections and related data for electron collisions with hydrogen molecules and molecular ions''},
  \href{http://dx.doi.org/10.1063/1.555856}{J. of Phys. and Chem. Ref. Data 19 (1990) 617--636.}





\bibitem{03-Bab} M.~Babutzka \etal ,
  \emph{``Monitoring of the properties of the KATRIN Windowless Gaseous Tritium Source''},
  \href{http://arxiv.org/abs/1205.5421v1}{http://arxiv.org/abs/1205.5421v1}

\bibitem{03-Kae12} W.~K{\"a}fer,
  \emph{``Sensitivity Studies for the KATRIN experiment''},
  \href{http://digbib.ubka.uni-karlsruhe.de/volltexte/1000026021}{PhD thesis, KIT, 2012}

\bibitem{03-Wan} N.~Wandkowsky \etal ,
  \emph{``Simulation of background from trapped electrons following radon $\alpha$-decays in the KATRIN pre-spectrometer''},
  to be published

\bibitem{03-Pra} M.~Prall \etal ,
  \emph{``The KATRIN Pre-Spectrometer at reduced Filter Energy''},
  \href{http://arxiv.org/abs/1203.2444v1}{acc. for publ. in New J. of Phys.}

\bibitem{03-Bor11} B. Bornschein \etal ,
  \emph{``Between Fusion and Cosmology – the future of the Tritium Laboratory Karlsruhe''},
  \href{http://www.new.ans.org/pubs/journals/fst/a_12604}{Fus. Sci. and Techn. 60 (2011) 1088--1091}



\bibitem{03-Glu12} F. Gl\"uck,
  \emph{``Tritium Gas flow in the KATRIN source tube},
  to be published

\bibitem{03-Ass00} V.~N.~Aseev \etal ,
  \emph{`` Energy loss of 18 keV electrons in gaseous T and quench condensed D films''},
  \href{http://dx.doi.org/10.1007/s100530050525}{Eur. Phys. J. D 10 (2000) 39}

\bibitem{03-Dus62} S. Dushman,
  \emph{``Scientific Foundations of Vacuum Technique''}
  John Wiley \& Sons, New York, US. (1962)


\bibitem{03-Gro08} S. Grohmann \etal ,
  \emph{``Cryogenic design of the KATRIN source cryostat''},
  \href{http://dx.doi.org/10.1063/1.2908483 }{AIP Conf. Proc. 985 (2008) 1277}



\bibitem{03-Gro11} S.~Grohmann \etal , 
  \emph{``Precise temperature measurement at 30 K in the KATRIN source cryostat''},
  \href{http://dx.doi.org/10.1016/j.cryogenics.2011.05.001}{Cryogenics 51, 8 (2011) 438-445 }




\bibitem{03-Lon02} D.A. Long,
  \emph{``The Raman effect: A unified treatment of the theory of Raman scattering by molecules''},
  Wiley, Chichester, UK. (2002)


\bibitem{03-Lew08} R.J. Lewis \etal ,
  \emph{``Dynamic Raman spectroscopy of hydrogen isotopomer mixtures in-line at TILO''},
  \href{http://onlinelibrary.wiley.com/doi/10.1002/lapl.200810026/abstract}{Laser Phys. Lett. 5 522--531}

\bibitem{03-Stu10}M. Sturm \etal ,
  \emph{``Monitoring of all hydrogen isotopologues at tritium laboratory Karlsruhe using Raman spectroscopy''},
  \href{http://dx.doi.org/10.1134/S1054660X10030163}{Laser Phys. 20 (2010) 493-507}

\bibitem{03-Sch11} M.~Schl{\"osser} \etal ,
  \emph{``Design Implications for Laser Raman Measurement Systems for Tritium Sample-Analysis, Accountancy or Process-Control Applications''},
  \href{http://www.new.ans.org/pubs/journals/fst/a_12579}{Fus. Sci. and Techn. 60 3 976--981}

\bibitem{03-Fis11} S. Fischer \etal ,
  \emph{``Monitoring of Tritium Purity During Long-Term Circulation in the KATRIN Test Experiment LOOPINO Using Laser Raman Spectroscopy''},
  \href{http://www.new.ans.org/pubs/journals/fst/a_12567}{Fus. Sci. and Techn. 60 3 925--930}


\bibitem{03-beck12} M. Beck \etal ,
\emph{An angular selective electron source for the KATRIN experiment},
 to be published

\bibitem{03-Val09} K. Valerius \etal ,
 \emph{A UV LED-based fast-pulsed photoelectron source for time-of-flight studies},
{New J.\ Phys.\  {\bf 11} (2009) 063018}

\bibitem{03-hugenberg10} K.~Hugenberg  [KATRIN Collaboration],
\emph{An Angular Resolved Pulsed Uv Led Photoelectron Source For Katrin},
{Prog.\ Part.\ Nucl.\ Phys.\  64 (2010) 288}

\bibitem{03-Val11} K. Valerius \etal ,
 \emph{``Prototype of an angular-selective photoelectron calibration source for the KATRIN experiment''},
 \href{http://dx.doi.org/10.1088/1748-0221/6/01/P01002}{J. of Instr. 6 (2011) P01002}


\bibitem{03-Nas05} A. R. Nastoyashchii \etal ,
  \emph{``Effects of Plasma Phenomena on Neutrino Mass Measurements Process Using a Gaseous Tritium $\beta$-Source''},
  \href{http://www.new.ans.org/pubs/journals/fst/a_1028}{Fus. Sci. and Techn. 48 (2005) 743}

\bibitem{03-Luk11}S. Lukic \etal ,
  \emph{``Measurement of the gas-flow reduction factor of the KATRIN DPS2-F differential pumping section''},
  \href{http://dx.doi.org/10.1016/j.vacuum.2011.10.017}{Vacuum 86~(8)(2012) 1126--1133.}

\bibitem{03-Luo06} X.~Luo \etal ,
  \emph{``Monte Carlo simulation of gas flow through the KATRIN DPS2-F differential pumping system''},
  \href{http://dx.doi.org/10.1016/j.vacuum.2005.11.044}{Vacuum 80~(8) (2006) 864--869}



\bibitem{03-Gil10} W.~Gil \etal ,
  \emph{``The Cryogenic Pumping Section of the KATRIN Experiment''},
  \href{http://dx.doi.org/10.1109/TASC.2009.2038581}{Appl. Supercond., IEEE Trans. on Appl. supercod. 20~(3) (2010) 316--319}

\bibitem{03-Luo08} X.~Luo and C.~Day,
  \emph{''Test particle Monte Carlo study of the cryogenic pumping system of the Karlsruhe tritium neutrino experiment''},
  \href{http://dx.doi.org/10.1116/1.2956628}{J. of Vac. Sci. and Techn. A 26 (2008) 1319--1325}

\bibitem{03-Kos12} A.~Kosmider,
  \emph{``Tritium Retention Techniques in the KATRIN Transport Section and Commissioning of its DPS2-F Cryostat''},
   {PhD thesis, KIT, 2012}

\bibitem{03-Wol11} J. Wolf \etal ,
  \emph{``Investigation of turbo-molecular pumps in strong magnetic fields''},
  \href{http://dx.doi.org/10.1016/j.vacuum.2011.07.063}{Vacuum 86 (2011) 361-369}

\bibitem{03-Ubi09} M.~Ubieto-D{\'i}az \etal ,
  \emph{``A broad-band FT-ICR Penning trap system for KATRIN''},
  \href{http://dx.doi.org/10.1016/j.ijms.2009.07.003}{Int. J. of Mass Spect. 288 (2009) 1--5}


\bibitem{03-Gil11} W. Gil \etal,
 \emph{``Status of the Magnets of the Two Tritium Pumping Sections for KATRIN''},
  \href{http://dx.doi.org/10.1109/TASC.2011.2175353} {IEEE Trans. on Appl. Supercond. 22 (2012)}

\bibitem{03-kaz07} O. Kazachenko \etal ,
\emph{``TRAP - a cryo-pump for pumping tritium on pre-condensed argon}
O. Kazachenko et al.,
\href{http://dx.doi.org/10.1016/j.nima.2007.12.024}{Nucl. Instrum. Meth. A 587 (2008) 136}

\bibitem{03-Ost09} B.~Ostrick,
  \emph{``Eine kondensierte $^{83m}$Kr-Kalibrationsquelle f\"ur das KATRIN-Experiment''},
  \href{http://miami.uni-muenster.de/servlets/DerivateServlet/Derivate-5046/diss_ostrick.pdf}{PhD thesis in German language, University of M\"unster, 2009}

\bibitem{03-Mer12} S.~Mertens,
  \emph{``Study of Background Processes in the Electrostatic Spectrometers of the KATRIN Experiment''},
  \href{http://digbib.ubka.uni-karlsruhe.de/volltexte/1000027058}{PhD thesis, KIT, 2012}

\bibitem{03-Fra10} F.~Fr{\"a}nkle,
  \emph{``Background Investigations of the KATRIN Pre-Spectrometer''},
  \href{http://digbib.ubka.uni-karlsruhe.de/volltexte/1000019392}{PhD thesis, KIT, 2010}

\bibitem{03-osipowicz12} A. Osipowicz \etal ,
\emph{A mobile magnetic sensor unit for the KATRIN main spectrometer},
\href{http://dx.doi.org/10.1088/1748-0221/7/06/T06002}{JINST 7 (2012) T06002}

\bibitem{03-KAT} KATRIN Webpage,
  \href{http://www.katrin.kit.edu/}{http://www.katrin.kit.edu/}

\bibitem{03-dipl_schall01} J.-P. Schall,
  \emph{Untersuchungen zu Untergrundprozessen am Mainzer Neutrinomassenexperiment},
  Dipl. Thesis in German language, Mainz University, 2001

\bibitem{03-dipl_ostrick02} B. M\"uller,
  \emph{Umbau des Mainzer Neutrinomassenexperiments und Untergrunduntersuchungen im Hinblick auf KATRIN},
  Dipl. Thesis in German language, Mainz University, 2002

\bibitem{03-phd_flatt05} B. Flatt,
  \emph{Voruntersuchungen zu den Spektrometern des KATRIN-Experiments},
  PhD Thesis in German language, Mainz University, 2005

\bibitem{03-Val06} K.~Valerius,
  \emph{``Electromagnetic design and inner electrode for the KATRIN main spectrometer''},
  \href{http://dx.doi.org/10.1016/j.ppnp.2005.11.011} {Progr. Part. and Nucl. Phys. 57~(1) (2006) 58--60}

\bibitem{03-Val10} K.~Valerius,
  \emph{``The wire electrode system for the KATRIN main spectrometer''},
  \href{http://dx.doi.org/10.1016/j.ppnp.2009.12.032} {Progr. Part. and Nucl. Phys. 64(2) (2010) 291--293}

\bibitem{03-Pra12b} M.~Prall \etal ,
  \emph{``The wire electrode system for the KATRIN main spectrometer''},
  to be published

\bibitem{03-Ott06} E.~W. Otten, J. Bonn and C. Weinheimer,
  \emph{``The Q-value of tritium β-decay and the neutrino mass''},
  \href{http://dx.doi.org/10.1016/j.ijms.2006.01.035}{International Journal of Mass Spectrometry 251 Iss. 2-3 (2006) 173-178}

\bibitem{03-Kas04} J. Kaspar \etal ,
  \emph{``Effect of energy scale imperfections on results of neutrino mass measurements from $\beta$-decay},
  \href{http://dx.doi.org/10.1016/j.nima.2004.03.201}{Nucl. Instr. and Meth. A 527 (2004) 423}

\bibitem{03-Thu09} T. Th\"ummler \etal ,
  \emph{``Precision high voltage divider for the KATRIN experiment''},
  \href{http://dx.doi.org/10.1088/1367-2630/11/10/103007}{New J. Phys. 11 (2009) 103007}

\bibitem{03-ven10} D. Venos \etal ,
  \emph{Development of a super-stable datum point for monitoring the energy scale of electron spectrometers in the energy range up to 20 keV},
  \href{http://dx.doi.org/10.1007/s11018-010-9501-2}{Measurement Techniques 53 (2010) 305}

\bibitem{03-zboril11} M. Zbo\v{r}il,
  \emph{Solid electron sources for the energy scale monitoring in the KATRIN experiment},
  \href{http://miami.uni-muenster.de/servlets/DerivateServlet/Derivate-6323/diss_zboril.pdf}{PhD thesis, University of M\"unster, 2011}

\bibitem{03-ven05} D. Venos  \etal ,
\emph{Kr-83m radioactive source based on Rb-83 trapped in cation-exchange paper or in zeolite},
\href{http://dx.doi.org/10.1016/j.apradiso.2005.04.011}{Appl. Rad. Isotopes 63 (2005) 323}

\bibitem{03-rasulbaev08} M. Rasulbaev \etal ,
\emph{Production of Rb-83 for the KATRIN experiment},
\href{http://dx.doi.org/10.1016/j.apradiso.2008.04.020}{Appl. Rad. Isotopes, 66 (2008) 1838}

\bibitem{03-ven06} D. Venos \etal ,
\emph{Precise energy of the weak 32-keV gamma transition observed in Kr-83m decay},
\href{http://dx.doi.org/10.1016/j.nima.2005.12.213}{Nucl. Instrum. Meth. A 560 (2006) 352}

\bibitem{03-marx01} R. Marx,
\emph{New concept of PTBs standard divider for direct voltages of up to 100 kV},
\href{http://dx.doi.org/10.1109/19.918158}{IEEE Trans. Instrum. Meas. 50 (2001) 426}

\bibitem{03-krieger11} A. Krieger \etal ,
\emph{Calibration of the ISOLDE acceleration voltage using a high-precision
  voltage divider and applying collinear fast beam laser spectroscopy},
\href{http://dx.doi.org/10.1016/j.nima.2010.12.145}{Nucl.\ Instrum.\ Meth.\  A {\bf 632} (2011) 23}

\bibitem{03-auger10} J. Abraham \etal ,
\emph{The fluorescence detector of the Pierre Auger Observatory},
\href{http://dx.doi.org/10.1016/j.nima.2010.04.023,}{Nucl. Instum. Meth. A 620 (2010) 227}

\bibitem{03-How} M. Howe and J. Wilkerson,
  \href{http://orca.physics.unc.edu/~markhowe/Orca_Help/Home.html}{ORCA webpage at UNC}

\bibitem{03-vanDevender12} B.A. VanDevender \etal ,
\emph{Performance of a TiN-coated monolithic silicon pin-diode array under mechanical stress},
\href{http://dx.doi.org/10.1016/j.nima.2012.01.033}{Nucl. Instrum. Meth. A 673 (2012) 46}



\bibitem{03-Dos06} N.~Doss \etal ,
  \emph{``Molecular effects in investigations of tritium molecule $\beta$-decay endpoint experiments''},
  \href{http://prc.aps.org/abstract/PRC/v73/i2/e025502}{Phys. Rev. C 73 (2006) 025502}

\bibitem{03-Dos08} N.~Doss and J.~Tennyson,
  \emph{``Excitations to the electronic continuum of $3\mathrm{HeT}^+$ in investigations of the $\mathrm{T}_2$ beta-decay experiments''},
  \href{http://dx.doi.org/10.1088/0953-4075/41/12/125701}{J. Phys. B 41 (2008) 125701}

\bibitem{03-Veg} H.J. de Vega \etal ,
  \emph{``Search of keV Sterile Neutrino Warm Dark Matter in the Rhenium and Tritium beta decays''},
  \href{http://arxiv.org/abs/1109.3452}{http://arxiv.org/abs/1109.3452}

\bibitem{03-Zacher12} M. Zacher,
\emph{Design of the high-field region of the KATRIN spectrometers and a pulsed angular-selective UV laser photoelectron source for investigating their transmission functions},
PhD thesis, University of M\"unster, expected to be finished in 2012

\bibitem{03-Mer10} S.~Mertens (for the KATRIN Collab.),
  \emph{``Electromagnetic design of the spectrometer section of the KATRIN experiment''},
  \href{http://dx.doi.org/10.1016/j.ppnp.2009.12.033}{Prog. in Part. and Nucl. Phys. 64(2) (2010) 294--296}

\bibitem{03-Bec10} M.~Beck \etal ,
  \emph{``Effect of a sweeping conductive wire on electrons stored in a Penning-like trap between the KATRIN spectrometers''},
  \href{http://dx.doi.org/10.1140/epja/i2010-10959-1}{Eur. Phys. J. A 44 (2010) 499}

\bibitem{03-Hillen11} B. Hillen ,
  \emph{Untersuchung von Methoden zur Unterdr\"uckung des Spektrometeruntergrunds beim KATRIN Experiment},
  \href{http://miami.uni-muenster.de/servlets/DerivateServlet/Derivate-6261/diss_hillen.pdf}{PhD thesis in German language, University of M\"unster, 2011}

\bibitem{03-Fra11} F.M.~Fr\"ankle \etal ,
  \emph{``Radon induced background processes in the KATRIN pre-spectrometer''},
  \href{http://dx.doi.org/10.1016/j.astropartphys.2011.06.009}{Astropart. Phys. 35 (2011) 128-134}

\bibitem{03-Mer12b} S.~Mertens \etal ,
  \emph{``Background due to stored electrons following nuclear decays in the KATRIN spectrometers and its impact on the neutrino mass sensitivity''},
  \href{http://dx.doi.org/10.1016/j.astropartphys.2012.10.005}{Astropart.Phys. 41 (2013) 52-62}

\bibitem{03-Mer12c} S.~Mertens \etal ,
  \emph{``Stochastic Heating by ECR as a Novel Means of Background Reduction in the KATRIN Spectrometers"},
  \href{http://dx.doi.org/10.1088/1748-0221/7/08/P08025}{JINST 7 (2012) P08025}

\bibitem{03-Jer10} M.~Jerkins \etal ,
  \emph{``Using cold atoms to measure neutrino mass''},
  \href{http://dx.doi.org/10.1088/1367-2630/12/4/043022}{New J. Phys. 12 (2010) 043022}

\bibitem{03-Ott11} E.W. ~Otten,
  \emph{``Comment on 'Using cold atoms to measure neutrino mass' ''},
  \href{http://dx.doi.org/10.1088/1367-2630/13/7/078001}{New J. Phys. 13 (2011) 078001}

  \bibitem{03-Mon09} B. Monreal and J. Formaggio,
  \emph{``Relativistic cyclotron radiation detection of tritium decay electrons as a new technique for measuring the neutrino mass''},
  \href{http://dx.doi.org/10.1103/PhysRevD.80.051301}{Phys. Rev. D 80 (2009) 051301(R)}

\bibitem{03-For11a} J. Formaggio,
  \emph{``Project 8: Using Radio-Frequency Techniques to Measure Neutrino Mass''},
  \href{http://de.arxiv.org/abs/1101.6077v1}{arXiv:1101.6077v1 [nucl-ex] (2011)}

\bibitem{03-Nuc08} A.~Nucciotti,
  \emph{``The MARE Project''},
  \href{http://dx.doi.org/10.1007/s10909-008-9718-5}{J. of Low Temp. Phys. 151 (2008) 597--602}

\bibitem{03-Gal12} M. Galeazzi \etal ,
  \emph{The Electron Capture Decay of $^{163}$Ho to measure the Electron Neutrino Mass with sub-eV Accuracy},
  \href{http://arxiv.org/abs/1202.4763v3}{arXiv:1202.4763 [physics.ins-det] (2012)}

\bibitem{03-Fer12} E. Ferri,
  \emph{``MARE-1 in Milan: Status and Perspectives''},
  \href{http://dx.doi.org/10.1007/s10909-011-0421-6}{J. of Low Temp. Phys. 167 (2012) 1035}

\bibitem{03-Ran12} P.C.-O. Ranitzsch \etal ,
  \emph{``Development of metallic magnetic calorimeters for high precision measurements of calorimetric $^{187}$Re and $^{163}$Ho spectra''},
  \href{http://dx.doi.org/10.1007/s10909-012-0556-0}{J. of Low Temp. Phys. 167 (2012) 1004}

\bibitem{03-Kra08} S. Kraft-Bermuth \etal ,
  \emph{``Development and Characterization of Microcalorimeters for a Next Generation 187Re Beta-Decay Experiment''},
  \href{http://dx.doi.org/10.1007/s10909-008-9715-8}{J. of Low Temp. Phys. 151 (2008) 619-622}

\bibitem{03-Nuc11} A. Nucciotti for the MARE collaboration,
  \emph{The MARE experiment and its capabilities to measure the mass of light (active) and heavy (sterile) neutrinos},
  presentation at the Workshop CIAS Meudon 2011


\bibitem{03-Fav12} M. Faverzani \etal ,
  \emph{``Developments of Microresonators Detectors for Neutrino Physics in Milan''},
  \href{http://dx.doi.org/10.1007/s10909-012-0538-2}{J. of Low Temp. Phys. 167 (2012) 1041}

\bibitem{03-Vac08} R. Vaccarone \etal ,
  \emph{``The Design of a Frequency Multiplexed Ir-Au TES Array''},
  \href{http://dx.doi.org/10.1007/s10909-008-9768-8}{J. of Low Temp. Phys. 151 (2008) 921-926}

\bibitem{03-Gas09} L. Gastaldo \etal ,
  \emph{``Low Temperature Magnetic Calorimeters For Neutrino Mass Direct Measurement''},
  \href{http://dx.doi.org/10.1063/1.3292415}{AIP Conf. Proc. 1185 (2009) 607-611}

\bibitem{03-Pie12} C. Pies \etal ,
  \emph{``maXs: Microcalorimeter Arrays for High-Resolution X-Ray Spectroscopy at GSI/FAIR''},
  \href{http://dx.doi.org/10.1007/s10909-012-0557-z}{J. Low Temp. Phys. 167 (2012) 269}

\bibitem{03-Lus11} M. Lusignoli and M. Vignati,
  \emph{``Relic antineutrino capture on 163Ho decaying nuclei''},
  \href{http://dx.doi.org/10.1016/j.physletb.2011.01.047}{Phys. Lett. B 697 (2011) 11–14}

\bibitem{03-Ben81} C.L. Bennet \etal ,
  \emph{The X-ray spectrum following $^{163}$Ho $M$ electron capture},
  \href{http://dx.doi.org/10.1016/0370-2693(81)91137-0}{Phys. Lett. B 107 (1981) 19-22}

\bibitem{03-Rav84} H. L. Ravn,
  \emph{The N/M electron-capture ratio of the neutrino-mass probe $^{163}$Ho},
  in \emph{Massive Neutrinos in Astrophysics and in Particle Physics}, Proceedings of the Fourth Moriond Workshop, La Plagne-Savoie-France (1984) 287-294

\bibitem{03-Har92} F.X. Hartmann and R.A. Naumann,
  \emph{High temperature gas proportional detector techniques and application to the neutrino mass limit using $^{163}$Ho},
  \href{http://dx.doi.org/10.1016/0168-9002(92)90102-A}{Nucl. Instr. and Meth. A 313 (1992) 237}

\bibitem{03-Gat97} F. Gatti \etal ,
  \emph{``Calorimetric measurement of the 163Ho spectrum by means of a cryogenic detector''},
  \href{http://dx.doi.org/10.1016/S0370-2693(97)00239-6}{Phys. Lett. B 398 (1997) 415}



\bibitem{03-Bur08} A. Burck \etal ,
  \emph{``Microstructured Magnetic Calorimeter with Meander-Shaped Pickup Coil''},
  \href{http://dx.doi.org/10.1007/s10909-007-9659-4}{J. of Low Temp. Phys. 151 (2008) 337}

\bibitem{03-Gas11} L. Gastaldo \etal ,
  \emph{163-Ho electron capture decay: high precision measurement of the calorimetric spectrum},
  presentation at the spring meeting of the DPG in Karlsruhe, 2011

\bibitem{03-Aud03} G. Audi \etal,
  \emph{``The Ame2003 atomic mass evaluation: (II). Tables, graphs and references''},
  \href{http://dx.doi.org/10.1016/j.nuclphysa.2003.11.003}{Nucl. Phys. A 729 (2003) 337}

\bibitem{03-Bla10} K. Blaum, Yu. N. Novikov and G. Werth,
  \emph{``Penning traps as a versatile tool for precise experiments in fundamental physics''},
  \href{http://dx.doi.org/10.1080/00107510903387652}{Contemp. Phys. 51 (2010) 149}


\end{thebibliography}
\end{document}